\documentclass{aa} 

\usepackage{graphicx}
\usepackage{txfonts}
\usepackage{xcolor}
\usepackage{tabularx}
\usepackage{chemformula}
\usepackage{longtable}
\usepackage{xspace}

\usepackage[pdfpagelabels=false]{hyperref}	% Hyperlinks
\hypersetup{colorlinks=true,linkcolor=blue,citecolor=blue,filecolor=blue,urlcolor=blue,}
\newcommand{\nodata}{--}
\newcommand{\teff}{$T_\mathrm{eff}$}
\newcommand{\logg}{$\log{g}$}
\newcommand{\bear}{\textsc{BeAR}\xspace}
\newcommand{\fc}{\textsc{FastChem}\xspace}

\begin{document} 

   %\title{Atmospheric Retrieval Study Continued: T-Y Sequence}
   \title{Clouds and Chemistry Across the Brown Dwarf T–Y Sequence: Insights from JWST Atmospheric Retrievals}
   %\titlerunning{Atmospheric Retrieval Study Continued: T-Y Sequence}
   \titlerunning{Clouds and Chemistry Across the Brown Dwarf T–Y Sequence: Insights from JWST Atmospheric Retrievals}
   \authorrunning{Lueber et al.}

   \author{A. Lueber\inst{1,2}, D. Kitzmann\inst{3,2}, and K. Heng\inst{1,4,5,6}
          }

   \institute{Faculty of Physics, Ludwig Maximilian University, Scheinerstrasse 1, D-81679, Munich, Bavaria, Germany,
    \and Center for Space and Habitability, University of Bern, Gesellschaftsstrasse 6, CH-3012 Bern, Switzerland,\\
    \email{anna.lueber@unibe.ch}
    \and Space Research and Planetary Sciences, Physics Institute, University of Bern, Gesellschaftsstrasse 6, 3012 Bern,
    \and ARTORG Center for Biomedical Engineering Research, University of Bern, Murtenstrasse 50, CH-3008, Bern, CH,
    \and University College London, Department of Physics \& Astronomy, Gower St, London, WC1E 6BT, United Kingdom,
    \and Astronomy \& Astrophysics Group, Department of Physics, University of Warwick, Coventry CV4 7AL, United Kingdom
    }

   \date{Received ; accepted}

  \abstract{The James Webb Space Telescope (JWST) offers exceptional spectral resolution and wavelength coverage, which are essential for studying the coldest brown dwarfs, particularly Y dwarfs. These objects are at the cold end of the sub-stellar sequence and exhibit atmospheric phenomena such as cloud formation, chemical disequilibrium, and radiative-convective coupling. %Building on the work of \cite{Lueber2022ApJ...930..136L},
  We examine a curated sample of 22 late-T to Y dwarfs through Bayesian atmospheric retrieval (nested sampling) and supervised machine learning (random forests). Bayesian model comparison indicates that cloud-free models are generally favored for the hottest objects in the sample (T6--T8). Conversely, later-type dwarfs exhibit varying preferences, with both gray-cloud and cloud-free models providing comparable fits. The atmospheric parameters retrieved are consistent across the applied methodologies. Evidence of vertical mixing and disequilibrium chemistry is found in several objects; notably, the Y1 dwarf WISEPAJ1541-22 favors a gray cloud model and shows elevated abundances of both \ch{CO} and \ch{CO2} compared to equilibrium chemistry calculations. As anticipated, the abundances of \ch{H2O}, \ch{CH4}, and \ch{NH3} increase with decreasing effective temperature over the T–Y sequence.}


  \maketitle

\section{Introduction}

The study of brown dwarfs has been greatly facilitated by accumulating a large amount of spectral data in recent years. Numerous observational campaigns have been conducted using ground-based and space-based telescopes, resulting in a wealth of information about the spectra of these objects (see e.g., \citealt{Kirkpatrick2021ApJS}). This large amount of data motivated several population studies intending to gain a holistic understanding of the atmospheres of brown dwarfs \citep{Line2015ApJ, Line2017ApJ...848...83L, Burningham2017MNRAS, Zalesky2019ApJ...877...24Z, Zalesky2022ApJ...936...44Z, Gonzales2020ApJ...905...46G, Gonzales2022ApJ...938...56G, Lueber2022ApJ...930..136L, Calamari2024ApJ...963...67C}. However, these efforts have so far been constrained by the existing limitations in data quality and near-infrared wavelength coverage. In the investigation conducted by \citet{Lueber2022ApJ...930..136L}, a comprehensive examination of brown dwarfs spanning the entire L-T spectral sequence was carried out using 19 SpeX Prism spectral standards \citep{Burgasser2014ASInC..11....7B}. Despite thorough analysis, some important research questions remained unanswered. For instance, the absence of a preference for cloudy models in the case of L dwarfs was unexpected, with both cloud-free and cloudy models providing to be equally consistent with the archival SpeX data from the perspective of Bayesian model comparison. A primary contributing factor to this ambiguity could be attributed to the restricted wavelength coverage of SpeX spectroscopy.

The launch of the \textit{James Webb Space Telescope} (JWST, \citealt{JWST2023PASP..135f8001G}), has now helped to shift these constraints. The instruments onboard JWST enable a major advance in spectral resolution and wavelength coverage, which are crucial for the coldest brown dwarfs due to their very low temperatures and faintness in the near-infrared. This study aims to expand the analysis performed by \cite{Lueber2022ApJ...930..136L} to the colder Y dwarf regime, which represents the coldest and least understood members of the brown dwarf population. Therefore, we utilize the curated data set of \cite{Beiler2024ApJ...973..107B}, who observed 23 brown dwarfs with the near-infrared spectrograph (NIRSpec, \citealt{NIRSpec2022A&A...661A..80J}) and the mid-infrared instrument (MIRI, \citealt{MIRI2015PASP..127..584R}) of the JWST and collected low-resolution spectroscopy with both instruments and broadband photometry with MIRI (GO 2302, PI: Cushing). 

Y dwarfs are crucial for understanding the lower temperature limits of sub-stellar atmospheres, where the interplay of complex processes like cloud formation, chemical disequilibrium, and radiative-convective dynamics become dominant \citep{Cushing2011ApJ...743...50C, Kirkpatrick2012ApJ...753..156K, Morley2012ApJ...756..172M, Morley2014ApJ...787...78M, Lacy2023ApJ...950....8L, Leggett2023ApJ...959...86L, Leggett2024arXiv240906158L}. To address these complexities, we utilize standard Bayesian nested-sampling atmospheric retrieval techniques \citep{Kitzmann2020ApJ...890..174K} across the T-Y spectral sequence, analogous to the approach taken by \cite{Lueber2022ApJ...930..136L}. Additionally, we compare our nested-sampling retrievals with those performed using the supervised machine learning method of the random forest \citep{Marquez2018NatAs...2..719M}. This method offers a data-driven approach that complements traditional Bayesian analysis, allowing us to explore the predictive power of different atmospheric features in a non-parametric fashion.

Machine learning approaches such as random forests facilitate an additional layer of analysis through the concept of "feature importance" \citep{Marquez2018NatAs...2..719M}, which quantifies the relative contribution of each input feature--in this case, each data point across wavelength--to the prediction of a given model parameter. We present an analysis of feature importance using a custom model grid constructed for this study. By evaluating which wavelengths contribute most significantly to model parameter predictions (e.g., temperature or chemical abundances), we gain deeper insights into the physical processes governing Y dwarf atmospheres. This feature importance analysis helps highlight key factors that may have been overlooked in prior studies, especially when dealing with the more complex, cooler atmospheric conditions found in Y dwarfs. Thus, by combining the Bayesian and random forest retrieval analysis, we aim to resolve some of the ambiguities identified in earlier work and contribute to a more comprehensive understanding of the sub-stellar population.

\section{Spectral Sample}

\begin{figure}[h]
    \centering
    \includegraphics[width=\linewidth]{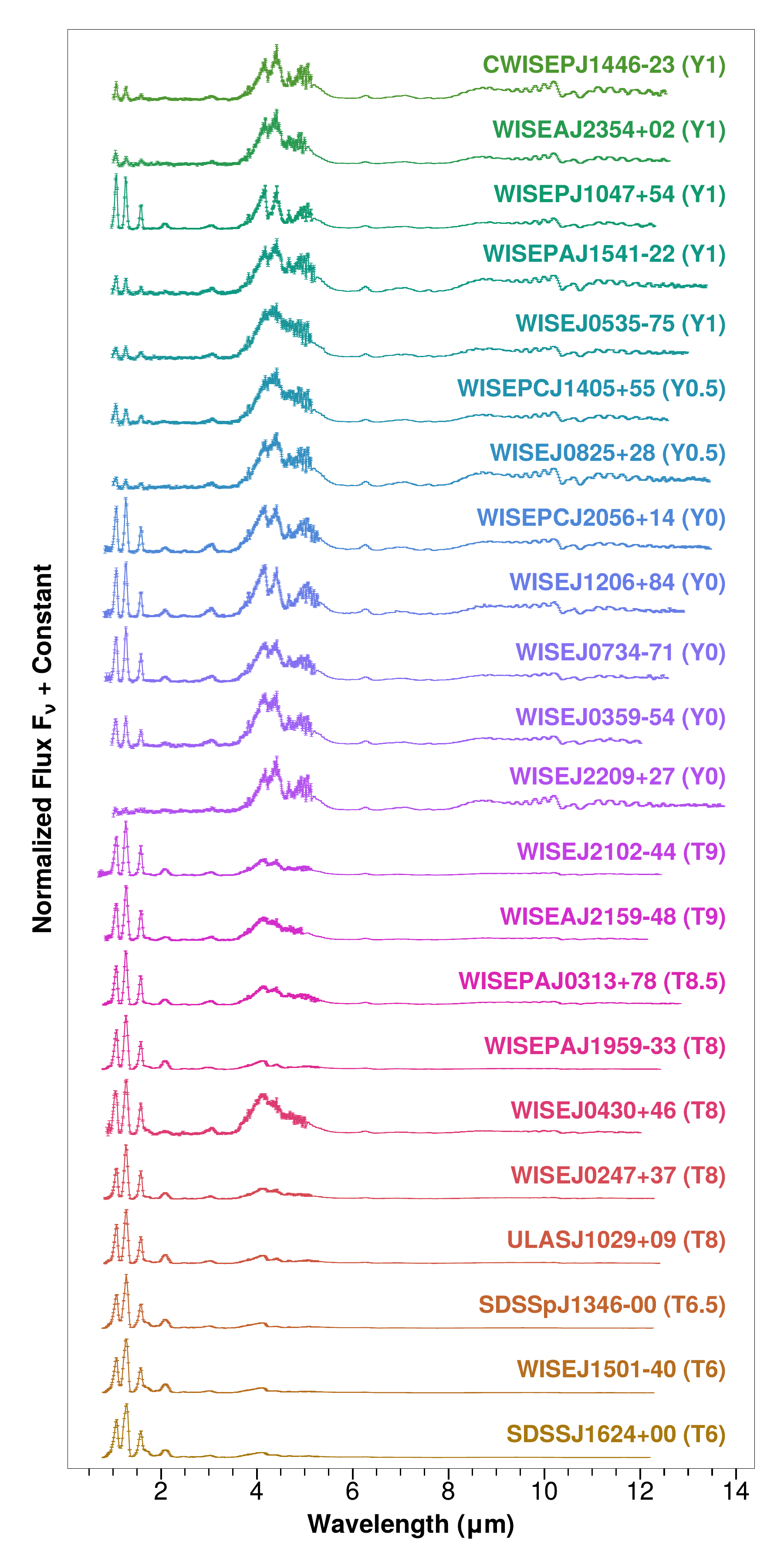}
    \caption{Spectral sample of 22 T and Y dwarfs originally published in \cite{Beiler2024ApJ...973..107B}. All spectra are normalized at their maximum flux value and offset by a constant.}
    \label{fig:data_sequence}
\end{figure}

In this study, we use the curated dataset from \cite{Beiler2024ApJ...973..107B}, which includes observations of 23 brown dwarfs with the JWST's near-infrared spectrograph (NIRSpec, \citealt{NIRSpec2022A&A...661A..80J}) and mid-infrared instrument (MIRI, \citealt{MIRI2015PASP..127..584R}). This dataset provides low-resolution spectroscopy from both instruments, covering the $\sim$1-12$\,\mu$m wavelength range at $\lambda/\Delta\lambda \approx 100$ (GO 2302, PI: Cushing). The data were not reprocessed; for a detailed description of the reduction procedures, refer to \cite{Beiler2024ApJ...973..107B}. WISEPA J2018-74 (T7) lacks NIRSpec observations and is therefore excluded from our analysis. Our set of 22 brown dwarfs are shown in Fig.~\ref{fig:data_sequence} and their known properties are given in Tab.~\ref{tab:objects}. 

\begin{table*}[t]
\renewcommand{\arraystretch}{1.1}
\caption{Set of brown dwarfs and their observational characteristics used in this study.}
\label{tab:objects}
\centering
%\resizebox{\columnwidth}{!}{%
\begin{tabular}{lcccc}
\hline\hline
Name & Short Name & Spectral Type & Parallax (mas) & References \\
\hline
SDSS J162414.37+002915.6 & SDSSJ1624+00  & T6   & 91.8 $\pm$ 1.2  & 1, 2, 3\\
WISE J150115.92--400418.4 & WISEJ1501-40 & T6   & 72.8 $\pm$ 2.3  & 4, 5, 6\\
SDSSp J134646.45--003150.4 & SDSSpJ1346-00  & T6.5 & 69.2 $\pm$ 2.3  & 7, 2, 3\\
ULAS J102940.52+093514.6  & ULASJ1029+09 & T8   & 68.6 $\pm$ 1.7  & 8, 9, 10\\
WISE J024714.52+372523.5 & WISEJ0247+37  & T8   & 64.8 $\pm$ 2.0  & 11, 11, 10\\
WISE J043052.92+463331.6 & WISEJ0430+46  & T8   & 96.1 $\pm$ 2.9  & 11, 11, 6\\
WISEPA J195905.66--333833.7 & WISEPAJ1959-33 & T8   & 83.9 $\pm$ 2.0  & 12, 12, 6\\
WISEPA J031325.96+780744.2 & WISEPAJ0313+78 & T8.5 & 135.6 $\pm$ 2.8 & 12, 12, 6\\
WISEA J215949.54--480855.2 & WISEAJ2159-48 & T9   & 73.9 $\pm$ 2.6  & 5, 5, 6\\
WISE J210200.15--442919.5  & WISEJ2102-44  & T9   & 92.9 $\pm$ 1.9  & 13, 13, 14\\
WISEA J220905.73+271143.9  & WISEJ2209+27 & Y0   & 161.7 $\pm$ 2.6 & 12, 15, 6\\
WISE J035934.06--540154.6  & WISEJ0359-54 & Y0   & 73.6 $\pm$ 2.0  & 13, 13, 6\\
WISE J073444.02--715744.0  & WISEJ0734-71 & Y0   & 74.5 $\pm$ 1.7  & 13, 13, 6\\
WISE J120604.38+840110.6   & WISEJ1206+84  & Y0   & 84.7 $\pm$ 2.1  & 16, 16, 6\\
WISEPC J205628.90+145953.3 & WISEPCJ2056+14  & Y0   & 140.8 $\pm$ 2.0 & 17, 17, 6\\
WISE J082507.35+280548.5   & WISEJ0825+28 & Y0.5 & 152.6 $\pm$ 2.0 & 16, 16, 6\\
WISEPC J140518.40+553421.4 & WISEPCJ1405+55 & Y0.5 & 158.2 $\pm$ 2.6 & 17, 16, 6\\
WISE J053516.80--750024.9  & WISEJ0535-75 & Y1   & 68.7 $\pm$ 2.0  & 13, 13, 6\\
WISEPA J154151.66--225025.2 & WISEPAJ1541-22 & Y1   & 166.9 $\pm$ 2.0 & 17, 16, 6\\
CWISEP J104756.81+545741.6 & CWISEPJ1047+54 & Y1   & 68.1 $\pm$ 4.9  & 18, 19, 19\\
WISEA J235402.79+024014.1 & WISEAJ2354+02 & Y1   & 130.6 $\pm$ 3.3 & 16, 16, 6\\
CWISEP J144606.62--231717.8 & CWISEPJ1446-23 & Y1   & 103.8 $\pm$ 5.0 & 18, 19, 19\\
\hline
\end{tabular}
%} %from Beiler et al. (2024): Tab 1
\tablebib{
(1) \citet{Strauss1999ApJ...522L..61S},
(2) \citet{Burgasser2006ApJ...637.1067B},
(3) \citet{Tinney2003AJ....126..975T},
(4) \citet{Tinney2012ApJ...759...60T},
(5) \citet{Tinney2018ApJS..236...28T},
(6) \citet{Kirkpatrick2021ApJS},
(7) \citet{Tsvetanov2000ApJ...531L..61T},
(8) \citet{Burningham2013MNRAS.433..457B},
(9) \citet{Thompson2013PASP..125..809T},
(10) \citet{Best2020AJ....159..257B},
(11) \citet{Mace2013ApJS..205....6M},
(12) \citet{Kirkpatrick2011ApJS..197...19K},
(13) \citet{Kirkpatrick2012ApJ...753..156K},
(14) \citet{Tinney2014ApJ...796...39T},
(15) \citet{Cushing2014ASSL..401..113C},
(16) \citet{Schneider2015ApJ...804...92S},
(17) \citet{Cushing2011ApJ...743...50C},
(18) \citet{Meisner2020ApJ...889...74M},
(19) \citet{Beiler2024ApJ...973..107B}
}
\end{table*}

\section{Atmospheric Retrieval Approach}

\subsection{Bayesian framework}
\label{sec:bear_description}

In this investigation, we employed the open-source Bern Atmospheric Retrieval (\bear) code, which represents an updated version of the earlier Helios-r2 code described by \cite{Kitzmann2020ApJ...890..174K}. \bear features enhanced capabilities and additional forward models. For our analysis, we used the emission spectroscopy forward model available in \bear. The code and comprehensive documentation are publicly accessible under the GNU General Public License (GPL) and can be found on GitHub\footnote{\url{https://github.com/newstrangeworlds/bear}}. \bear incorporates the \texttt{MultiNest} \citep{Feroz2008MNRAS, Feroz2009MNRAS} algorithm for Bayesian nested sampling \citep{Skilling2004AIPC, Skilling2006AIPC..872..321S} to explore the multi-dimensional parameter space of the models, compute posterior distributions, and to calculate the Bayesian evidence. For this study, we employ the emission spectroscopy forward model of \bear, which, as originally presented by \citet{Kitzmann2020ApJ...890..174K}, solves the radiative transfer equation by using the short characteristics method as described by \citet{Olson1987JQSRT..38..325O}. 

The one-dimensional, plane-parallel model atmosphere comprises 99 layers (100 levels) extending from 100~bar to 1~mbar. The temperature-pressure profile is characterized by a finite element approach that ensures smoothness and continuity \citep{Kitzmann2020ApJ...890..174K}. In the current study, it is described by two regimes. 

First, the radiative-convective boundary (RCB) is parametrized, defining the depth at which the atmosphere transitions from being dominated by convective energy transport in deeper atmospheric layers to being dominated by radiative energy transport. Below the RCB (at higher pressures), the thermal profile is described by following an adiabatic gradient, determined by the adiabatic index 
\begin{equation}
  \gamma = c_\mathrm{p}/c_\mathrm{V} = 1 + 2/n\ , 
\end{equation} 
where $c_p$ and $c_v$ are the heat capacities at constant pressure and at constant volume, respectively. The number of degrees of freedom are denoted as $n$. We approximate the adiabatic temperature-pressure profile
\begin{equation}
  \left( \frac{d \ln T}{d \ln p} \right)_{\mathrm{ad}} = \frac{\gamma - 1}{\gamma}
\end{equation}
with a constant value of $\gamma$, thus neglecting the potential temperature dependence of $c_\mathrm{p}$ and $c_\mathrm{V}$. The constant value of $\gamma$, however, still depends on the chemical composition and degrees of freedom of the gas. For an ideal, diatomic gas (e.g., molecular hydrogen (\ch{H2})), $\gamma =7/5 = 1.4$, whereas for monatomic gases, $\gamma = 5/3 \approx 1.67$. Both parameters, $\gamma$ and the RCB are treated as free parameters of the model.

The adiabatic temperature profile starts at the bottom of the atmosphere at a temperature $T_0$, which is used a free parameter of the model. Above the RCB, four parameters $b_i$ — equally distributed in $\log p$-space — are used to parametrize the atmosphere \citep{Kitzmann2020ApJ...890..174K}. The temperature at these points $i$ is determined with $T_i = b_i T_{i-1}$.

The opacities (absorption cross sections per unit mass) for various atoms and molecules were computed using the open-source \texttt{HELIOS-K} calculator\footnote{\url{https://github.com/exoclime/HELIOS-K}} \citep{Grimm2015ApJ, Grimm2021ApJS..253...30G}, and are available through the \texttt{DACE} platform\footnote{\url{https://dace.unige.ch}} \citep{Grimm2021ApJS..253...30G}. Our study included the following molecules: water (\ch{H2O}), carbon monoxide (\ch{CO}), carbon dioxide (\ch{CO2}), hydrogen sulfide (\ch{H2S}), sulphur dioxide (\ch{SO2}), methane (\ch{CH4}), ammonia (\ch{NH3}), and phosphine (\ch{PH3}). Line lists were obtained from the \texttt{ExoMol} database \citep{Barber2006MNRAS, Yurchenko2011MNRAS.413.1828Y, Yurchenko2014MNRAS.440.1649Y, Azzam2016MNRAS} and the \texttt{HITEMP} database \citep{Rothman2010JQSRT.111.2139R}. Collision-induced absorption (CIA) coefficients for \ch{H2}–\ch{H2} and \ch{H2}–\ch{He} pairs are based on \citet{Abel2011} and \citet{Abel2012JChPh}, respectively. 
For the alkali metals sodium (\ch{Na}) and potassium (\ch{K}), the absorption coefficients were calculated based on the \citet{1995KurCD..23.....K} line list data. As described by \citet{Kitzmann2020ApJ...890..174K}, the line profiles of the important resonance lines of \ch{Na} and \ch{K} are based on the studies by  \citet{Allard2016A&A} and \citet{Allard2019A&A}. 

\bear incorporates several cloud descriptions, including a gray cloud layer and a non-gray cloud approximation based on \cite{Kitzmann2018MNRAS}. The \bear framework allows for combining cloud models in a single retrieval, although this was not applied in this study. For detailed information, we refer the reader to the \bear user guide\footnote{\url{https://newstrangeworlds.github.io/BeAR/}}.

The gray cloud layer provides a simple parametrization with a wavelength-independent vertical optical depth $\tau_\mathrm{c}$ at a given cloud top pressure $p_\mathrm{t}$. The bottom pressure of the cloud is determined via $p_\mathrm{b} = b_\mathrm{c} \cdot p_\mathrm{t}$, where $b_\mathrm{c}$ is a free parameter.

The non-gray cloud model uses analytical fits to Mie efficiencies for a more detailed description of wavelength-dependent effects. Following \citet{Kitzmann2018MNRAS} it is described by three parameters: $Q_0$, the size parameter where the extinction coefficients peak, $a_0$, the power-law index for small particles, and the particle size $a$. Furthermore, the cloud layer is characterized by a vertical optical depth at a wavelength of 1 $\mu$m, as well as the cloud-top pressure $p_\mathrm{t}$ and the corresponding $b_\mathrm{c}$ to determine the vertical extent of the cloud layer.

The chemical composition is characterized by vertically constant mixing ratios $x_i$ for the chemical species considered in the retrieval. The background atmosphere is filled by a combination of \ch{H2} and He, with their abundance ratio given by their solar abundances. 

The distance $d$ of the brown dwarfs is derived from the parallax measurements listed in Tab.~\ref{tab:objects}. We assume an a priori radius of 1 Jupiter radius for the brown dwarfs but introduce a flux scaling factor $f$ to scale photospheric flux to the one measured by the observer. Besides the radius contribution, $f$ also partially captures inadequacies of the simplified forward model to describe the brown dwarf atmosphere, including the effect of a reduced emitting surface due to a potentially heterogeneous atmosphere. However, assuming that the scaling factor only includes deviations with respect to the assumed a priori radius, it can be used to derive the radius of the brown dwarf via $R = \sqrt{f} \, R_\mathrm{Jup}$ (Section 2.2 of \citealt{Kitzmann2020ApJ...890..174K}).

All parameters and their prior distributions are shown in Tab.~\ref{tab:priors}. Following \cite{Lueber2022ApJ...930..136L}. We apply uniform prior distributions for the cloud optical depth to ensure proper constraints. Negative values are allowed in the prior range to sample the boundary condition of 0 adequately. Negative optical depths are replaced with 0 internally to avoid nonphysical results. Bayesian model comparison was then conducted using standard methods for calculating the Bayesian evidence and the log Bayes factor $\ln{B_{ij}}$ \citep{Trotta2008ConPh..49...71T}, which are direct outputs of the nested sampling algorithm. The Bayes factor is used to assess statistical preference between the cloud models, with values of $\ln{B_{ij}}$ of 1, 2.5, or larger than 5 indicating "weak", "moderate", and "strong" evidence, respectively (Table 2 of \citealt{Trotta2008ConPh..49...71T}). However, it is important to note that Bayesian model comparison may not always rule out non-physical scenarios, as discussed by, for example, \citep{Fisher2019ApJ}.

\begin{table}
\renewcommand{\arraystretch}{1.1}
\caption{Summary of retrieval parameters and prior distributions for the free chemistry approach used in the cloud-free and non-gray cloud models.}
\label{tab:priors}
\centering
%\scriptsize
\resizebox{\columnwidth}{!}{
\begin{tabular}{lcc} %{p{0.55\columnwidth}>{\centering}p{0.17\columnwidth}>{\centering\arraybackslash}p{0.22\columnwidth}}
\hline \hline
Parameter &\multicolumn{2}{c}{Prior}\\
\cline{2-3} & Type & Value\\
\hline
  \logg \, (cm/s$^2$) & Gaussian &  2.0 -- 5.5\\
  $d$ (pc)    & Gaussian &  measured\\
  $f$ & uniform &  0.2 -- 2.0 \\
  $x_i$ & log-uniform &   $10^{-12}$ -- $10^{-2}$ \\
  $10^{\epsilon}$$^{\,*}$  & log-uniform  & $0.01 \cdot \min \sigma_i^2$ -- $100 \cdot \max \sigma_i^2$ \\ 
\hline
\textit{Temperature profile} & &   \\  
  $T_0$ (K)   & uniform & 500 -- 3000\\
  $b_{i=1,\dots,4}$ & uniform &  0.1 -- 0.95  \\
  RCB (bar) & log-uniform &  1.0 -- $10^2$\\
  $\gamma$ & uniform &  1.0 -- 2.0 \\
\hline
\textit{Gray clouds} & &   \\
$p_{\mathrm{t}}$ (bar)  & log-uniform & $10^{-3}$ -- $10^{2}$\\
$b_{\mathrm{c}}$  & log-uniform & 1 -- 10\\
$\tau_\mathrm{c}$  & uniform &  -10 -- 20\\
\hline
\textit{Non-gray clouds} & &   \\
$p_{\mathrm{t}}$ (bar)  & log-uniform &  $10^{-3}$ -- $10^{2}$\\
$b_{\mathrm{c}}$  & log-uniform & 1 -- 10\\
$\tau_{\mathrm{c}}$  & uniform &  -10 -- 20 \\
$Q_0$  & uniform & 1 -- 100 \\
$a_0$  & uniform &  3 -- 6 \\
$a$ ($\mu$m) & log-uniform &  $10^{-7}$ -- 0.1 \\
\hline
\end{tabular} }
\footnotesize{$^*$ We consider additional variance through $s_i^2=\sigma_i^2+10^\epsilon$, where $\sigma_i$ are the reported observational uncertainties (see \citealt{Kitzmann2020ApJ...890..174K}).}\\
% \footnotesize{$^{**}$ The only, actually retrieved temperature $\rm T_0$ is the one at the bottom of the modeled atmosphere. For all other temperatures, we retrieve a parameter $\rm b_i$, such that, e.g., $T_1= b_1 \cdot T_0$.}
\end{table}

\subsection{Machine learning framework}

In addition to retrievals within a Bayesian framework, we apply atmospheric retrievals using the random forest supervised machine learning algorithm \citep{Ho1998random, Breiman2001random}, implemented through the \texttt{HELA} code \citep{Marquez2018NatAs...2..719M}. It combines the use of decision trees (e.g., \citealt{breiman1984classification}) with bootstrapping techniques (e.g., \citealt{efron1994introduction}), and can be applied to both discrete and continuous training sets. 

A key advantage of random forest retrievals is their computational efficiency and ability to quantify feature importance, ranking the relative influence of individual spectral data points (features) in constraining specific model parameters. This interpretability has made random forests particularly appealing for atmospheric retrievals in studies of exoplanets and brown dwarfs \citep{Marquez2018NatAs...2..719M, Oreshenko2020AJ, Fisher2020AJ....159..192F, Guzman2020AJ....160...15G, Fisher2022ApJ...934...31F, Lueber2023ApJ...954...22L, Lueber2024A&A...687A.110L}.

The random forest method \texttt{HELA} uses pre-calculated grids of model spectra %\citep{Tremblin2015ApJ, Morley2012ApJ...756..172M, Malik2019AJ....157..170M, Marley2021ApJ...920...85M, Leggett2021ApJ...918...11L, Leggett2025ApJ...979..145L, Mukherjee2024ApJ...963...73M}
as individual training data within the Approximate Bayesian Computation framework \citep{Sisson2018handbook}. In our approach, all model spectra, which will be described in the next section, are binned to the overlapping wavelength ranges of the spectra by \citet{Beiler2024ApJ...973..107B}, which were obtained with the \textit{James Webb Space Telescope}. The NIRSpec PRISM and MIRI LRS spectra are covering the $\sim$1-12$\,\mu$m wavelength range at $\lambda/\Delta\lambda \approx100$. For each random forest spectral fitting, 3000 regression trees are utilized, allowing each tree to grow without a limit on depth. The trees are pruned only when additional splits result in a variance reduction smaller than 0.01. Furthermore, the maximum number of features allowed for splitting at each node was restricted to the square root of the total number of spectral data points utilized.

The performance of the trained model is typically assessed using real versus predicted (RvP) plots. These comparisons are quantitatively evaluated using the coefficient of determination, 
\begin{equation}
  R^2 = 1 - S_{\mathrm{res}}/S_{\mathrm{tot}} \ ,
\end{equation}
where $S_{\mathrm{res}}$ is the sum of squared residuals between real ($y_i$) and predicted ($f_i$) values: 
\begin{equation}
  S_ {\mathrm{res}} = \sum_ {i=1} ^ {n} (y_i - f_i)^2 \ ,
\end{equation}
and $S_ {\mathrm{tot}}$ represents the total variance of the real values:
\begin{equation}
  S_ {\text {tot}} = \sum_ {i=1} ^ {n} (y_i - \bar {y})^2 \ .
\end{equation}
The $R^2$ value indicates how well the model generalizes to unseen data: $R^2 = 1$ indicates perfect predictions, $R^2 = 0$ suggests no correlation, and $R^2 < 0$ indicates predictions that are worse than random chance.

\subsubsection{Applied Model Grids}
\label{sec:model_grids}

Here, the model grids used in this work will be briefly summarized. We reduced the number of models by constraining them to ranges
200~K $\le$ {\teff} $\le$ 800~K, 3.0 $\le$ {\logg} $\le$ 6.0 (in units of cm/s$^2$), $-$1.0 $\le$ [M/H] $\le$ +1.0, and and 0.5 $\le$ C/O $\le$ 1.5. These constraints are, in addition to the parameter limits set in the original model grids, described below. The exact parameter ranges are listed in Tab.~\ref{tab:models}. 

\begin{table*}[t]
\renewcommand{\arraystretch}{1.1}
\scriptsize
\centering
\caption{Parameter Ranges for Spectral Model Grids.}
\label{tab:models}
\resizebox{\textwidth}{!}{
\begin{tabular}{lcccccccccc}
\hline
Model & \teff & \logg & [M/H] & C/O & $\gamma$ & $\log\kappa_{zz}$ & $f_\mathrm{sed}$ & \# models & Ref. \\
 & (K) & (cm/s$^2$) & (dex) &  &  & (cm$^2$/s) &  &  & \\
\hline
%Morley 2012 & 400 -- 800 & 4.0 -- 5.5 & 1.0 & \nodata & \nodata & \nodata & 2 -- 5 & 103 & 1 \\
%Saumon 2012 & 400 -- 800 & 3.5 -- 5.5 & 1.0 & \nodata & \nodata & \nodata & \nodata & 73 & 2 \\
%Tremblin 2015 & 200 -- 800 & 3.5 -- 5.0 & $-$1.0 -- +0.5 & \nodata & 0 -- 1.3 & 0 -- 9 & \nodata & 500 & 3 \\
HELIOS & 200 -- 800 & 3.6 -- 6.0 & 1.0 & 0.5 & \nodata & \nodata & \nodata & 91 & 1 \\
Sonora Bobcat & 200 -- 800 & 3.75 -- 5.5 & $-$0.5 -- +0.5 & 0.5 -- 1.5\tablefootmark{a} & \nodata & \nodata & \nodata & 462 & 2 \\
Sonora Elf Owl & 275 -- 800 & 3.0 -- 5.5 & $-$1.0 -- +1.0 & 0.5 -- 1.5\tablefootmark{a} & \nodata & 2.0 --9.0 & \nodata & 8640 & 3 \\
Lacy \& Burrows & 200 -- 600 & 3.5 -- 5.0 & $-$0.5 -- +0.5 & \nodata & \nodata & 2.0 --6.0 & \nodata & 1443 & 4 \\
ATMO2020++ without PH3 & 250 -- 800 & 3.5 -- 5.5 & $-$1.0 -- +0.3 & \nodata & 0.12 -- +12.5 & 4.0 -- 8.0 & \nodata & 200 & 5 \\
\hline
\end{tabular}}
%\footnotesize{$^1$: C/O ratio relative to solar abundance, with (C/O)$_\odot$ = 0.458 \citep{2009LanB...4B..712L}.}\\
\tablebib{%(1)~\citet{Morley2012ApJ...756..172M}; (2)~\citet{Saumon2012ApJ...750...74S}; (3)~\citet{Tremblin2015ApJ};
(1)~\citet{Malik2019AJ....157..170M}; (2)~\citet{Marley2021ApJ...920...85M}; (3)~\citet{Mukherjee2024ApJ...963...73M}; (4)~\citet{Lacy2023ApJ...950....8L}; (5)~\citet{Leggett2025ApJ...979..145L}.}
\tablefoot{
\tablefoottext{a}{The C/O ratio is relative to the solar abundance one by \citet{2009LanB...4B..712L}, with (C/O)$_\odot$ = 0.458.}
}
\end{table*}

\paragraph{HELIOS}
The open-source radiative transfer code \texttt{HELIOS} \citep{Malik2017AJ....153...56M, Malik2019AJ....157..170M} utilizes an improved two-stream method \citep{HK2017, Heng2018ApJS..237...29H} that incorporates non-isotropic scattering and convective adjustment. The equilibrium gas-phase chemistry is computed with the original version of the open-source \fc equilibrium chemistry code\footnote{\url{https://github.com/NewStrangeWorlds/FastChem}} \citep{Stock2018MNRAS.479..865S, Stock2022MNRAS.517.4070S}, which did not include condensation \citep{Kitzmann2024MNRAS}. Within the cloud-free HELIOS atmosphere grid, the removal of gas-phase species due to condensation is approximated using stability curves of various condensates (see Appendix B in \citet{Malik2017AJ....153...56M} for details). The elemental abundances for the chemistry calculations are based on those of the Sun's photosphere from the compilation by \cite{Asplund2009ARA&A..47..481A}. 

%\subsubsection*{ATMO2020++}\\

\paragraph{Sonora Bobcat}
This atmospheric grid by \cite{Marley2021ApJ...920...85M}, known as Sonora Bobcat, consists of cloud-free models developed for radiative-convective equilibrium atmospheres using a layer-by-layer convective adjustment method. This enables solutions with distinct convective zones. Rainout chemistry incorporates condensation from the gas phase, as described by \cite{Lodders2006asup.book....1L}, but does not consider explicit cloud opacity. The models adopt the $k$-distribution approach for calculating wavelength-dependent gas absorption \citep{Goody1989JQSRT..42..539G}, utilizing updated opacities and equilibrium constants for gas-phase chemistry (for additional details, see \citealt{Marley2021ApJ...920...85M}). 

\paragraph{Sonora Elf Owl} 
The Sonora Elf Owl models \citep{Mukherjee2024ApJ...963...73M} are the latest additions to the Sonora series, which includes the Bobcat \citep{Marley2021ApJ...920...85M}, Cholla \citep{Karalidi2021ApJ...923..269K}, and Diamondback \citep{Morley2024ApJ} models. The Cholla models introduced self-consistent disequilibrium chemistry, though limited to solar composition, while the Diamondback models incorporated the effects of clouds and metallicity on atmospheric structure. The Elf Owl models further advance with a self-consistent, cloud-free 1D radiative-convective equilibrium framework and include mixing-induced disequilibrium chemistry beyond solar composition and account for non-solar C/O ratios. The Elf Owl model features additional free parameters such as the vertical eddy diffusion coefficient, $\kappa_{zz}$, and explores broader parameter spaces compared to previous versions. The updated model version 2 with corrected disequilibrium \ch{CO2} abundance and with \ch{PH3} contributions removed has been applied (see \citealt{Beiler2024ApJ...973...60B} and \citealt{Wogan2025RNAAS...9..108W}). %Notably, \ch{PH3} abundance is now set to chemical equilibrium due to the lack of its detection in most brown dwarf atmospheres.

\paragraph{Lacy \& Burrows}
The atmospheric model grid by \cite{Lacy2023ApJ...950....8L} consists of 1D radiative–convective equilibrium models produced with coolTLUSTY for plane-parallel atmospheres. Chemical abundances are calculated assuming either equilibrium or non-equilibrium chemistry driven by vertical mixing, following \cite{Hubeny2007ApJ...669.1248H}, and encompass processes such as condensation and rainout. For effective temperatures below 400~K, the models incorporate water clouds using a parameterized, spatially uniform prescription.

\paragraph{ATMO2020++ without PH3}
The ATMO2020++ grid without PH3 \citep{Leggett2025ApJ...979..145L} is an updated version of the original ATMO2020 models\footnote{\url{https://opendata.erc-atmo.eu/}} \citep{Phillips2020A&A...637A..38P}, incorporating empirical modifications to better reproduce the observed spectra of late-T and Y dwarfs. In particular, the temperature gradient in convective regions has been reduced -- by adopting a lower effective adiabatic index -- to create cooler deep atmospheres, where the near-infrared flux originates, and warmer upper layers, which emit in the mid-infrared. Although this modification is not derived from first principles, it is physically motivated and may reflect processes such as the inhibition of convection by condensate formation or rapid rotation, though these explanations remain unproven (see \citealt{Leggett2025ApJ...979..145L} for discussion). The model is tailored for solar metallicity and metal-poor atmospheres. Disequilibrium chemistry is utilized with a vertical eddy diffusion coefficient, $\kappa_{zz}$, which is set at $\kappa_{zz}=10^{5}$ cm$^2$/s for \logg\,= 5.0 and is scaled by $10^{(2\cdot(5-\log{g}))}$ for other surface gravities. The scaling $\kappa_{zz} \propto g^{-2}$ reflects the change in atmospheric scale height with gravity and its impact on vertical mixing efficiency and chemical disequilibrium.

%The ATMO2020++ grid without PH3 \citep{Leggett2025ApJ...979..145L} is an updated version of the original ATMO2020 models\footnote{\url{https://opendata.erc-atmo.eu/}} \citep{Phillips2020A&A...637A..38P}, in which the temperature gradient has been decreased due to convective instabilities arising from chemical transitions like \ch{CO}/\ch{CH4} and \ch{N2}/\ch{NH3}. This modification integrates an adjustment of the temperature gradient by adopting an adiabatic index. The model is tailored for solar metallicity and metal-poor atmospheres. Disequilibrium chemistry is utilized with a vertical eddy diffusion coefficient, $\kappa_{zz}$, which is set at $\kappa_{zz}=10^{5}$ cm$^2$/s for \logg\,= 5.0 and is scaled by $10^{(2\cdot(5-\log{g}))}$ for other surface gravities. The scaling $\kappa_{zz} \propto g^{-2}$ reflects the idea that lower surface gravity results in more efficient vertical mixing, thereby enhancing out-of-equilibrium chemistry.

\subsection{Feature Importance Analysis of JWST NIRSpec+MIRI modes: Case study for the Y1-dwarf WISEPAJ1541-22}

\begin{figure*}[h]
    \centering
    \includegraphics[width=0.96\linewidth]{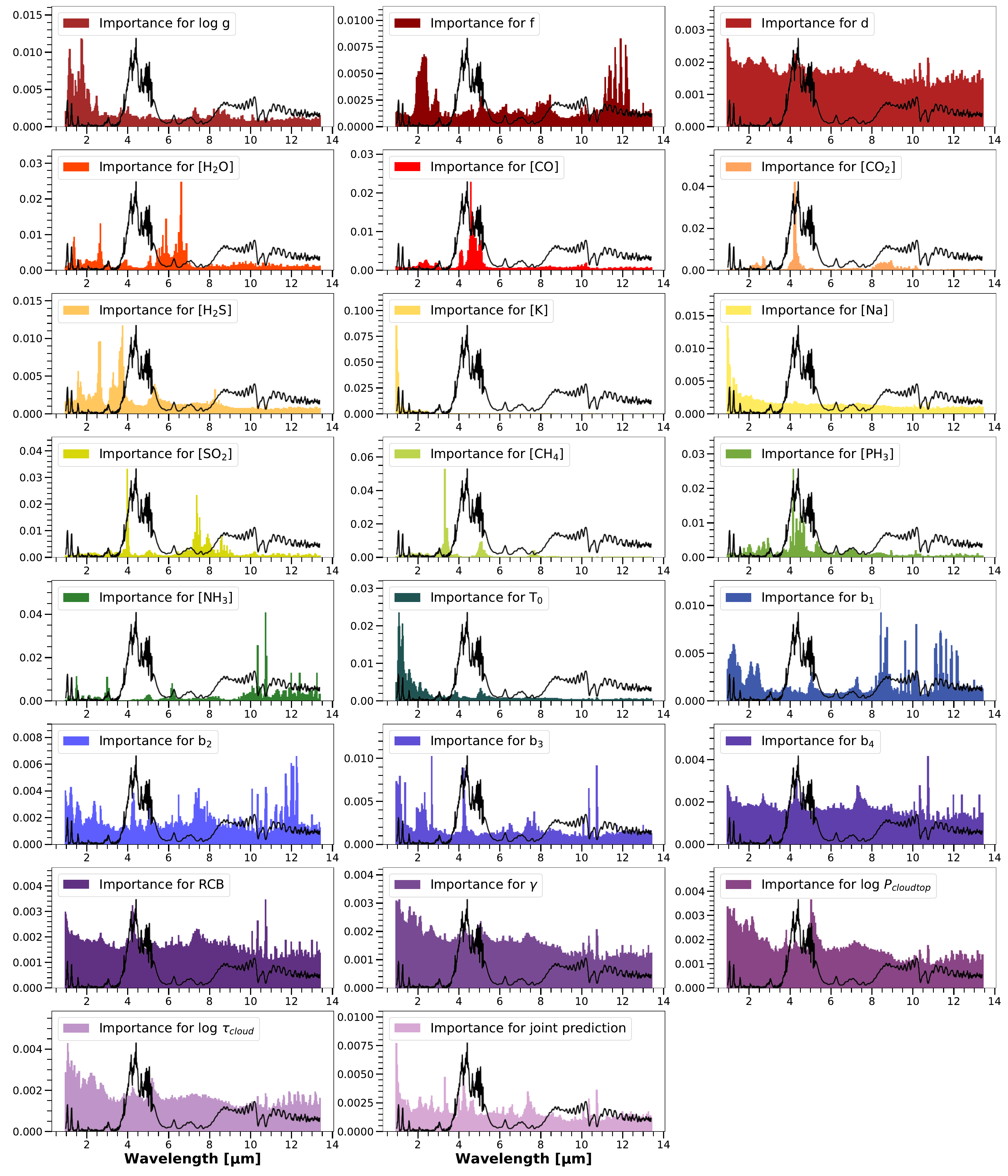}
    \vspace{-10pt}
    \caption{Feature importance across wavelengths for all parameters included in our generated model grid for case study Y1-dwarf WISEPAJ1541-22, as well as their joint retrieval. The significance of a feature is quantified by the normalized decrease in variance it generates throughout the training process. Specifically, it denotes the total reduction in variance achieved whenever that feature is employed for partitioning nodes in a decision tree within the ensemble. For visual guidance, the rescaled spectrum of WISEPAJ1541-22 is overlaid as solid black lines.}
    \label{fig:grid_feature}
\end{figure*}

To investigate feature importance using the random forest approach, we generated a model grid of 100,000 synthetic spectra by using the forward model of \bear. Most parameters were sampled from uniform distributions, while volume mixing ratios followed log-uniform distributions. For the current study, no noise was being considered in the creation of the grid. 

The grid was constructed using the following prior ranges: surface gravity \logg$\,$ between 2.0 and 5.5, scaling factor $f$ from 0.2 to 2, and distance $d$ specific to the considered object, sampled from a range of uncertainties around its known value in parsecs (5.99 $\pm$ 0.07 pc, for the Y1-dwarf WISEPAJ1541-22). Molecular log abundances were drawn from a uniform distribution between $10^{-12}$ and $10^{-2}$ for the ten key molecules: \ch{H2O}, \ch{CO}, \ch{CO2}, \ch{H2S}, \ch{K}, \ch{Na}, \ch{SO2}, \ch{CH4}, \ch{PH3}, and \ch{NH3}. 

The temperature structure was defined by a base temperature T$_0$ at the bottom of the modeled atmosphere, sampled between 500 and 3000~K, an adiabatic index $\gamma$ between 1.0 and 2.0, a radiative-convective boundary 10$^2 \leq$ RCB $\leq$ 1.0 bar, and subsequent layers above set by applying factors $b_1$ to $b_4$ that scale the temperature at each level between 0.1 and 0.95 of the layer below. Cloud top pressures $p_\mathrm{t}$ and (vertical) gray optical depths $\tau_\mathrm{c}$ were sampled within the ranges of 10$^{-3}$ to 10$^2$ bar and -10 to 20, respectively.

This grid enabled us to train a random forest model, which was evaluated using real-versus-predicted (RvP) assessments, quantified by the coefficient of determination $R^2$ (see Fig.~\ref{fig:grid_RvP} in Appendix \ref{sec:appendix_figures}). The temperature at the base of the atmosphere, $T_0$, demonstrates high self-consistency, while the predictability diminishes with each additional temperature–pressure profile coefficient $b_i$. This decline is due to the emission spectrum becoming increasingly insensitive to temperature variations in the upper atmosphere.

At higher abundances, most chemical species show strong predictability. Exceptions include species such as sodium and potassium, which lack the broad absorption bands typical of many molecules. At lower abundances, detection becomes more challenging, leading to reduced sensitivity in the RvP plots.

The contribution of different wavelength regions to parameter inference was analyzed using the feature importance plots shown in Fig.~\ref{fig:grid_feature}. These highlight how specific spectral features relate to physical and chemical processes, such as pressure broadening, vibrational transitions, and disequilibrium chemistry.

The surface gravity \logg \, exhibits strong spectral importance, particularly at shorter wavelengths, due to the broad pressure-broadened wings of alkali metal resonance lines, and a distinct peak near 2.0 $\mu$m, likely associated with collision-induced \ch{H2} absorption (CIA) \citep{Linsky1969ApJ...156..989L, Burgasser2002ApJ...564..421B}. Because CIA arises from collisions between gas-phase molecules, it is highly pressure-sensitive. While the feature importance analysis suggests that \logg \, contributes mainly at shorter wavelengths, this does not imply that it has no impact at longer wavelengths; in fact, gravity significantly affects \ch{CO} and \ch{CO2} absorption around 5~$\mu$m, as shown in \cite{Lacy2023ApJ...950....8L} and \cite{Leggett2025ApJ...979..145L}. The flux scaling factor $f$, acting as a proxy for planetary radius (see Sect. \ref{sec:bear_description}), shows peaks in the water absorption bands between 1.8 -- 3.0 $\mu$m and in the mid-infrared beyond 11 $\mu$m.

Water vapor (\ch{H2O}) dominates the opacity across many wavelengths, with strong contributions at 1.4 $\mu$m and 2.7 $\mu$m, as well as broader importance between 5.0 -- 7.0 $\mu$m, underscoring its central role in shaping both continuum and absorption features. Carbon-bearing species like CO and \ch{CO2} peak at 4.6 $\mu$m and 4.2 $\mu$m, respectively, consistent with their presence in warmer or vertically mixed atmospheres.

Alkali metals Na and K are influential in the 0.7 -- 1.0 $\mu$m range due to their broad absorption profiles, which also help constrain gravity. Sulfur-bearing species like \ch{H2S} show slight contributions between 2.5 -- 4.0 $\mu$m, though typically masked by stronger absorbers. Sulfur dioxide (\ch{SO2}), which is usually absent under equilibrium conditions, exhibits mid-infrared importance near 7.3 -- 8.7 $\mu$m, where it could possibly indicate disequilibrium processes.

Methane (\ch{CH4}), a key absorber in cooler atmospheres, is prominent around 3.3~$\mu$m, reflecting its increasing opacity at lower temperatures. Ammonia (\ch{NH3}) contributes notably between 10 -- 11 $\mu$m in the cooler atmospheres of late T and Y dwarfs. Phosphine (\ch{PH3}) displays localized importance near 4.3~$\mu$m, with minor features across the spectrum, consistent with expectations for reducing atmospheric conditions.

The base temperature $T_0$ is primarily constrained by features below 2.0 $\mu$m and is closely linked to the surface gravity due to its effect on the pressure–temperature profile. The thermal structure parameters $b_1$ through $b_4$ exhibit a progressively weaker spectral influence. The parameters $\rm b_1$ and $\rm b_2$ that describe the deeper atmospheric temperature gradient, still show some distinct spectral feature importance peaks indicating sensitivity at certain wavelengths. The parameters $\rm b_3$ to $\rm b_4$, on the other hand, describe the temperature-pressure profile in the upper layers. Here, however, the spectrum is almost unaffected by the form of the profile. Consequently, no distinct peaks are visible in the feature importance plots. Note that the feature importance is a normalized measure for each parameter, hence absolute values cannot be directly compared across different parameters.

Only limited insights can be gained from the feature importance of the adiabatic index ($\gamma$), the radiative–convective boundary (RCB), and cloud model parameters such as cloud-top pressure ($p_\mathrm{t}$) and gray optical depth ($\tau_\mathrm{c}$). These parameters exhibit slightly decreasing importance from the near- to mid-infrared but show localized significance around 5 $\mu$m. We speculate that the increased importance of the data at around 5 microns is likely due to the region’s sensitivity to cloud properties affecting emergent flux.

By interpreting feature importance plots, we effectively reverse-engineer the spectroscopic fingerprints that encode the model parameters. This data-driven approach provides valuable insights into opacity physics, thermal structure, chemical composition, not to precisely retrieve parameters for individual objects, but to better understand which spectral features are most informative, guiding future retrieval efforts. It also allows us to understand how different model grids emphasize each of these aspects differently across wavelength.

\section{Results and Discussion}
The primary aim of this study is to revisit and adapt long-standing research questions from the initial investigation by \citet{Lueber2022ApJ...930..136L}, specifically tailoring them to explore the unique characteristics of late T- and Y-dwarfs. While previous work provided insights, key questions remained unresolved or require re-evaluation within the context of the colder and less luminous objects analyzed here. This study seeks to bridge that gap, advancing our understanding of these enigmatic, ultra-cool brown dwarfs.

\subsection{Retrieval of the Y1 dwarf WISEPAJ1541-22}

\begin{figure*}
    \centering
    \includegraphics[width=\linewidth]{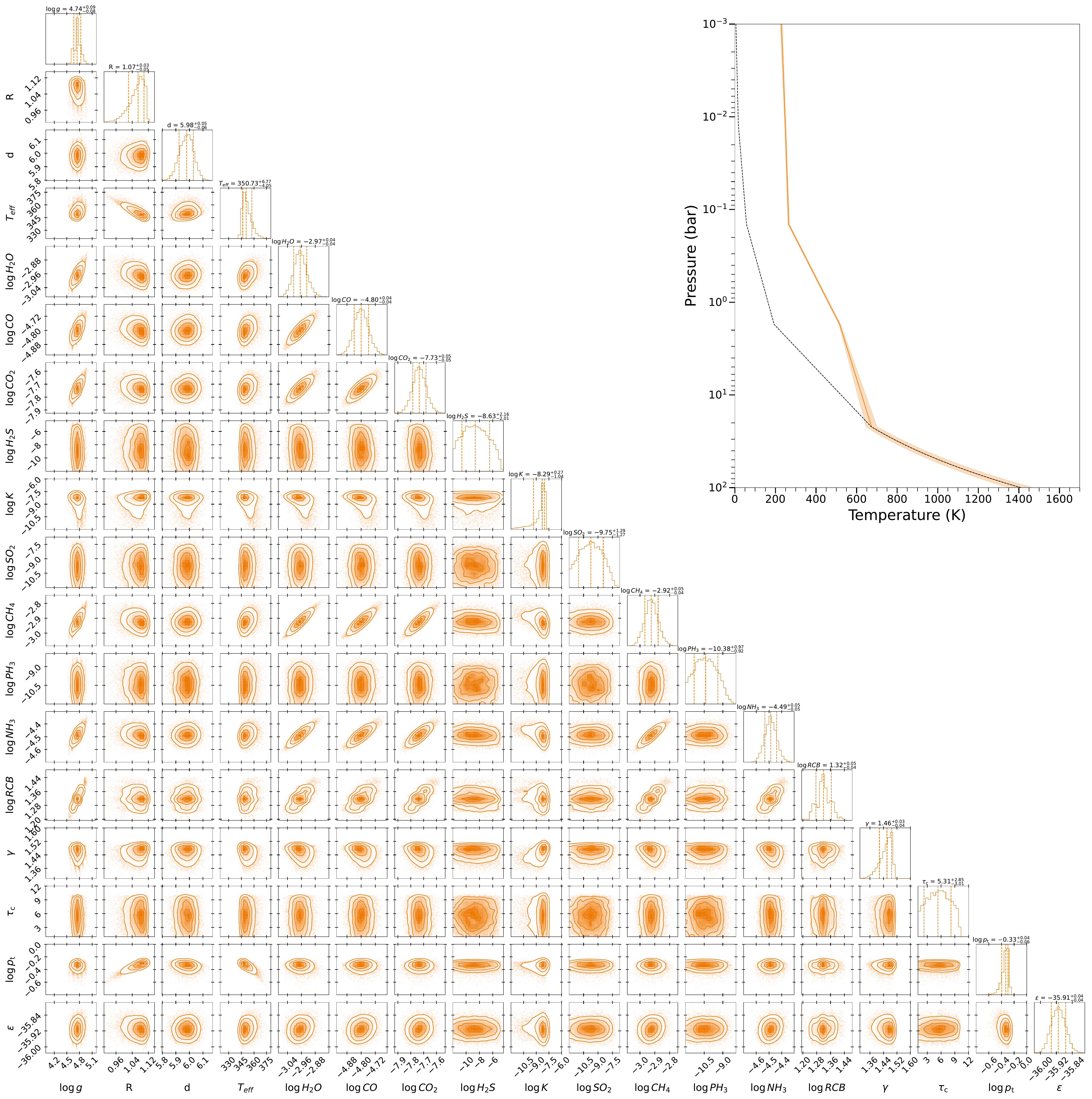}%{Figures/corner_constrained.pdf}
    \caption{Joint posterior distributions from the free-chemistry retrieval analyses of the case study Y1-dwarf WISEPAJ1541-22, using a gray-cloud model. The vertical dashed lines within the histograms indicate the median parameter values and their 1-$\sigma$ uncertainties. Accompanying the montage of joint posterior distributions is the retrieved median temperature–pressure profile along with its associated 1-$\sigma$ uncertainties. The adiabatic profile (black dashed line) is indicated for comparison.}
    \label{fig:Case_corner}
\end{figure*}

\begin{figure}
    \centering
    \includegraphics[width=\linewidth]{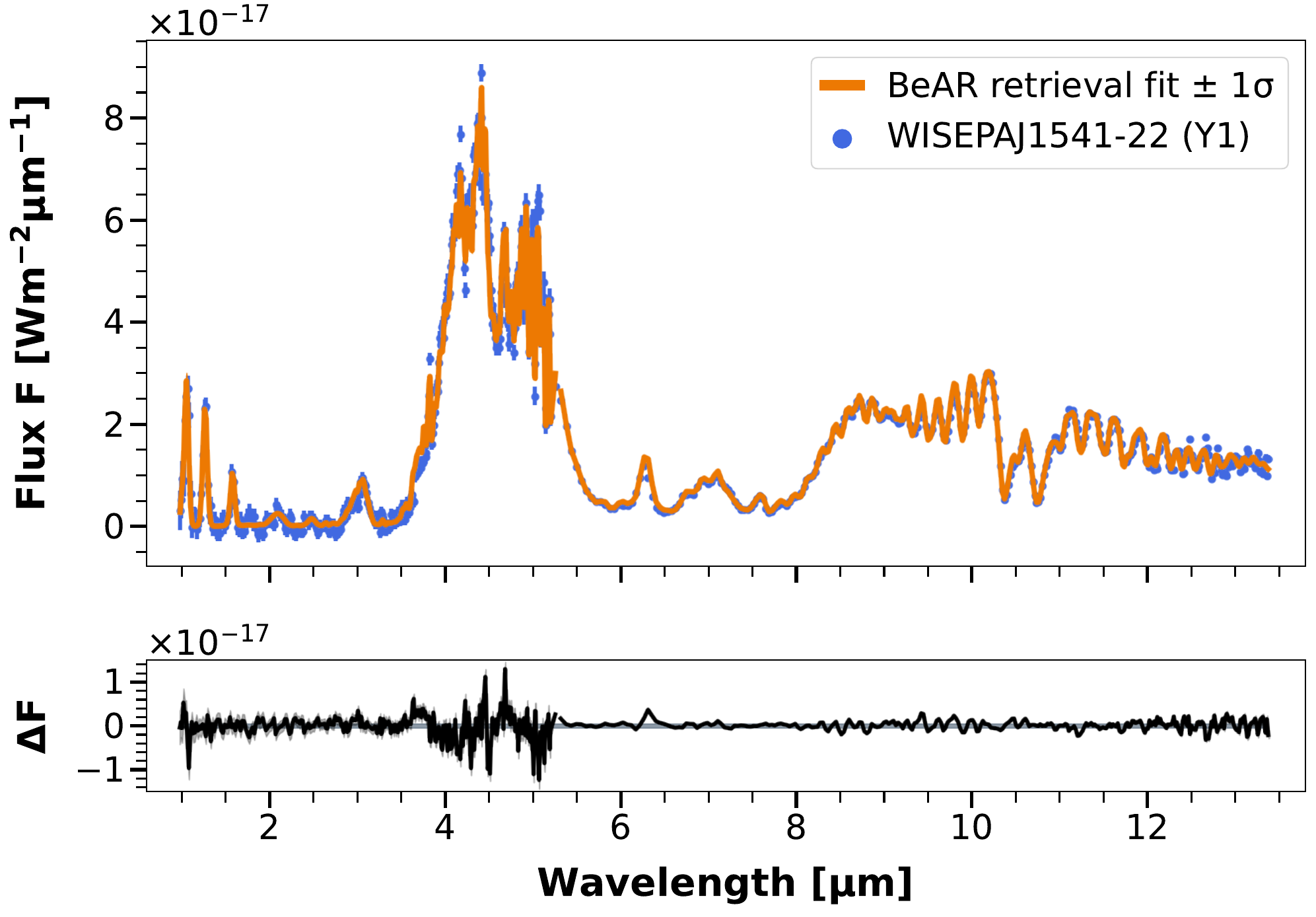}
    \caption{Posterior retrieval median fit $F$ (orange line) and residuals $\Delta$F (black line) associated with the free-chemistry retrieval analyses of the case study Y1-dwarf WISEPAJ1541-22, using a gray-cloud model. The retrieved error inflation is represented by a gray horizontal bar. The JWST NIRSpec and MIRI data are shown as blue dots with associated uncertainties.}
    \label{fig:Case_BF_spectrum}
\end{figure}

Before we present the overall trends in the retrieved properties across the T-Y sequence as a function of the effective temperature, we first discuss the results for the Y1 dwarf WISEPAJ1541-22 in more detail. The posterior distributions for this object are shown in Fig.~\ref{fig:Case_corner}, while the posterior median spectra and the comparison to the observed spectrum are depicted in Fig.~\ref{fig:Case_BF_spectrum}. To account for underestimated uncertainties or unmodeled systematics, we considered additional variance using $s_i^2=\sigma_i^2+10^\epsilon$ (see \citealt{Kitzmann2020ApJ...890..174K}). The retrieved error inflation parameter $\epsilon = -35.91_{-0.04}^{+0.04}$ corresponds to a flux level of $\sqrt{10^{\epsilon}} = (1.11 \pm 0.05) \cdot 10^{-18} \rm \, Wm^{-2}\mu m^{-1}$, comparable to the largest reported uncertainty for WISEPAJ1541-22. This results in an overall increase in uncertainty by a factor of $\sim$2.

For this object we find a strong preference for a gray cloud over a cloud-free model by performing Bayesian model comparison ($\ln B_{ij}=11.12$, see Tab.~\ref{tab:BeAR_Bayes_table}). Additionally, we find high abundances of both \ch{CO} and \ch{CO2} compared to equilibrium chemistry calculations. The posterior distributions in Fig.~\ref{fig:Case_corner} illustrate degenerate joint posteriors among the retrieved molecular abundances of \ch{H2O}, \ch{CO}, \ch{CO2}, \ch{CH4}, and \ch{NH3}, as well as their association with the retrieved surface gravity \logg. Notably, the retrieved thermal profile indicates that the cloud top is located at $\log p_{\mathrm{t}} = -0.33_{-0.06}^{+0.04}$~bar ($p_{\mathrm{t}} = 0.47_{-0.06}^{+0.04}$~bar). This positioning lies near the anticipated condensation temperature of water vapor at $\sim$300~K, but remains too warm for condensation to occur at this pressure level. Water clouds, if present, would form at higher altitudes where the temperature decreases further. Nonetheless,
%This positioning is in proximity to the anticipated condensation temperature of water vapor at $\sim$300~K, suggesting that the cloud within this Y1 dwarf atmosphere may plausibly consist of water droplets. The
the optical depth of the cloud layer ($\tau_\mathrm{c}=5.31_{-3.01}^{+2.85}$) supports the existence of an optically dense and vertically extended cloud layer, in alignment with the predicted onset of water cloud formation for late-T and Y dwarfs. Furthermore, the constraint of the adiabatic index $\gamma$ of $1.46_{-0.04}^{+0.03}$ is consistent with a mixture of molecular hydrogen and helium.

The simultaneous presence of both \ch{CO} and \ch{CH4} at such low effective temperatures (\teff \, = $350_{-4}^{+7}$~K) likely reflects vertical mixing and chemical quenching within the deep atmosphere. Collectively, these findings firmly position WISEPAJ1541-22, within the transition regime where clouds constitute a significant atmospheric component, thereby providing empirical support for theoretical predictions concerning the evolution of condensate clouds within Y dwarf atmospheres.

Based on the retrieved abundances, we derive an overall metallicity of [M/H] = $0.22 \pm 0.03$ by summing the elemental abundances relative to hydrogen and normalizing to their solar values. The derived metallicity corresponds to approximately 1.7 times the solar elemental abundance (using solar values from \citealt{Asplund2009ARA&A..47..481A}: C/H = $2.69\cdot10^{-4}$, N/H = $6.76\cdot10^{-5}$, O/H = $4.90\cdot10^{-4}$). However, this enrichment is not uniform across all elements. Carbon, for instance, is enriched by a factor of 2.7, and oxygen by a factor of 1.3, while nitrogen shows a mild depletion, of $\sim 0.3$ with respect to its solar value. We note, however, that the latter elemental abundance is based on ammonia alone. Nitrogen might still be present in other molecules, such as molecular nitrogen (\ch{N2}), that have no distinct spectral features in the wavelength range of the observations used here. The resulting carbon-to-oxygen (C/O) ratio is $1.12 \pm 0.14$, corresponding to $(\mathrm{C/O})/(\mathrm{C/O})_\odot = 2.0 \pm 0.4$, relative to the solar value of $\mathrm{C/O}_\odot = 0.55 \pm 0.09$ \citep{Asplund2009ARA&A..47..481A}.
%which is higher than the solar value of $0.55 \pm 0.09$ \citep{Asplund2009ARA&A..47..481A}.

The prior distribution of the C/O ratio, derived from combinations of molecular gas abundance priors, exhibits a characteristic double-peaked structure, with maxima at $(\mathrm{C/O})/(\mathrm{C/O})_\odot \approx 1$ and $(\mathrm{C/O})/(\mathrm{C/O})_\odot \approx 1.8$, 
%C/O $\approx 0.5$ and C/O $\approx 1$, 
regardless of whether the priors of molecules are log-uniform or broad Gaussian \citep{Line2013ApJ}. The former is caused by the abundance of \ch{CO2}, which would result in a solar-like C/O ratio of 0.5, while the latter is determined by \ch{CO}. Consequently, the prior distribution for the C/O ratio is directly affected by the choice of molecules included in a retrieval.

\begin{figure}
    \centering
    \includegraphics[width=\linewidth]{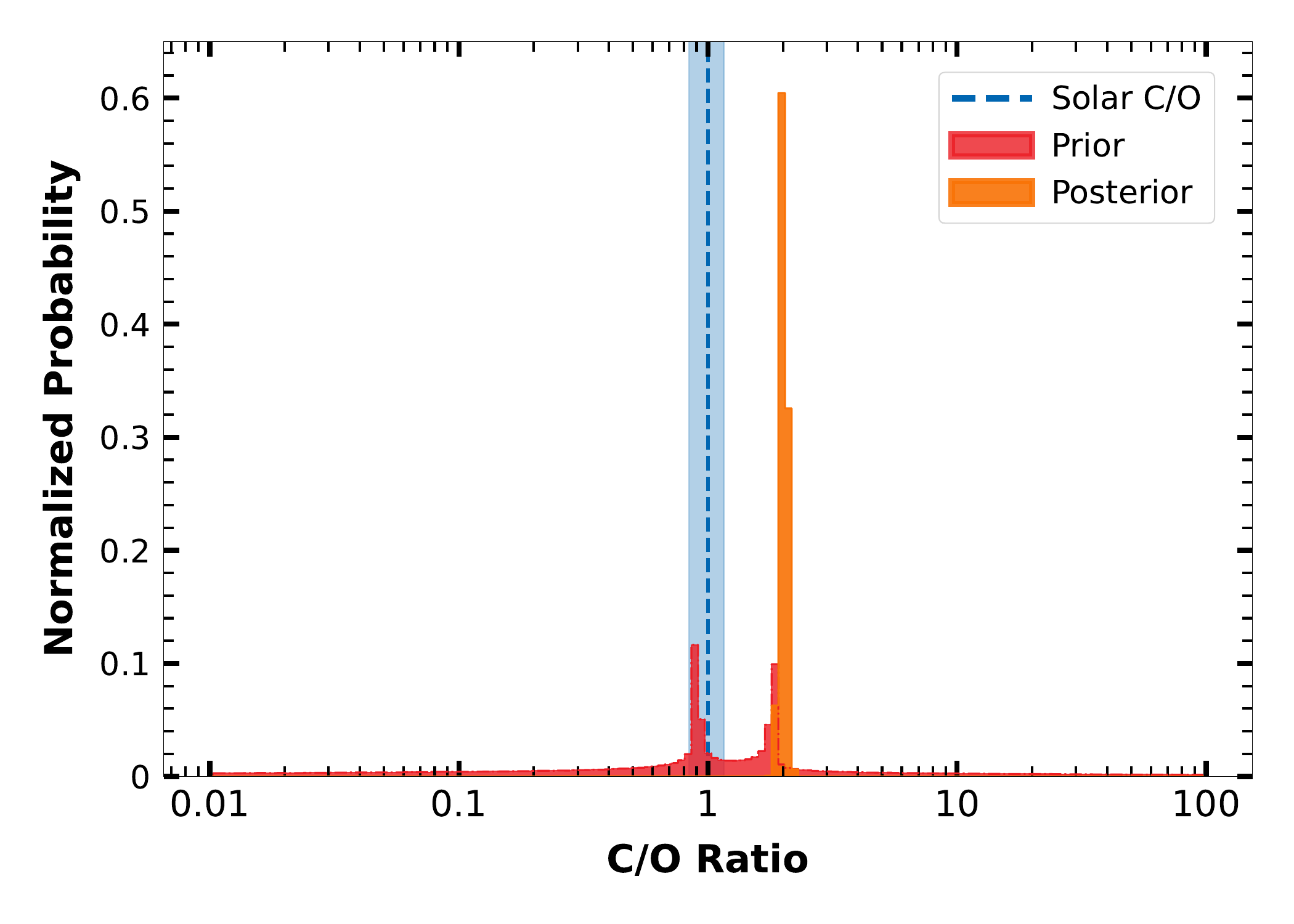}
    \caption{Normalized carbon-to-oxygen (C/O) ratio probabilities, expressed relative to the solar value of $0.55 \pm 0.09$ \citep{Asplund2009ARA&A..47..481A}, for case study Y1-dwarf WISEPAJ1541-22. Prior (red) and posterior (orange) probability distributions are derived from combinations of molecular gas abundance distributions. The solar value is indicated as dashed blue line and shaded region.}
    \label{fig:C/O}
\end{figure}

Comparing this analytically constructed prior to the posterior distribution derived from the constrained molecular abundances in the atmosphere of WISEPAJ1541-22, Fig.~\ref{fig:C/O} shows a clear divergence: the posterior does not exhibit the double-peaked structure characteristic of the prior. This indicates that the data provide meaningful constraints on the C/O ratio. In particular, the absence of the prior’s double peaks in the posterior suggests that JWST observations are sufficient to break prior degeneracies and extract genuine information about the C/O ratio. This result underscores the power of the JWST in constraining elemental abundance ratios in brown dwarf atmospheres.

Using a grid-fit retrieval with the \texttt{ATMO2020++} atmosphere model grid, \citet{Tu2024ApJ...976...82T} derived a much lower metallicity of [M/H] = $-0.21 \pm 0.01$. It is important to note, however, that the metallicity values derived from retrieval calculations assume that all major carriers of each element are adequately constrained. Species with limited spectral signatures, such as molecular nitrogen, may not significantly contribute to the observed spectrum and could therefore be absent from the metallicity estimates. As a result, our derived N/H value should be interpreted as a lower limit.

The apparent enrichment of carbon relative to oxygen may partly result from condensation processes. In the deeper layers of the atmosphere, some oxygen is sequestered into oxygen-bearing condensates, such as for example forsterite (\ch{Mg2SiO4}). As a result, the gas-phase C/O ratio may still not accurately reflect the atmosphere’s bulk composition.

\begin{figure}
    \centering
    \includegraphics[width=0.77\linewidth]{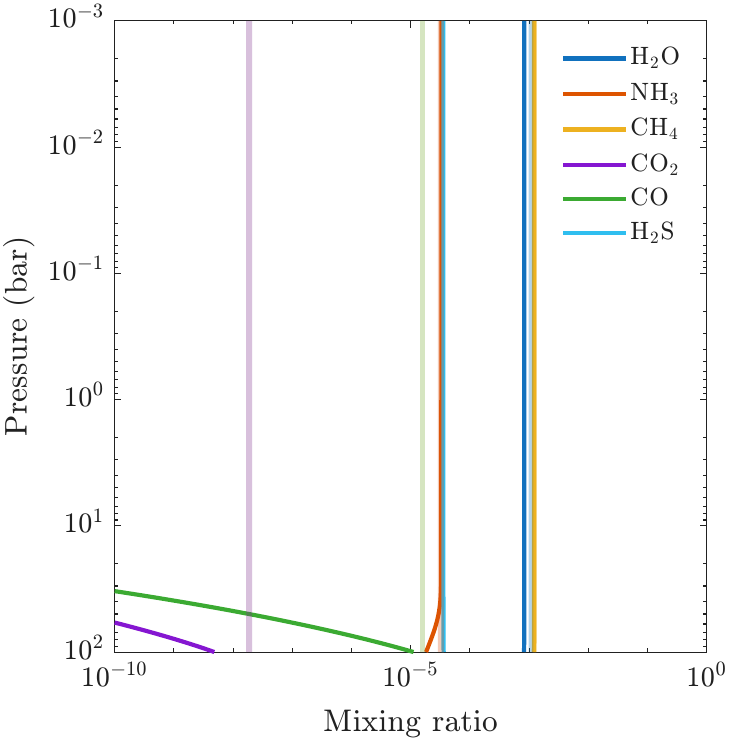}\\
    \includegraphics[width=0.8\linewidth]{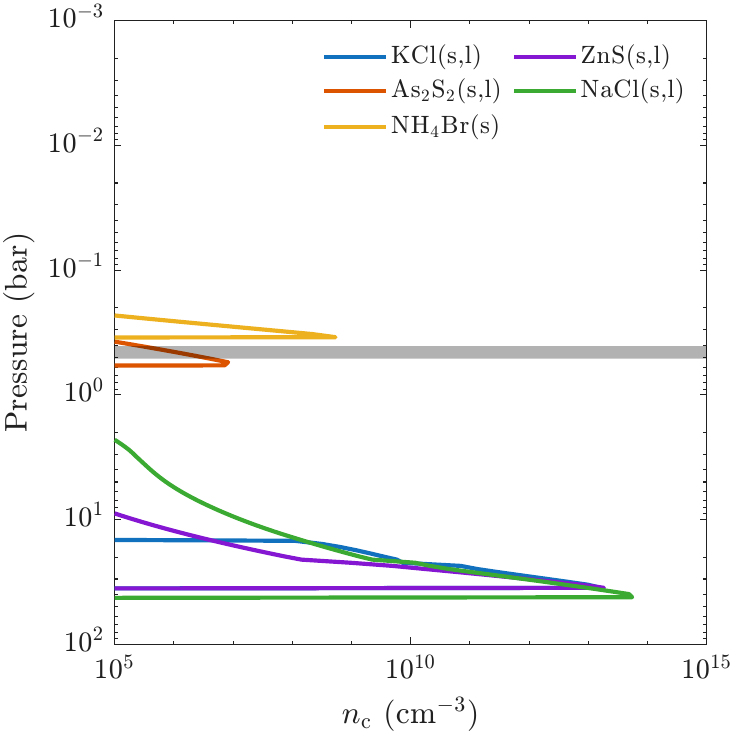}
    \caption{Chemical composition for Y1-dwarf WISEPAJ1541-22 calculated with \fc. The upper panel shows the volume mixing ratios of major gas-phase species. Retrieved abundances from the \bear retrieval and their 1$\sigma$ confidence intervals are depicted by the shaded areas. Stable condensates are shown in the lower panel. Their number densities $n_\mathrm{c}$ refer to the effective molecule number contained in the condensed phase and do not correspond to an actual cloud particle number density. The cloud top position and its 1$\sigma$ confidence interval constrained by the retrieval calculations ($p_{\mathrm{t}} = 0.47_{-0.06}^{+0.04}$~bar) are marked by the horizontal, gray area. Note that the corresponding cloud bottom pressure ($p_{\mathrm{b}} = 1.55_{-0.21}^{+0.16}$~bar) lies approximately one atmospheric scale height beneath the cloud top pressure.
    %The cloud position ($p_{\mathrm{t}}$) and its 1$\sigma$ confidence interval constrained by the retrieval calculations are marked by the horizontal, gray area.
    }
    \label{fig:Case_fastchem}
\end{figure}

To assess the consistency of our results from a chemical perspective, we performed additional calculations using the \fc chemistry code \citep{Stock2018MNRAS.479..865S, Stock2022MNRAS.517.4070S}. \fc is a chemical equilibrium model that includes condensate formation \citep{Kitzmann2024MNRAS} and incorporates the commonly used rainout approximation, particularly relevant for brown dwarf atmospheres. For this analysis, we used the temperature–pressure profile and elemental abundances for C, O and N, obtained from the retrieval results to compute the gas-phase chemical composition and the potential formation of condensates as a function of pressure. The results are shown in Fig.~\ref{fig:Case_fastchem}. These calculations employed an expanded set of chemical elements, gas-phase species, and condensates recently added to \fc (Kitzmann et al., in prep.).

The gas-phase mixing ratios for \ch{H2O}, \ch{CH4}, and \ch{NH3} predicted by \fc are consistent with the retrieved abundances. These species also appear to exhibit nearly constant mixing ratios (i.e., isoprofiles), supporting the common retrieval assumption of uniform vertical abundances. In contrast, the equilibrium abundances of \ch{CO2} and \ch{CO} are significantly lower than the retrieved values, suggesting the influence of vertical mixing. As shown in Fig.~\ref{fig:Case_fastchem}, this mixing likely originates from pressure levels around $\sim10^2$ bar, where the retrieved abundances intersect the chemical equilibrium curves.

%\fc also predicts hydrogen sulfide to be one of the primary sulfur-bearing species. In our retrieval analysis, \ch{H2S} was only constrained to an upper limit on its mixing ratio, between approximately $10^{-4}$ and $10^{-5}$. This upper limit is close to the abundance predicted by \fc, suggesting that \ch{H2S} may not be directly constrained by the available spectral data.

The bottom panel of Fig.~\ref{fig:Case_fastchem} displays the stable condensate species predicted by \fc using the rainout approach. Three commonly considered condensates, potassium chloride (\ch{KCl}), zinc sulfide (\ch{ZnS}), and sodium chloride (\ch{NaCl}), appear to form too deep in the atmosphere compared to the retrieved cloud pressure of approximately 0.5 bar. Notably, both \ch{H2O} and \ch{NaS2} are not predicted to be present by \fc.
%While condensates forming deeper than the retrieved cloud pressure can likely be excluded, species forming at higher altitudes cannot be definitively ruled out given the simplicity of our gray cloud model. Nonetheless, the high abundance of gaseous water from the retrievals suggests that water has not yet condensed in the atmospheric layers probed by spectroscopy.
Interestingly, two species predicted to be stable near 0.5 bar are arsenic(II) sulfide (\ch{As2S2(s,l)}) and ammonium bromide (\ch{NH4Br(s)}). Whether these species are sufficiently abundant to form optically thick cloud layers remains uncertain. A definitive assessment would require detailed cloud-formation modeling, which is beyond the scope of this study.

The atmospheric retrieval supports the presence of an optically thick cloud layer, and while chemical equilibrium calculations point to plausible condensate candidates, no further, definitive conclusion may be drawn.

\subsection{Are cloudy models required to fit the spectra of late-T and Y-dwarfs?}

Clouds have long been considered essential for accurately modeling the spectra of late T and Y dwarfs, particularly because they help replicate the spectral reddening observed in these ultra-cool objects. The work by \cite{Morley2012ApJ...756..172M, Morley2014ApJ...787...78M} emphasizes this point, demonstrating that water and sulfide clouds, which become optically significant at effective temperatures below 350–375~K, can align model spectra more closely with observational data.

\begin{table}[h]
\renewcommand{\arraystretch}{1.1}
\scriptsize
\centering
\caption{Summary of Bayesian statistics for the retrieval analysis of the curated set of brown dwarfs. The preferred model with its degree of preference according to \cite{Trotta2008ConPh..49...71T}, and $|\ln B_{ij}|$ (with $i$ = cloud free: CF, $j$ = gray cloud: G) are indicated.}
\label{tab:BeAR_Bayes_table}
\resizebox{\columnwidth}{!}{
\begin{tabular}{ccccc}
\hline 
Object & Spectral & Preferred & Degree of & $\left|\ln B_{ij}\right|$ \\
Name & Type & Model & Preference &  (CF vs G) \\
\hline
SDSSJ1624+00 & T6 & cloud-free & inconclusive & 0.88 \\
WISEJ1501-40 & T6 & cloud-free & inconclusive & 0.63 \\
SDSSpJ1346-00 & T6.5 & cloud-free & weak & 1.28 \\
ULASJ1029+09 & T8 & cloud-free & weak & 2.01 \\
WISEJ0247+37 & T8 & cloud-free & strong & 10.52 \\
WISEJ0430+46 & T8 & gray cloud & moderate & 3.17 \\
WISEPAJ1959-33 & T8 & cloud-free & inconclusive & 0.04 \\
WISEPAJ0313+78 & T8.5 & gray cloud & strong & 73.82 \\
WISEAJ2159-48 & T9 & gray cloud & strong & 12.44 \\
WISEJ2102-44 & T9 & gray cloud & strong & 45.87 \\
WISEJ2209+27 & Y0 & cloud-free & weak & 1.56 \\
WISEJ0359-54 & Y0 & cloud-free & weak & 1.05 \\
WISEJ0734-71 & Y0 & cloud-free & strong & 46.49 \\
WISEJ1206+84 & Y0 & gray cloud & strong & 21.12 \\
WISEPCJ2056+14 & Y0 & gray cloud & strong & 9.07 \\
WISEJ0825+28 & Y0.5 & cloud-free & inconclusive & 0.88 \\
WISEPCJ1405+55 & Y0.5 & cloud-free & strong & 21.12 \\
WISEJ0535-75 & Y1 & cloud-free & inconclusive & 0.82 \\
WISEPAJ1541-22 & Y1 & gray cloud & strong & 11.12 \\
CWISEPJ1047+54 & Y1 & gray cloud & strong & 15.82 \\
WISEAJ2354+02 & Y1 & cloud-free & inconclusive & 0.39 \\
CWISEPJ1446-23 & Y1 & cloud-free & weak & 1.71 \\
\hline
\end{tabular}}
\end{table}

However, findings by \cite{Tremblin2015ApJ} challenge this necessity, presenting a cloud-free alternative that successfully reproduces the spectra of Y dwarfs when accounting for factors such as vertical mixing and \ch{NH3} quenching. \cite{Tremblin2015ApJ} suggest that modifying the atmospheric temperature gradient to incorporate fingering convection--a process driven by condensation and subsequent chemical quenching--can explain the reddening previously attributed to clouds.

\cite{Leggett2017ApJ...842..118L} examined the atmospheric properties of Y-type brown dwarfs and found that non-equilibrium cloud-free models accurately replicate the near-infrared spectra and mid-infrared photometry of warmer Y dwarfs with effective temperatures between 425~K and 450~K. While Y dwarfs are often modeled with cloud-free atmospheres due to their low temperatures, \cite{Leggett2017ApJ...842..118L} proposed that thin cloud layers composed of sulfides and other condensates might still influence the spectra of even colder Y dwarfs, particularly in the near-infrared range.

Recently, \cite{Lacy2023ApJ...950....8L} developed a set of self-consistent model atmospheres for Y dwarfs and highlighted that disequilibrium in \ch{CH4}-\ch{CO} and \ch{NH3}-\ch{N2} chemistry, along with the presence of water clouds, could improve, although still not fully resolve, the agreement between models and observations of Y dwarfs.

Knowing that Y dwarfs are rapid rotators \citep{Hsu2021ApJS..257...45H}, rotational dynamics are expected to substantially impact atmospheric energy transport, as recent atmospheric circulation models of brown dwarfs and exoplanets demonstrated \citep{Tan2021MNRAS.502..678T}. Recent enhancements to General Circulation Model (GCM) simulations have included the integration of consistent chemical equilibrium, advanced cloud microphysics, and fully coupled three-dimensional radiative transfer, enabling a more accurate representation of the atmospheric conditions of the coldest brown dwarfs \citep{Lee2025A&A...695A.111L}.

These findings imply that turbulence, vertical mixing, and disequilibrium chemistry dominate the shaping of the spectra of late-T and Y dwarfs. The presence or absence of water clouds is still under active debate. In the context of our Bayesian model comparison analysis, cloud-free models are preferred for the hottest objects of the curated sample (T6 - T8), while later type objects vary in their preference for gray-cloud and cloud-free models, sometimes providing equally good fits to the data (Tab.~\ref{tab:BeAR_Bayes_table}). Non-gray models were never preferred by the data in the atmospheric retrieval analysis of our curated set of 22 late-T and Y-dwarfs. However, strong indications can be found for the presence of vertical mixing and disequilibrium chemistry. While our retrieval framework captures the effects of these processes, it does not rule out alternative explanations such as fingering convection proposed by \citet{Tremblin2015ApJ}, which may produce similar observational signatures.

\subsection{What are the trends in the retrieved surface gravity, effective temperature, or radii across the T-Y transition?}

\begin{figure*}[ht]
    \centering
    \includegraphics[width=\linewidth]{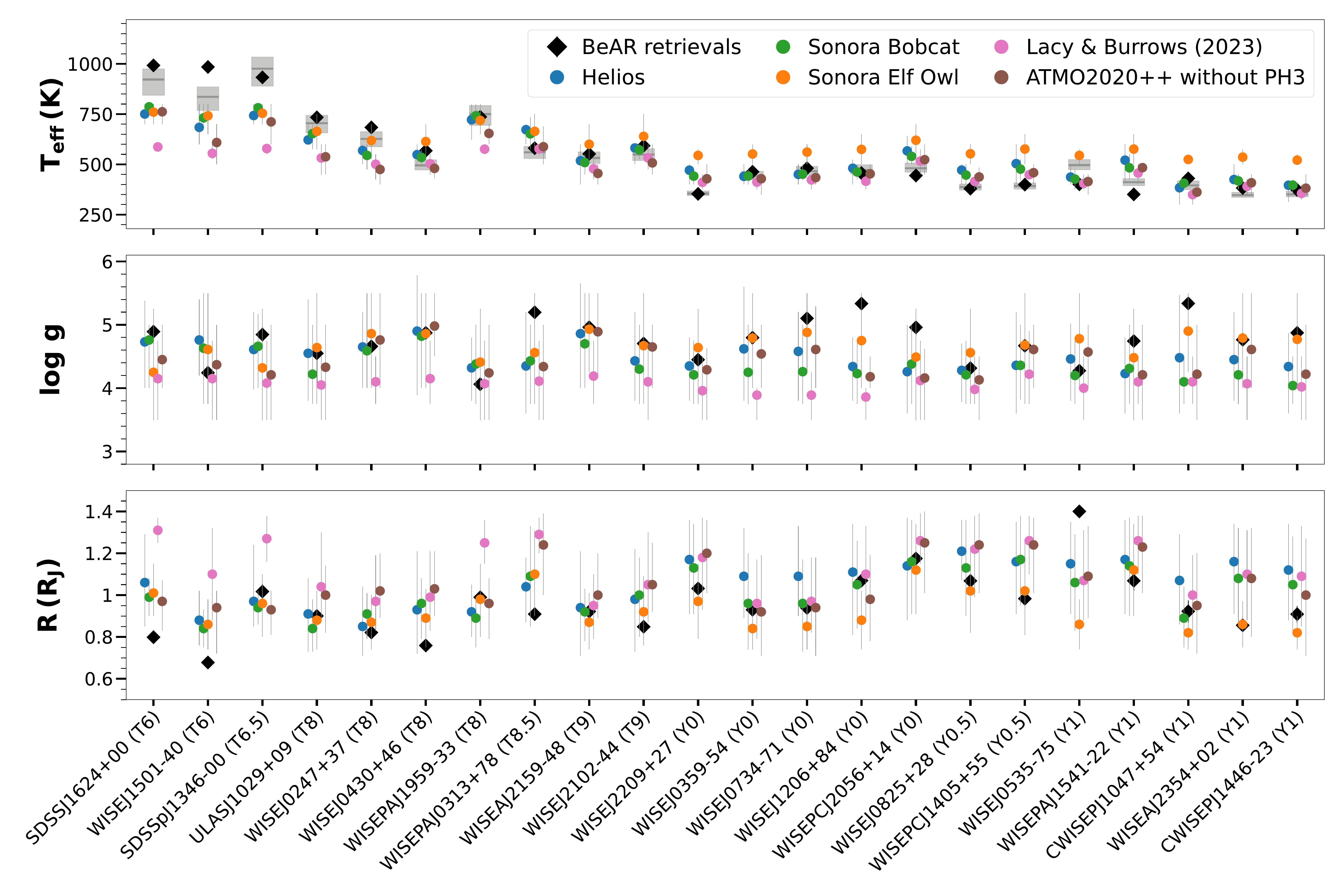}
    \caption{Comparison of retrieved posterior outcomes for effective temperature (top), surface gravity (middle), and radii (bottom) from our suite of brown dwarf retrievals across the T–Y sequence. The brown dwarfs are ordered with respect to their spectral class, from T6 to Y1 (see Tab. \ref{tab:objects} for details). For each object within the curated sample, all permutations of the theoretical grid models, which were used for training the \texttt{HELA} random forest routine, are shown, as well as the posterior parameters retrieved by using the nested sampling \bear framework. For each posterior, the 1$\sigma$ uncertainties are shown, but are too small to be visible for \bear posteriors. Gray bars in the top panel indicate the effective temperature values determined by \cite{Beiler2024ApJ...973..107B}, using integrated luminosities and radii from evolutionary models (see their Table 4).
    }
    \label{fig:sequence_post}
\end{figure*}

In this section, we examine trends in the retrieved physical properties of brown dwarfs across the T--Y spectral sequence. Key parameters, including effective temperature ($T_{\mathrm{eff}}$), surface gravity ($\log g$), and radius ($R$), are presented in Fig.~\ref{fig:sequence_post} for the complete sample analyzed in this study. We compare results from atmospheric retrievals using the Bayesian atmospheric retrieval code \bear with those obtained from random forest machine-learning models trained on various atmospheric grids, as introduced in Sect.~\ref{sec:model_grids}. The posterior spectral fits and all retrieved temperature-pressure profiles are shown in Figs.~\ref{fig:bestfits} and~\ref{fig:tp} in Appendix \ref{sec:appendix_figures}.

Across the T--Y transition, the retrieved values align well with predictions from brown dwarf evolutionary models (e.g., \citealt{Filippazzo2015ApJ, Kirkpatrick2021ApJS, Marley2021ApJ...920...85M}). As shown in Fig.~\ref{fig:sequence_post}, effective temperatures decrease from $\sim$1000\,K in late-T dwarfs to $\sim$400\,K in Y dwarfs \citep{Cushing2011ApJ...743...50C, Leggett2017ApJ...842..118L}. As brown dwarfs cool over Gyr timescales, their mass, radius, and effective temperature are intrinsically linked through evolutionary cooling tracks \citep{Marley2021ApJ...920...85M}. While evolutionary models predict a modest increase in surface gravity as a result of long-term contraction, the associated change in radius is small, and no clear trend in $\log g$ is observed in our retrieval outcomes. Notably, retrievals often yield broader $\log g$ distributions for Y dwarfs \citep{Zalesky2022ApJ...936...44Z}, likely due to degeneracies between temperature, gravity, and cloud opacity, compounded by limited spectral resolution in observations of these faint objects. Additionally, comparing retrieved radii with model predictions, which suggest nearly constant values between 0.8 and 1.2\,$R_{\mathrm{Jup}}$, is complicated by uncertainties in object age and cloud structure \citep{Burrows2011ApJ}. For a local sample of cool and faint brown dwarfs, ages of a few Gyr—comparable to the Sun—are expected, consistent with recent population studies \citep{Best2024ApJ...967..115B}. For example, at an age of $\sim$4~Gyr, evolutionary models predict that a 750~K T dwarf has $\log g \approx 5.1$ and $R \approx 0.088~R_{\mathrm{Jup}}$, while a 400~K Y dwarf has $\log g \approx 4.6$ and $R \approx 0.101~R_{\mathrm{Jup}}$, broadly consistent with the trends shown in Fig.~\ref{fig:sequence_post}. In some cases, the retrieved radii are anomalously large, as observed for Y1 dwarf WISEJ0535-75. These discrepancies may indicate unresolved binarity, a younger age, or systematic errors in distance estimates or atmospheric modeling. However, we generally do not face the well-known radius problem, where retrieved radii tend to be un-physically small.

Both retrieval approaches, nested sampling and random forest, produced broadly consistent results. Discrepancies in retrieved temperatures were found primarily for the hottest objects in the sample, including the T6 dwarfs SDSSJ1624+00 and WISEJ1501-40, and the T6.5 dwarf SDSSpJ1346-00. These differences are attributed to the upper temperature limit (800\,K) of the atmospheric model grid used in the machine learning model. Moreover, the random forest method generally produced more conservative estimates, characterized by larger uncertainties. This behavior likely results from limitations in the size and resolution of the training grid (see e.g., \citealt{Lueber2023ApJ...954...22L}).

Comparisons with previous studies \citep{Beiler2024ApJ...973..107B, Tu2024ApJ...976...82T} show good agreement in effective temperature estimates, while greater variability is seen in retrieved surface gravities, which are inherently more model-dependent given the strong degeneracy between surface gravity and metallicity when fitting spectra of late-T and Y dwarfs. One notable exception is the Y1 dwarf CWISEPJ1047+54. \citet{Tu2024ApJ...976...82T} report a remarkably low surface gravity of $\log g = 2.50 \pm 0.01$ from their \texttt{ATMO2020++} grid fit, whereas our \bear retrieval yields a significantly higher value of $\log g = 5.34 \pm 0.08$. \citet{Tu2024ApJ...976...82T} caution that their posterior converged at the edge of the model grid and should be interpreted with care. This sensitivity to model assumptions is further illustrated by the large differences in surface gravity and metallicity found by \citet{Tu2024ApJ...976...82T} when comparing \texttt{ATMO2020++} and \texttt{Sonora Elf Owl} models, underscoring the impact of model-dependent degeneracies. While the random forest trained on a variety of different grids also predict smaller $\log g$ values for this object than \bear, these values are still considerably higher than the estimate by \citet{Tu2024ApJ...976...82T}. We note that $\log g \gtrsim 5$ for Y dwarfs would imply implausibly old ages greater than 15~Gyr according to evolutionary models, likely indicating residual degeneracies in the models rather than physical constraints. In contrast, effective temperature estimates for this object differ by only $\sim$50\,K, with \citet{Tu2024ApJ...976...82T} reporting $T_{\mathrm{eff}} = 381.0 \pm 0.2$ \,K and our \bear retrieval yielding $T_{\mathrm{eff}} = 430 \pm 6$\,K using a gray-cloud model. Despite this discrepancy, the overall consistency across retrieval techniques underscores the robustness of the inferred atmospheric parameters within the bounds of current model limitations.

\subsection{What can we learn about the chemistry of late T and Y dwarf atmospheres?}

\begin{figure*}[ht]
    \centering
    \includegraphics[width=0.88\linewidth]{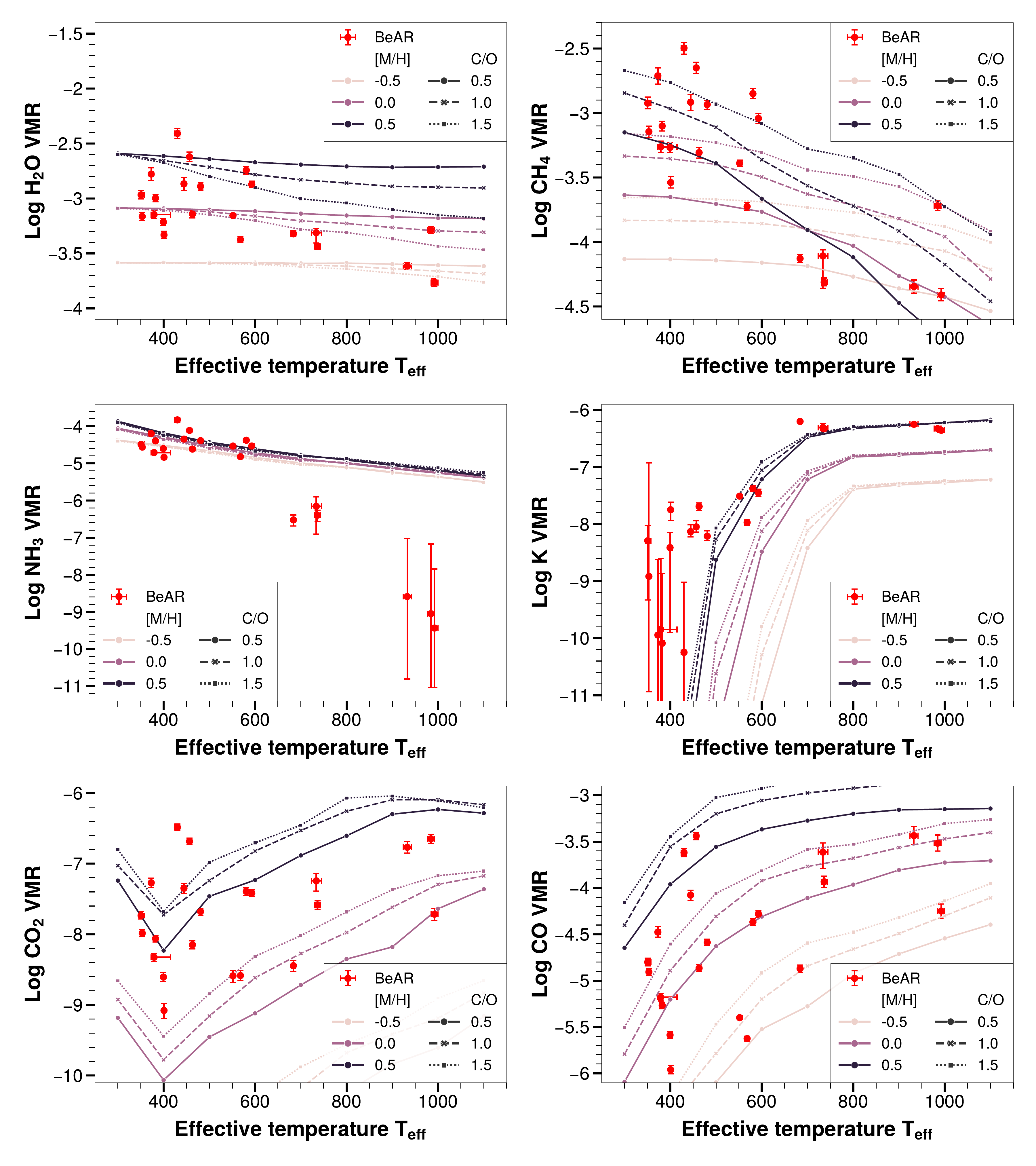}
    \caption{Posterior \bear retrieval values for molecular log-volume mixing ratios (log VMR) of \ch{H2O}, \ch{CH4}, \ch{NH3}, \ch{K}, \ch{CO2}, and \ch{CO}, plotted against the retrieved effective Temperature \teff \, are shown in red. For visual comparison, modeled volume mixing ratios from the Sonora Elf Owl model grid \citep{Mukherjee2024ApJ...963...73M} are overlaid for a surface gravity of \logg = 5.0 (cgs units) and vertical eddy diffusion coefficient $\log\kappa_{zz}=8.0$ with varying metallicities ([M/H]) and carbon-to-oxygen (C/O) ratios relative to solar.}
    \label{fig:Molecules_vs_Teff}
\end{figure*}

The atmospheres of late T and Y dwarfs exhibit complex chemical interactions influenced by disequilibrium processes and vertical mixing, as well as by condensate rainout, both of which have been established by early observational and theoretical studies (e.g. \citealt{Noll1997ApJ...489L..87N, Saumon2000ApJ...541..374S, Leggett2007ApJ...655.1079L, geballe09, Lodders1999ApJ...519..793L, Visscher2006ApJ...648.1181V, Visscher2010ApJ...716.1060V, Morley2012ApJ...756..172M}). These mechanisms significantly affect the abundances of key molecules such as \ch{CO}, \ch{CH4}, \ch{CO2}, \ch{NH3}, and \ch{PH3}. \citet{Zahnle2014ApJ...797...41Z} demonstrated that vertical mixing enhances \ch{CO} abundances beyond equilibrium predictions, even in the colder atmospheres of late T and Y dwarfs, where \ch{CH4} would otherwise dominate. This disequilibrium effect is particularly pronounced in low-gravity objects, as it transports \ch{CO} from deeper, hotter regions into the observable atmosphere, thereby challenging traditional equilibrium-based models. It is further enhanced at higher metallicities, which increases vertical mixing \citep{Zahnle2014ApJ...797...41Z}.

Observational studies support these theoretical predictions. \citet{Miles2020AJ....160...63M} detected CO in late T and Y dwarfs at levels inconsistent with equilibrium chemistry. Similarly, \citet{Hubeny2007ApJ...669.1248H} showed that vertical mixing alters both the thermal structure and spectral features, highlighting the importance of including disequilibrium chemistry in atmospheric modeling.

\citet{Beiler2024ApJ...973...60B} examined discrepancies in the predicted abundances of \ch{PH3} and \ch{CO2}, noting that existing models often underestimate \ch{CO2} while overestimating \ch{PH3} in cool atmospheres. This suggests an incomplete understanding of phosphorus chemistry in late T and Y dwarfs. \citet{Morley2014ApJ...787...78M} further predicted that PH$_3$, although difficult to detect under equilibrium conditions, could become more abundant through disequilibrium processes, leading to a prominent mid-infrared spectral feature near 4.3\,$\mu$m in Y dwarfs with $T_\mathrm{eff} \sim 450$\,K.

Our retrieval analysis provides additional evidence for chemical disequilibrium, as illustrated in Fig.~\ref{fig:Molecules_vs_Teff} and summarized in Tabs.~\ref{tab:BeAR_Post1_CF}--\ref{tab:BeAR_Post2_G} of Appendix \ref{sec:appendix_tables}. Both \ch{CH4} and \ch{CO} are retrieved simultaneously, consistent with the scenario described by \citet{Zahnle2014ApJ...797...41Z}, in which vertical mixing transports \ch{CO} into the upper atmosphere. Furthermore, our results confirm the findings of \citet{Beiler2024ApJ...973...60B}: \ch{PH3} is absent or weakly present in the observed atmospheres, while \ch{CO2} appears in higher-than-expected concentrations in several objects compared to equilibrium chemistry calculations. In this low-temperature regime, \ch{CO} and \ch{CO2} abundances are highly sensitive to the location of the chemical quench level, such that small variations in vertical mixing or thermal structure can lead to significant changes in the retrieved volume mixing ratios across the sample.

Trends in the retrieved abundances of \ch{H2O}, \ch{CH4}, and \ch{NH3} show an anticipated decline with increasing effective temperature ($T_\mathrm{eff}$), in agreement with the results of \citet{Zalesky2019ApJ...877...24Z, Zalesky2022ApJ...936...44Z}. Notably, \ch{NH3} in the near-infrared is only detectable in Y dwarfs, reinforcing its role as a spectral signature of the Y-dwarf class.

A clear depletion trend is also observed for potassium, decreasing from the hotter late T dwarfs to the cooler Y dwarfs. This trend aligns with predictions from the Sonora Elf Owl model grid \citep{Mukherjee2024ApJ...963...73M} and is driven by rainout chemistry. Around 1300\,K, aluminum and silicate condensates begin to form, depleting the available elemental reservoirs. Subsequent, additional elements are depleted as well, including sodium and potassium near 700\,K. There they condense into \ch{Na2S(s)} and KCl(s), respectively \citep{Zalesky2022ApJ...936...44Z}.

\section{Summary and Conclusions}

In this study, we performed a retrieval analysis on a curated sample of 22 late-T and Y dwarfs, originally observed by \citet{Beiler2024ApJ...973..107B} using the NIRSpec and MIRI instruments aboard the \textit{James Webb Space Telescope}. Our main findings are summarized as follows:

\begin{enumerate}
    \item Clouds play a critical role in modeling the spectra of late-T and Y dwarfs, with their impact modulated by vertical mixing and disequilibrium chemistry. Bayesian model comparison indicates that cloud-free models are generally favored for the hottest objects in the sample (T6--T8). In contrast, later-type dwarfs show varying preferences, with both gray-cloud and cloud-free models yielding comparable fits. Several objects exhibit clear signatures of disequilibrium chemistry; notably, the Y1 dwarf WISEPAJ1541-22 favors a gray-cloud model and shows elevated abundances of both CO and \ch{CH4}.
    
    \item Retrieved physical parameters across the T--Y spectral transition are consistent with predictions from evolutionary models. Specifically, effective temperatures decrease across the sample, while radii remain within the expected range of 0.8--1.2\,$R_{\mathrm{Jup}}$. The results are generally consistent within uncertainties between the Bayesian nested-sampling atmospheric retrieval and the supervised machine learning random forest approach. Notably, we do not encounter the well-known radius problem in atmospheric retrievals, where retrieved radii are often constrained to be un-physically small.
    
    %\item A feature importance analysis using the HELA random forest method provides insight into the spectroscopic features that constrain model parameters. This data-driven approach highlights the role of opacity sources, temperature--pressure profiles, and chemical abundances in shaping observable spectra.
    
    \item Retrievals from nested sampling and random forest methods show strong overall agreement. However, the random forest results tend to be more conservative, with broader uncertainties--an outcome primarily reflecting the method’s intrinsic tendency to average over parameter space. Additionally, limitations in grid resolution and spacing may further contribute to this broadening.
    
    \item Atmospheres of late-T and Y dwarfs show evidence of chemical disequilibrium, particularly due to vertical mixing. Both \ch{CH4} and CO are retrieved, with CO likely transported from deeper atmospheric layers. As physically expected, retrieved abundances of \ch{H2O}, \ch{CH4}, and \ch{NH3} decrease with increasing effective temperature, consistent with trends reported by \citet{Zalesky2019ApJ...877...24Z, Zalesky2022ApJ...936...44Z}.
\end{enumerate}

\begin{acknowledgements}
We acknowledge partial financial support from the Swiss National Science Foundation (via grant No. 192022 awarded to K.H.), the European Research Council (ERC) Geoastronomy Synergy Grant (via grant No. 101166936 awarded to K.H.), as well as administrative support from the Center for Space and Habitability (CSH). D.K. acknowledges the support from the Swiss National Science Foundation under the grant 200021-231596. We also like to thank Adam Burgasser for his valuable input. 
\end{acknowledgements}

\bibliographystyle{aa}
\bibliography{references.bib}

@ARTICLE{Leggett2025ApJ...979..145L,
       author = {{Leggett}, S.~K. and {Tremblin}, Pascal},
        title = "{Redshifting the Study of Cold Brown Dwarfs and Exoplanets: The Mid-infrared Wavelength Region as an Indicator of Surface Gravity and Mass}",
      journal = {\apj},
         year = 2025,
        month = feb,
       volume = {979},
       number = {2},
          eid = {145},
        pages = {145},
          doi = {10.3847/1538-4357/ad8fa6},
archivePrefix = {arXiv},
       eprint = {2411.03549},
 primaryClass = {astro-ph.SR},
       adsurl = {https://ui.adsabs.harvard.edu/abs/2025ApJ...979..145L}
}

@ARTICLE{Line2013ApJ,
       author = {{Line}, Michael R. and {Wolf}, Aaron S. and {Zhang}, Xi and {Knutson}, Heather and {Kammer}, Joshua A. and {Ellison}, Elias and {Deroo}, Pieter and {Crisp}, Dave and {Yung}, Yuk L.},
        title = "{A Systematic Retrieval Analysis of Secondary Eclipse Spectra. I. A Comparison of Atmospheric Retrieval Techniques}",
      journal = {\apj},
         year = 2013,
        month = oct,
       volume = {775},
       number = {2},
          eid = {137},
        pages = {137},
          doi = {10.1088/0004-637X/775/2/137},
archivePrefix = {arXiv},
       eprint = {1304.5561},
 primaryClass = {astro-ph.EP},
       adsurl = {https://ui.adsabs.harvard.edu/abs/2013ApJ...775..137L}
}

@ARTICLE{Beiler2024ApJ...973..107B,
       author = {{Beiler}, Samuel A. and {Cushing}, Michael C. and {Kirkpatrick}, J. Davy and {Schneider}, Adam C. and {Mukherjee}, Sagnick and {Marley}, Mark S. and {Marocco}, Federico and {Smart}, Richard L.},
        title = "{Precise Bolometric Luminosities and Effective Temperatures of 23 Late-T and Y Dwarfs Obtained with JWST}",
      journal = {\apj},
         year = 2024,
        month = oct,
       volume = {973},
       number = {2},
          eid = {107},
        pages = {107},
          doi = {10.3847/1538-4357/ad6301},
archivePrefix = {arXiv},
       eprint = {2407.08518},
 primaryClass = {astro-ph.SR},
       adsurl = {https://ui.adsabs.harvard.edu/abs/2024ApJ...973..107B}
}

@ARTICLE{Marquez2018NatAs...2..719M,
       author = {{M{\'a}rquez-Neila}, Pablo and {Fisher}, Chloe and {Sznitman}, Raphael and {Heng}, Kevin},
        title = "{Supervised machine learning for analysing spectra of exoplanetary atmospheres}",
      journal = {Nature Astronomy},
         year = 2018,
        month = jun,
       volume = {2},
        pages = {719-724},
          doi = {10.1038/s41550-018-0504-2},
archivePrefix = {arXiv},
       eprint = {1806.03944},
 primaryClass = {astro-ph.EP},
       adsurl = {https://ui.adsabs.harvard.edu/abs/2018NatAs...2..719M}
}

@article{Ho1998random,
  title={The random subspace method for constructing decision forests},
  author={Ho, Tin Kam},
  journal={IEEE transactions on pattern analysis and machine intelligence},
  volume={20},
  number={8},
  pages={832--844},
  year={1998},
  publisher={Ieee}
}

@ARTICLE{Linsky1969ApJ...156..989L,
       author = {{Linsky}, Jeffrey L.},
        title = "{On the Pressure-Induced Opacity of Molecular Hydrogen in Late-Type Stars}",
      journal = {\apj},
         year = 1969,
        month = jun,
       volume = {156},
        pages = {989},
          doi = {10.1086/150030},
       adsurl = {https://ui.adsabs.harvard.edu/abs/1969ApJ...156..989L}
}

@article{Breiman2001random,
  title={Random forests},
  author={Breiman, Leo},
  journal={Machine learning},
  volume={45},
  number={1},
  pages={5--32},
  year={2001},
  publisher={Springer}
}

@book{Sisson2018handbook,
  title={Handbook of approximate Bayesian computation},
  author={Sisson, Scott A and Fan, Yanan and Beaumont, Mark},
  year={2018},
  publisher={CRC Press}
}

@ARTICLE{Stock2022MNRAS.517.4070S,
       author = {{Stock}, Joachim W. and {Kitzmann}, Daniel and {Patzer}, A. Beate C.},
        title = "{FASTCHEM 2 : an improved computer program to determine the gas-phase chemical equilibrium composition for arbitrary element distributions}",
      journal = {\mnras},
         year = 2022,
        month = dec,
       volume = {517},
       number = {3},
        pages = {4070-4080},
          doi = {10.1093/mnras/stac2623},
archivePrefix = {arXiv},
       eprint = {2206.08247},
 primaryClass = {astro-ph.EP},
       adsurl = {https://ui.adsabs.harvard.edu/abs/2022MNRAS.517.4070S}
}

@ARTICLE{Asplund2009ARA&A..47..481A,
       author = {{Asplund}, Martin and {Grevesse}, Nicolas and {Sauval}, A. Jacques and {Scott}, Pat},
        title = "{The Chemical Composition of the Sun}",
      journal = {\araa},
         year = 2009,
        month = sep,
       volume = {47},
       number = {1},
        pages = {481-522},
          doi = {10.1146/annurev.astro.46.060407.145222},
archivePrefix = {arXiv},
       eprint = {0909.0948},
 primaryClass = {astro-ph.SR},
       adsurl = {https://ui.adsabs.harvard.edu/abs/2009ARA&A..47..481A}
}

@ARTICLE{Kitzmann2024MNRAS,
       author = {{Kitzmann}, Daniel and {Stock}, Joachim W. and {Patzer}, A. Beate C.},
        title = "{FASTCHEM COND: equilibrium chemistry with condensation and rainout for cool planetary and stellar environments}",
      journal = {\mnras},
         year = 2024,
        month = jan,
       volume = {527},
       number = {3},
        pages = {7263-7283},
          doi = {10.1093/mnras/stad3515},
archivePrefix = {arXiv},
       eprint = {2309.02337},
 primaryClass = {astro-ph.EP},
       adsurl = {https://ui.adsabs.harvard.edu/abs/2024MNRAS.527.7263K}
}

@ARTICLE{Stock2018MNRAS.479..865S,
   author = {{Stock}, J.~W. and {Kitzmann}, D. and {Patzer}, A.~B.~C. and
	{Sedlmayr}, E.},
    title = "{FastChem: A computer program for efficient complex chemical equilibrium calculations in the neutral/ionized gas phase with applications to stellar and planetary atmospheres}",
  journal = {\mnras},
archivePrefix = "arXiv",
   eprint = {1804.05010},
 primaryClass = "astro-ph.EP",
     year = 2018,
    month = sep,
   volume = 479,
    pages = {865-874},
      doi = {10.1093/mnras/sty1531},
   adsurl = {http://adsabs.harvard.edu/abs/2018MNRAS.479..865S}
}

@ARTICLE{Olson1987JQSRT..38..325O,
   author = {{Olson}, G.~L. and {Kunasz}, P.~B.},
    title = "{Short characteristic solution of the non-LTE transfer problem by operator perturbation. I. The one-dimensional planar slab.}",
  journal = {\jqsrt},
     year = 1987,
   volume = 38,
    pages = {325-336},
      doi = {10.1016/0022-4073(87)90027-6},
   adsurl = {http://adsabs.harvard.edu/abs/1987JQSRT..38..325O}
}

@ARTICLE{Line2015ApJ,
   author = {{Line}, M.~R. and {Teske}, J. and {Burningham}, B. and {Fortney}, J.~J. and
	{Marley}, M.~S.},
    title = "{Uniform Atmospheric Retrieval Analysis of Ultracool Dwarfs. I. Characterizing Benchmarks, Gl 570D and HD 3651B}",
  journal = {\apj},
archivePrefix = "arXiv",
   eprint = {1504.06670},
 primaryClass = "astro-ph.SR",
     year = 2015,
    month = jul,
   volume = 807,
      eid = {183},
    pages = {183},
      doi = {10.1088/0004-637X/807/2/183},
   adsurl = {http://adsabs.harvard.edu/abs/2015ApJ...807..183L}
}

@ARTICLE{Grimm2015ApJ,
   author = {{Grimm}, S.~L. and {Heng}, K.},
    title = "{HELIOS-K: An Ultrafast, Open-source Opacity Calculator for Radiative Transfer}",
  journal = {\apj},
archivePrefix = "arXiv",
   eprint = {1503.03806},
 primaryClass = "astro-ph.EP",
     year = 2015,
    month = aug,
   volume = 808,
      eid = {182},
    pages = {182},
      doi = {10.1088/0004-637X/808/2/182},
   adsurl = {http://adsabs.harvard.edu/abs/2015ApJ...808..182G}
}

@INPROCEEDINGS{Skilling2004AIPC,
   author = {{Skilling}, J.},
    title = "{Nested Sampling}",
booktitle = {American Institute of Physics Conference Series},
     year = 2004,
   series = {24th International Workshop on Bayesian Inference and Maximum Entropy Methods in Science and Engineering, American Institute of Physics Conference Series},
   volume = 735,
   editor = {{Fischer}, R. and {Preuss}, R. and {Toussaint}, U.~V.},
    month = nov,
    pages = {395-405},
      doi = {10.1063/1.1835238},
   adsurl = {http://adsabs.harvard.edu/abs/2004AIPC..735..395S}
}

@ARTICLE{Feroz2008MNRAS,
   author = {{Feroz}, F. and {Hobson}, M.~P.},
    title = "{Multimodal nested sampling: an efficient and robust alternative to Markov Chain Monte Carlo methods for astronomical data analyses}",
  journal = {\mnras},
archivePrefix = "arXiv",
   eprint = {0704.3704},
     year = 2008,
    month = feb,
   volume = 384,
    pages = {449-463},
      doi = {10.1111/j.1365-2966.2007.12353.x},
   adsurl = {http://adsabs.harvard.edu/abs/2008MNRAS.384..449F}
}

@ARTICLE{Feroz2009MNRAS,
   author = {{Feroz}, F. and {Hobson}, M.~P. and {Bridges}, M.},
    title = "{MULTINEST: an efficient and robust Bayesian inference tool for cosmology and particle physics}",
  journal = {\mnras},
archivePrefix = "arXiv",
   eprint = {0809.3437},
     year = 2009,
    month = oct,
   volume = 398,
    pages = {1601-1614},
      doi = {10.1111/j.1365-2966.2009.14548.x},
   adsurl = {http://adsabs.harvard.edu/abs/2009MNRAS.398.1601F}
}

@ARTICLE{Noll1997ApJ...489L..87N,
       author = {{Noll}, Keith S. and {Geballe}, T.~R. and {Marley}, Mark S.},
        title = "{Detection of Abundant Carbon Monoxide in the Brown Dwarf Gliese 229B}",
      journal = {\apjl},
         year = 1997,
        month = nov,
       volume = {489},
       number = {1},
        pages = {L87-L90},
          doi = {10.1086/310954},
       adsurl = {https://ui.adsabs.harvard.edu/abs/1997ApJ...489L..87N}
}

@ARTICLE{Saumon2000ApJ...541..374S,
       author = {{Saumon}, D. and {Geballe}, T.~R. and {Leggett}, S.~K. and {Marley}, M.~S. and {Freedman}, R.~S. and {Lodders}, K. and {Fegley}, Jr., B. and {Sengupta}, S.~K.},
        title = "{Molecular Abundances in the Atmosphere of the T Dwarf GL 229B}",
      journal = {\apj},
         year = 2000,
        month = sep,
       volume = {541},
       number = {1},
        pages = {374-389},
          doi = {10.1086/309410},
archivePrefix = {arXiv},
       eprint = {astro-ph/0003353},
 primaryClass = {astro-ph},
       adsurl = {https://ui.adsabs.harvard.edu/abs/2000ApJ...541..374S}
}

@ARTICLE{Leggett2007ApJ...655.1079L,
       author = {{Leggett}, S.~K. and {Saumon}, D. and {Marley}, M.~S. and {Geballe}, T.~R. and {Golimowski}, D.~A. and {Stephens}, D. and {Fan}, X.},
        title = "{3.6-7.9 {\ensuremath{\mu}}m Photometry of L and T Dwarfs and the Prevalence of Vertical Mixing in their Atmospheres}",
      journal = {\apj},
         year = 2007,
        month = feb,
       volume = {655},
       number = {2},
        pages = {1079-1094},
          doi = {10.1086/510014},
archivePrefix = {arXiv},
       eprint = {astro-ph/0610214},
 primaryClass = {astro-ph},
       adsurl = {https://ui.adsabs.harvard.edu/abs/2007ApJ...655.1079L}
}

@ARTICLE{Lodders1999ApJ...519..793L,
       author = {{Lodders}, Katharina},
        title = "{Alkali Element Chemistry in Cool Dwarf Atmospheres}",
      journal = {\apj},
         year = 1999,
        month = jul,
       volume = {519},
       number = {2},
        pages = {793-801},
          doi = {10.1086/307387},
       adsurl = {https://ui.adsabs.harvard.edu/abs/1999ApJ...519..793L}
}

@ARTICLE{Visscher2006ApJ...648.1181V,
       author = {{Visscher}, Channon and {Lodders}, Katharina and {Fegley}, Jr., Bruce},
        title = "{Atmospheric Chemistry in Giant Planets, Brown Dwarfs, and Low-Mass Dwarf Stars. II. Sulfur and Phosphorus}",
      journal = {\apj},
         year = 2006,
        month = sep,
       volume = {648},
       number = {2},
        pages = {1181-1195},
          doi = {10.1086/506245},
archivePrefix = {arXiv},
       eprint = {astro-ph/0511136},
 primaryClass = {astro-ph},
       adsurl = {https://ui.adsabs.harvard.edu/abs/2006ApJ...648.1181V}
}

@ARTICLE{Visscher2010ApJ...716.1060V,
       author = {{Visscher}, Channon and {Lodders}, Katharina and {Fegley}, Jr., Bruce},
        title = "{Atmospheric Chemistry in Giant Planets, Brown Dwarfs, and Low-mass Dwarf Stars. III. Iron, Magnesium, and Silicon}",
      journal = {\apj},
         year = 2010,
        month = jun,
       volume = {716},
       number = {2},
        pages = {1060-1075},
          doi = {10.1088/0004-637X/716/2/1060},
archivePrefix = {arXiv},
       eprint = {1001.3639},
 primaryClass = {astro-ph.EP},
       adsurl = {https://ui.adsabs.harvard.edu/abs/2010ApJ...716.1060V}
}

@ARTICLE{Wogan2025RNAAS...9..108W,
       author = {{Wogan}, Nicholas F. and {Mang}, James and {Batalha}, Natasha E. and {Zahnle}, Kevin and {Mukherjee}, Sagnick and {Visscher}, Channon and {Fortney}, Jonathan J. and {Marley}, Mark S. and {Morley}, Caroline V.},
        title = "{The Sonora Substellar Atmosphere Models. V. A Correction to the Disequilibrium Abundance of CO$_{2}$ for Sonora Elf Owl}",
      journal = {Research Notes of the American Astronomical Society},
         year = 2025,
        month = may,
       volume = {9},
       number = {5},
          eid = {108},
        pages = {108},
          doi = {10.3847/2515-5172/add407},
archivePrefix = {arXiv},
       eprint = {2505.03994},
 primaryClass = {astro-ph.EP},
       adsurl = {https://ui.adsabs.harvard.edu/abs/2025RNAAS...9..108W}
}

@ARTICLE{Goody1989JQSRT..42..539G,
       author = {{Goody}, Richard and {West}, Robert and {Chen}, Luke and {Crisp}, David},
        title = "{The correlated-k method for radiation calculations in nonhomogeneous atmospheres.}",
      journal = {\jqsrt},
         year = 1989,
        month = dec,
       volume = {42},
        pages = {539-550},
          doi = {10.1016/0022-4073(89)90044-7},
       adsurl = {https://ui.adsabs.harvard.edu/abs/1989JQSRT..42..539G}
}

@ARTICLE{Karalidi2021ApJ...923..269K,
       author = {{Karalidi}, Theodora and {Marley}, Mark and {Fortney}, Jonathan J. and {Morley}, Caroline and {Saumon}, Didier and {Lupu}, Roxana and {Visscher}, Channon and {Freedman}, Richard},
        title = "{The Sonora Substellar Atmosphere Models. II. Cholla: A Grid of Cloud-free, Solar Metallicity Models in Chemical Disequilibrium for the JWST Era}",
      journal = {\apj},
         year = 2021,
        month = dec,
       volume = {923},
       number = {2},
          eid = {269},
        pages = {269},
          doi = {10.3847/1538-4357/ac3140},
archivePrefix = {arXiv},
       eprint = {2110.11824},
 primaryClass = {astro-ph.EP},
       adsurl = {https://ui.adsabs.harvard.edu/abs/2021ApJ...923..269K}
}

@ARTICLE{Morley2024ApJ,
       author = {{Morley}, Caroline V. and {Mukherjee}, Sagnick and {Marley}, Mark S. and {Fortney}, Jonathan J. and {Visscher}, Channon and {Lupu}, Roxana and {Gharib-Nezhad}, Ehsan and {Thorngren}, Daniel and {Freedman}, Richard and {Batalha}, Natasha},
        title = "{The Sonora Substellar Atmosphere Models. III. Diamondback: Atmospheric Properties, Spectra, and Evolution for Warm Cloudy Substellar Objects}",
      journal = {\apj},
         year = 2024,
        month = nov,
       volume = {975},
       number = {1},
          eid = {59},
        pages = {59},
          doi = {10.3847/1538-4357/ad71d5},
archivePrefix = {arXiv},
       eprint = {2402.00758},
 primaryClass = {astro-ph.SR},
       adsurl = {https://ui.adsabs.harvard.edu/abs/2024ApJ...975...59M}
}

@ARTICLE{2009LanB...4B..712L,
       author = {{Lodders}, K. and {Palme}, H. and {Gail}, H. -P.},
        title = "{Abundances of the Elements in the Solar System}",
      journal = {Landolt B{\"o}rnstein},
         year = 2009,
        month = jan,
       volume = {4B},
        pages = {712},
          doi = {10.1007/978-3-540-88055-4_34},
archivePrefix = {arXiv},
       eprint = {0901.1149},
 primaryClass = {astro-ph.EP},
       adsurl = {https://ui.adsabs.harvard.edu/abs/2009LanB...4B..712L}
}

@ARTICLE{Fisher2020AJ....159..192F,
       author = {{Fisher}, Chloe and {Hoeijmakers}, H. Jens and {Kitzmann}, Daniel and {M{\'a}rquez-Neila}, Pablo and {Grimm}, Simon L. and {Sznitman}, Raphael and {Heng}, Kevin},
        title = "{Interpreting High-resolution Spectroscopy of Exoplanets using Cross-correlations and Supervised Machine Learning}",
      journal = {\aj},
         year = 2020,
        month = may,
       volume = {159},
       number = {5},
          eid = {192},
        pages = {192},
          doi = {10.3847/1538-3881/ab7a92},
archivePrefix = {arXiv},
       eprint = {1910.11627},
 primaryClass = {astro-ph.EP},
       adsurl = {https://ui.adsabs.harvard.edu/abs/2020AJ....159..192F}
}

@ARTICLE{Guzman2020AJ....160...15G,
       author = {{Guzm{\'a}n-Mesa}, Andrea and {Kitzmann}, Daniel and {Fisher}, Chloe and {Burgasser}, Adam J. and {Hoeijmakers}, H. Jens and {M{\'a}rquez-Neila}, Pablo and {Grimm}, Simon L. and {Mandell}, Avi M. and {Sznitman}, Raphael and {Heng}, Kevin},
        title = "{Information Content of JWST NIRSpec Transmission Spectra of Warm Neptunes}",
      journal = {\aj},
         year = 2020,
        month = jul,
       volume = {160},
       number = {1},
          eid = {15},
        pages = {15},
          doi = {10.3847/1538-3881/ab9176},
archivePrefix = {arXiv},
       eprint = {2004.10106},
 primaryClass = {astro-ph.EP},
       adsurl = {https://ui.adsabs.harvard.edu/abs/2020AJ....160...15G}
}

@ARTICLE{Fisher2022ApJ...934...31F,
       author = {{Fisher}, Chloe and {Heng}, Kevin},
        title = "{How Do We Optimally Sample Model Grids of Exoplanet Spectra?}",
      journal = {\apj},
         year = 2022,
        month = jul,
       volume = {934},
       number = {1},
          eid = {31},
        pages = {31},
          doi = {10.3847/1538-4357/ac7801},
archivePrefix = {arXiv},
       eprint = {2206.12194},
 primaryClass = {astro-ph.EP},
       adsurl = {https://ui.adsabs.harvard.edu/abs/2022ApJ...934...31F}
}

@ARTICLE{Lueber2023ApJ...954...22L,
       author = {{Lueber}, Anna and {Kitzmann}, Daniel and {Fisher}, Chloe E. and {Bowler}, Brendan P. and {Burgasser}, Adam J. and {Marley}, Mark and {Heng}, Kevin},
        title = "{Intercomparison of Brown Dwarf Model Grids and Atmospheric Retrieval Using Machine Learning}",
      journal = {\apj},
         year = 2023,
        month = sep,
       volume = {954},
       number = {1},
          eid = {22},
        pages = {22},
          doi = {10.3847/1538-4357/ace530},
archivePrefix = {arXiv},
       eprint = {2305.07719},
 primaryClass = {astro-ph.SR},
       adsurl = {https://ui.adsabs.harvard.edu/abs/2023ApJ...954...22L}
}

@ARTICLE{Lueber2024A&A...687A.110L,
       author = {{Lueber}, Anna and {Novais}, Aline and {Fisher}, Chloe and {Heng}, Kevin},
        title = "{Information content of JWST spectra of WASP-39b}",
      journal = {\aap},
         year = 2024,
        month = jul,
       volume = {687},
          eid = {A110},
        pages = {A110},
          doi = {10.1051/0004-6361/202348802},
archivePrefix = {arXiv},
       eprint = {2405.02656},
 primaryClass = {astro-ph.EP},
       adsurl = {https://ui.adsabs.harvard.edu/abs/2024A&A...687A.110L}
}

@ARTICLE{Burrows2011ApJ,
       author = {{Burrows}, Adam and {Heng}, Kevin and {Nampaisarn}, Thane},
        title = "{The Dependence of Brown Dwarf Radii on Atmospheric Metallicity and Clouds: Theory and Comparison with Observations}",
      journal = {\apj},
         year = 2011,
        month = jul,
       volume = {736},
       number = {1},
          eid = {47},
        pages = {47},
          doi = {10.1088/0004-637X/736/1/47},
archivePrefix = {arXiv},
       eprint = {1102.3922},
 primaryClass = {astro-ph.SR},
       adsurl = {https://ui.adsabs.harvard.edu/abs/2011ApJ...736...47B}
}

@ARTICLE{Phillips2020A&A...637A..38P,
       author = {{Phillips}, M.~W. and {Tremblin}, P. and {Baraffe}, I. and {Chabrier}, G. and {Allard}, N.~F. and {Spiegelman}, F. and {Goyal}, J.~M. and {Drummond}, B. and {H{\'e}brard}, E.},
        title = "{A new set of atmosphere and evolution models for cool T-Y brown dwarfs and giant exoplanets}",
      journal = {\aap},
         year = 2020,
        month = may,
       volume = {637},
          eid = {A38},
        pages = {A38},
          doi = {10.1051/0004-6361/201937381},
archivePrefix = {arXiv},
       eprint = {2003.13717},
 primaryClass = {astro-ph.SR},
       adsurl = {https://ui.adsabs.harvard.edu/abs/2020A&A...637A..38P}
}

@ARTICLE{Heng2018ApJS..237...29H,
       author = {{Heng}, Kevin and {Malik}, Matej and {Kitzmann}, Daniel},
        title = "{Analytical Models of Exoplanetary Atmospheres. VI. Full Solutions for Improved Two-stream Radiative Transfer, Including Direct Stellar Beam}",
      journal = {\apjs},
         year = 2018,
        month = aug,
       volume = {237},
       number = {2},
          eid = {29},
        pages = {29},
          doi = {10.3847/1538-4365/aad199},
archivePrefix = {arXiv},
       eprint = {1804.04961},
 primaryClass = {astro-ph.EP},
       adsurl = {https://ui.adsabs.harvard.edu/abs/2018ApJS..237...29H}
}

@ARTICLE{geballe09,
   author = {{Geballe}, T.~R. and {Saumon}, D. and {Golimowski}, D.~A. and
	{Leggett}, S.~K. and {Marley}, M.~S. and {Noll}, K.~S.},
    title = "{Spectroscopic Detection of Carbon Monoxide in Two Late-Type T Dwarfs}",
  journal = {\apj},
archivePrefix = "arXiv",
   eprint = {0901.2134},
 primaryClass = "astro-ph.SR",
     year = 2009,
    month = apr,
   volume = 695,
    pages = {844-854},
      doi = {10.1088/0004-637X/695/2/844},
   adsurl = {http://adsabs.harvard.edu/abs/2009ApJ...695..844G}
}

@ARTICLE{Filippazzo2015ApJ,
   author = {{Filippazzo}, J.~C. and {Rice}, E.~L. and {Faherty}, J. and
	{Cruz}, K.~L. and {Van Gordon}, M.~M. and {Looper}, D.~L.},
    title = "{Fundamental Parameters and Spectral Energy Distributions of Young and Field Age Objects with Masses Spanning the Stellar to Planetary Regime}",
  journal = {\apj},
archivePrefix = "arXiv",
   eprint = {1508.01767},
 primaryClass = "astro-ph.SR",
     year = 2015,
    month = sep,
   volume = 810,
      eid = {158},
    pages = {158},
      doi = {10.1088/0004-637X/810/2/158},
   adsurl = {http://adsabs.harvard.edu/abs/2015ApJ...810..158F}
}

@ARTICLE{Burgasser2006ApJ...637.1067B,
       author = {{Burgasser}, Adam J. and {Geballe}, T.~R. and {Leggett}, S.~K. and {Kirkpatrick}, J. Davy and {Golimowski}, David A.},
        title = "{A Unified Near-Infrared Spectral Classification Scheme for T Dwarfs}",
      journal = {\apj},
         year = 2006,
        month = feb,
       volume = {637},
       number = {2},
        pages = {1067-1093},
          doi = {10.1086/498563},
archivePrefix = {arXiv},
       eprint = {astro-ph/0510090},
 primaryClass = {astro-ph},
       adsurl = {https://ui.adsabs.harvard.edu/abs/2006ApJ...637.1067B}
}

@ARTICLE{Allard2016A&A,
   author = {{Allard}, N.~F. and {Spiegelman}, F. and {Kielkopf}, J.~F.},
    title = "{K-H$_{2}$ line shapes for the spectra of cool brown dwarfs}",
  journal = {\aap},
     year = 2016,
    month = may,
   volume = 589,
      eid = {A21},
    pages = {A21},
      doi = {10.1051/0004-6361/201628270},
   adsurl = {http://adsabs.harvard.edu/abs/2016A%26A...589A..21A}
}

@ARTICLE{Burningham2017MNRAS,
   author = {{Burningham}, B. and {Marley}, M.~S. and {Line}, M.~R. and {Lupu}, R. and
	{Visscher}, C. and {Morley}, C.~V. and {Saumon}, D. and {Freedman}, R.
	},
    title = "{Retrieval of atmospheric properties of cloudy L dwarfs}",
  journal = {\mnras},
archivePrefix = "arXiv",
   eprint = {1701.01257},
 primaryClass = "astro-ph.SR",
     year = 2017,
    month = sep,
   volume = 470,
    pages = {1177-1197},
      doi = {10.1093/mnras/stx1246},
   adsurl = {https://ui.adsabs.harvard.edu/abs/2017MNRAS.470.1177B}
}

@ARTICLE{Tremblin2015ApJ,
   author = {{Tremblin}, P. and {Amundsen}, D.~S. and {Mourier}, P. and {Baraffe}, I. and
	{Chabrier}, G. and {Drummond}, B. and {Homeier}, D. and {Venot}, O.
	},
    title = "{Fingering Convection and Cloudless Models for Cool Brown Dwarf Atmospheres}",
  journal = {\apjl},
archivePrefix = "arXiv",
   eprint = {1504.03334},
 primaryClass = "astro-ph.SR",
     year = 2015,
    month = may,
   volume = 804,
      eid = {L17},
    pages = {L17},
      doi = {10.1088/2041-8205/804/1/L17},
   adsurl = {https://ui.adsabs.harvard.edu/abs/2015ApJ...804L..17T}
}

@ARTICLE{Fisher2019ApJ,
   author = {{Fisher}, C. and {Heng}, K.},
    title = "{How Much Information Does the Sodium Doublet Encode? Retrieval Analysis of Non-LTE Sodium Lines at Low and High Spectral Resolutions}",
  journal = {\apj},
archivePrefix = "arXiv",
   eprint = {1906.07035},
 primaryClass = "astro-ph.EP",
     year = 2019,
    month = aug,
   volume = 881,
      eid = {25},
    pages = {25},
      doi = {10.3847/1538-4357/ab29e8},
   adsurl = {https://ui.adsabs.harvard.edu/abs/2019ApJ...881...25F}
}

@ARTICLE{Kitzmann2018MNRAS,
   author = {{Kitzmann}, D. and {Heng}, K.},
    title = "{Optical properties of potential condensates in exoplanetary atmospheres}",
  journal = {\mnras},
archivePrefix = "arXiv",
   eprint = {1710.04946},
 primaryClass = "astro-ph.EP",
     year = 2018,
    month = mar,
   volume = 475,
    pages = {94-107},
      doi = {10.1093/mnras/stx3141},
   adsurl = {https://ui.adsabs.harvard.edu/abs/2018MNRAS.475...94K}
}

@ARTICLE{Allard2019A&A,
       author = {{Allard}, N.~F. and {Spiegelman}, F. and {Leininger}, T. and
         {Molliere}, P.},
        title = "{New study of the line profiles of sodium perturbed by H$_{2}$}",
      journal = {\aap},
         year = "2019",
        month = "Aug",
       volume = {628},
          eid = {A120},
        pages = {A120},
          doi = {10.1051/0004-6361/201935593},
archivePrefix = {arXiv},
       eprint = {1908.01989},
 primaryClass = {astro-ph.SR},
       adsurl = {https://ui.adsabs.harvard.edu/abs/2019A&A...628A.120A}
}

@article{Barber2006MNRAS,
    author = {Barber, R. J. and Tennyson, J. and Harris, G. J. and Tolchenov, R. N.},
    title = "{A high-accuracy computed water line list}",
    journal = {Monthly Notices of the Royal Astronomical Society},
    volume = {368},
    number = {3},
    pages = {1087-1094},
    year = {2006},
    month = {04},
    issn = {0035-8711},
    doi = {10.1111/j.1365-2966.2006.10184.x},
    url = {https://doi.org/10.1111/j.1365-2966.2006.10184.x},
    eprint = {http://oup.prod.sis.lan/mnras/article-pdf/368/3/1087/18665515/mnras0368-1087.pdf},
}

@ARTICLE{Yurchenko2014MNRAS.440.1649Y,
   author = {{Yurchenko}, S.~N. and {Tennyson}, J.},
    title = "{ExoMol line lists - IV. The rotation-vibration spectrum of methane up to 1500 K}",
  journal = {\mnras},
archivePrefix = "arXiv",
   eprint = {1401.4852},
 primaryClass = "astro-ph.EP",
     year = 2014,
    month = may,
   volume = 440,
    pages = {1649-1661},
      doi = {10.1093/mnras/stu326},
   adsurl = {https://ui.adsabs.harvard.edu/abs/2014MNRAS.440.1649Y}
}

@ARTICLE{Yurchenko2011MNRAS.413.1828Y,
   author = {{Yurchenko}, S.~N. and {Barber}, R.~J. and {Tennyson}, J.},
    title = "{A variationally computed line list for hot NH$_{3}$}",
  journal = {\mnras},
archivePrefix = "arXiv",
   eprint = {1011.1569},
 primaryClass = "astro-ph.EP",
     year = 2011,
    month = may,
   volume = 413,
    pages = {1828-1834},
      doi = {10.1111/j.1365-2966.2011.18261.x},
   adsurl = {https://ui.adsabs.harvard.edu/abs/2011MNRAS.413.1828Y}
}

@ARTICLE{Azzam2016MNRAS,
   author = {{Azzam}, A.~A.~A. and {Tennyson}, J. and {Yurchenko}, S.~N. and 
	{Naumenko}, O.~V.},
    title = "{ExoMol molecular line lists - XVI. The rotation-vibration spectrum of hot H$_{2}$S}",
  journal = {\mnras},
archivePrefix = "arXiv",
   eprint = {1607.00499},
 primaryClass = "astro-ph.EP",
     year = 2016,
    month = aug,
   volume = 460,
    pages = {4063-4074},
      doi = {10.1093/mnras/stw1133},
   adsurl = {https://ui.adsabs.harvard.edu/abs/2016MNRAS.460.4063A}
}

@ARTICLE{Rothman2010JQSRT.111.2139R,
   author = {{Rothman}, L.~S. and {Gordon}, I.~E. and {Barber}, R.~J. and 
	{Dothe}, H. and {Gamache}, R.~R. and {Goldman}, A. and {Perevalov}, V.~I. and 
	{Tashkun}, S.~A. and {Tennyson}, J.},
    title = "{HITEMP, the high-temperature molecular spectroscopic database}",
  journal = {\jqsrt},
     year = 2010,
    month = oct,
   volume = 111,
    pages = {2139-2150},
      doi = {10.1016/j.jqsrt.2010.05.001},
   adsurl = {https://ui.adsabs.harvard.edu/abs/2010JQSRT.111.2139R}
}

@ARTICLE{1995KurCD..23.....K,
   author = {{Kurucz}, R. and {Bell}, B.},
    title = "{Atomic Line Data}",
  journal = {Atomic Line Data (R.L.~Kurucz and B.~Bell) Kurucz CD-ROM No.~23.~Cambridge, Mass.: Smithsonian Astrophysical Observatory, 1995.},
     year = 1995,
   volume = 23,
   adsurl = {https://ui.adsabs.harvard.edu/abs/1995KurCD..23.....K}
}

@ARTICLE{Oreshenko2020AJ,
       author = {{Oreshenko}, Maria and {Kitzmann}, Daniel and
         {M{\'a}rquez-Neila}, Pablo and {Malik}, Matej and {Bowler}, Brendan P. and
         {Burgasser}, Adam J. and {Sznitman}, Raphael and {Fisher}, Chloe E. and
         {Heng}, Kevin},
        title = "{Supervised Machine Learning for Intercomparison of Model Grids of Brown Dwarfs: Application to GJ 570D and the Epsilon Indi B Binary System}",
      journal = {\aj},
         year = "2020",
        month = "Jan",
       volume = {159},
       number = {1},
          eid = {6},
        pages = {6},
          doi = {10.3847/1538-3881/ab5955},
archivePrefix = {arXiv},
       eprint = {1910.11795},
 primaryClass = {astro-ph.SR},
       adsurl = {https://ui.adsabs.harvard.edu/abs/2020AJ....159....6O}
}

@ARTICLE{Malik2017AJ....153...56M,
       author = {{Malik}, Matej and {Grosheintz}, Luc and {Mendon{\c{c}}a}, Jo{\~a}o M. and
         {Grimm}, Simon L. and {Lavie}, Baptiste and {Kitzmann}, Daniel and
         {Tsai}, Shang-Min and {Burrows}, Adam and {Kreidberg}, Laura and
         {Bedell}, Megan and {Bean}, Jacob L. and {Stevenson}, Kevin B. and
         {Heng}, Kevin},
        title = "{HELIOS: An Open-source, GPU-accelerated Radiative Transfer Code for Self-consistent Exoplanetary Atmospheres}",
      journal = {\aj},
         year = "2017",
        month = "Feb",
       volume = {153},
       number = {2},
          eid = {56},
        pages = {56},
          doi = {10.3847/1538-3881/153/2/56},
archivePrefix = {arXiv},
       eprint = {1606.05474},
 primaryClass = {astro-ph.EP},
       adsurl = {https://ui.adsabs.harvard.edu/abs/2017AJ....153...56M}
}

@article{Abel2011,
author = {Abel, Martin and Frommhold, Lothar and Li, Xiaoping and Hunt, Katharine L. C.},
title = {Collision-Induced Absorption by H2 Pairs: From Hundreds to Thousands of Kelvin},
journal = {The Journal of Physical Chemistry A},
volume = {115},
number = {25},
pages = {6805-6812},
year = {2011},
doi = {10.1021/jp109441f},
    note ={PMID: 21207941}
}

@ARTICLE{Abel2012JChPh,
       author = {{Abel}, Martin and {Frommhold}, Lothar and {Li}, Xiaoping and
         {Hunt}, Katharine L.~C.},
        title = "{Infrared absorption by collisional H$_{2}$-He complexes at temperatures up to 9000 K and frequencies from 0 to 20 000 cm$^{-1}$}",
      journal = {\jcp},
         year = "2012",
        month = "Jan",
       volume = {136},
       number = {4},
        pages = {044319-044319},
          doi = {10.1063/1.3676405},
       adsurl = {https://ui.adsabs.harvard.edu/abs/2012JChPh.136d4319A}
}

@ARTICLE{HK2017,
       author = {{Heng}, Kevin and {Kitzmann}, Daniel},
        title = "{The theory of transmission spectra revisited: a semi-analytical method for interpreting WFC3 data and an unresolved challenge}",
      journal = {\mnras},
         year = 2017,
        month = sep,
       volume = {470},
       number = {3},
        pages = {2972-2981},
          doi = {10.1093/mnras/stx1453},
archivePrefix = {arXiv},
       eprint = {1702.02051},
 primaryClass = {astro-ph.EP},
       adsurl = {https://ui.adsabs.harvard.edu/abs/2017MNRAS.470.2972H}
}

@ARTICLE{Grimm2021ApJS..253...30G,
       author = {{Grimm}, Simon L. and {Malik}, Matej and {Kitzmann}, Daniel and {Guzm{\'a}n-Mesa}, Andrea and {Hoeijmakers}, H. Jens and {Fisher}, Chloe and {Mendon{\c{c}}a}, Jo{\~a}o M. and {Yurchenko}, Sergey N. and {Tennyson}, Jonathan and {Alesina}, Fabien and {Buchschacher}, Nicolas and {Burnier}, Julien and {Segransan}, Damien and {Kurucz}, Robert L. and {Heng}, Kevin},
        title = "{HELIOS-K 2.0 Opacity Calculator and Open-source Opacity Database for Exoplanetary Atmospheres}",
      journal = {\apjs},
         year = 2021,
        month = mar,
       volume = {253},
       number = {1},
          eid = {30},
        pages = {30},
          doi = {10.3847/1538-4365/abd773},
archivePrefix = {arXiv},
       eprint = {2101.02005},
 primaryClass = {astro-ph.EP},
       adsurl = {https://ui.adsabs.harvard.edu/abs/2021ApJS..253...30G}
}

@ARTICLE{Kitzmann2020ApJ...890..174K,
       author = {{Kitzmann}, Daniel and {Heng}, Kevin and {Oreshenko}, Maria and {Grimm}, Simon L. and {Apai}, D{\'a}niel and {Bowler}, Brendan P. and {Burgasser}, Adam J. and {Marley}, Mark S.},
        title = "{Helios-r2: A New Bayesian, Open-source Retrieval Model for Brown Dwarfs and Exoplanet Atmospheres}",
      journal = {\apj},
         year = 2020,
        month = feb,
       volume = {890},
       number = {2},
          eid = {174},
        pages = {174},
          doi = {10.3847/1538-4357/ab6d71},
archivePrefix = {arXiv},
       eprint = {1910.01070},
 primaryClass = {astro-ph.EP},
       adsurl = {https://ui.adsabs.harvard.edu/abs/2020ApJ...890..174K}
}

@ARTICLE{Leggett2017ApJ...842..118L,
       author = {{Leggett}, S.~K. and {Tremblin}, P. and {Esplin}, T.~L. and {Luhman}, K.~L. and {Morley}, Caroline V.},
        title = "{The Y-type Brown Dwarfs: Estimates of Mass and Age from New Astrometry, Homogenized Photometry, and Near-infrared Spectroscopy}",
      journal = {\apj},
         year = 2017,
        month = jun,
       volume = {842},
       number = {2},
          eid = {118},
        pages = {118},
          doi = {10.3847/1538-4357/aa6fb5},
archivePrefix = {arXiv},
       eprint = {1704.03573},
 primaryClass = {astro-ph.SR},
       adsurl = {https://ui.adsabs.harvard.edu/abs/2017ApJ...842..118L}
}

@ARTICLE{Hsu2021ApJS..257...45H,
       author = {{Hsu}, Chih-Chun and {Burgasser}, Adam J. and {Theissen}, Christopher A. and {Gelino}, Christopher R. and {Birky}, Jessica L. and {Diamant}, Sharon J.~M. and {Bardalez Gagliuffi}, Daniella C. and {Aganze}, Christian and {Blake}, Cullen H. and {Faherty}, Jacqueline K.},
        title = "{The Brown Dwarf Kinematics Project (BDKP). V. Radial and Rotational Velocities of T Dwarfs from Keck/NIRSPEC High-resolution Spectroscopy}",
      journal = {\apjs},
         year = 2021,
        month = dec,
       volume = {257},
       number = {2},
          eid = {45},
        pages = {45},
          doi = {10.3847/1538-4365/ac1c7d},
archivePrefix = {arXiv},
       eprint = {2107.01222},
 primaryClass = {astro-ph.SR},
       adsurl = {https://ui.adsabs.harvard.edu/abs/2021ApJS..257...45H}
}

@ARTICLE{Tan2021MNRAS.502..678T,
       author = {{Tan}, Xianyu and {Showman}, Adam P.},
        title = "{Atmospheric circulation of brown dwarfs and directly imaged exoplanets driven by cloud radiative feedback: effects of rotation}",
      journal = {\mnras},
         year = 2021,
        month = mar,
       volume = {502},
       number = {1},
        pages = {678-699},
          doi = {10.1093/mnras/stab060},
archivePrefix = {arXiv},
       eprint = {2005.12152},
 primaryClass = {astro-ph.EP},
       adsurl = {https://ui.adsabs.harvard.edu/abs/2021MNRAS.502..678T}
}

@ARTICLE{Zahnle2014ApJ...797...41Z,
       author = {{Zahnle}, Kevin J. and {Marley}, Mark S.},
        title = "{Methane, Carbon Monoxide, and Ammonia in Brown Dwarfs and Self-Luminous Giant Planets}",
      journal = {\apj},
         year = 2014,
        month = dec,
       volume = {797},
       number = {1},
          eid = {41},
        pages = {41},
          doi = {10.1088/0004-637X/797/1/41},
archivePrefix = {arXiv},
       eprint = {1408.6283},
 primaryClass = {astro-ph.EP},
       adsurl = {https://ui.adsabs.harvard.edu/abs/2014ApJ...797...41Z}
}

@ARTICLE{Miles2020AJ....160...63M,
       author = {{Miles}, Brittany E. and {Skemer}, Andrew J.~I. and {Morley}, Caroline V. and {Marley}, Mark S. and {Fortney}, Jonathan J. and {Allers}, Katelyn N. and {Faherty}, Jacqueline K. and {Geballe}, Thomas R. and {Visscher}, Channon and {Schneider}, Adam C. and {Lupu}, Roxana and {Freedman}, Richard S. and {Bjoraker}, Gordon L.},
        title = "{Observations of Disequilibrium CO Chemistry in the Coldest Brown Dwarfs}",
      journal = {\aj},
         year = 2020,
        month = aug,
       volume = {160},
       number = {2},
          eid = {63},
        pages = {63},
          doi = {10.3847/1538-3881/ab9114},
archivePrefix = {arXiv},
       eprint = {2004.10770},
 primaryClass = {astro-ph.EP},
       adsurl = {https://ui.adsabs.harvard.edu/abs/2020AJ....160...63M}
}

@ARTICLE{Hubeny2007ApJ...669.1248H,
       author = {{Hubeny}, Ivan and {Burrows}, Adam},
        title = "{A Systematic Study of Departures from Chemical Equilibrium in the Atmospheres of Substellar Mass Objects}",
      journal = {\apj},
         year = 2007,
        month = nov,
       volume = {669},
       number = {2},
        pages = {1248-1261},
          doi = {10.1086/522107},
archivePrefix = {arXiv},
       eprint = {0705.3922},
 primaryClass = {astro-ph},
       adsurl = {https://ui.adsabs.harvard.edu/abs/2007ApJ...669.1248H}
}

@ARTICLE{Beiler2024ApJ...973...60B,
       author = {{Beiler}, Samuel A. and {Mukherjee}, Sagnick and {Cushing}, Michael C. and {Kirkpatrick}, J. Davy and {Schneider}, Adam C. and {Kothari}, Harshil and {Marley}, Mark S. and {Visscher}, Channon},
        title = "{A Tale of Two Molecules: The Underprediction of CO$_{2}$ and Overprediction of PH$_{3}$ in Late T and Y Dwarf Atmospheric Models}",
      journal = {\apj},
         year = 2024,
        month = sep,
       volume = {973},
       number = {1},
          eid = {60},
        pages = {60},
          doi = {10.3847/1538-4357/ad6759},
archivePrefix = {arXiv},
       eprint = {2407.15950},
 primaryClass = {astro-ph.EP},
       adsurl = {https://ui.adsabs.harvard.edu/abs/2024ApJ...973...60B}
}

@ARTICLE{Burgasser2002ApJ...564..421B,
       author = {{Burgasser}, Adam J. and {Kirkpatrick}, J. Davy and {Brown}, Michael E. and {Reid}, I. Neill and {Burrows}, Adam and {Liebert}, James and {Matthews}, Keith and {Gizis}, John E. and {Dahn}, Conard C. and {Monet}, David G. and {Cutri}, Roc M. and {Skrutskie}, Michael F.},
        title = "{The Spectra of T Dwarfs. I. Near-Infrared Data and Spectral Classification}",
      journal = {\apj},
         year = 2002,
        month = jan,
       volume = {564},
       number = {1},
        pages = {421-451},
          doi = {10.1086/324033},
archivePrefix = {arXiv},
       eprint = {astro-ph/0108452},
 primaryClass = {astro-ph},
       adsurl = {https://ui.adsabs.harvard.edu/abs/2002ApJ...564..421B}
}

@ARTICLE{Line2017ApJ...848...83L,
       author = {{Line}, Michael R. and {Marley}, Mark S. and {Liu}, Michael C. and {Burningham}, Ben and {Morley}, Caroline V. and {Hinkel}, Natalie R. and {Teske}, Johanna and {Fortney}, Jonathan J. and {Freedman}, Richard and {Lupu}, Roxana},
        title = "{Uniform Atmospheric Retrieval Analysis of Ultracool Dwarfs. II. Properties of 11 T dwarfs}",
      journal = {\apj},
         year = 2017,
        month = oct,
       volume = {848},
       number = {2},
          eid = {83},
        pages = {83},
          doi = {10.3847/1538-4357/aa7ff0},
archivePrefix = {arXiv},
       eprint = {1612.02809},
 primaryClass = {astro-ph.SR},
       adsurl = {https://ui.adsabs.harvard.edu/abs/2017ApJ...848...83L}
}

@ARTICLE{Zalesky2019ApJ...877...24Z,
       author = {{Zalesky}, Joseph A. and {Line}, Michael R. and {Schneider}, Adam C. and {Patience}, Jennifer},
        title = "{A Uniform Retrieval Analysis of Ultra-cool Dwarfs. III. Properties of Y Dwarfs}",
      journal = {\apj},
         year = 2019,
        month = may,
       volume = {877},
       number = {1},
          eid = {24},
        pages = {24},
          doi = {10.3847/1538-4357/ab16db},
archivePrefix = {arXiv},
       eprint = {1903.11658},
 primaryClass = {astro-ph.SR},
       adsurl = {https://ui.adsabs.harvard.edu/abs/2019ApJ...877...24Z}
}

@ARTICLE{Gonzales2020ApJ...905...46G,
       author = {{Gonzales}, Eileen C. and {Burningham}, Ben and {Faherty}, Jacqueline K. and {Cleary}, Colleen and {Visscher}, Channon and {Marley}, Mark S. and {Lupu}, Roxana and {Freedman}, Richard},
        title = "{Retrieval of the d/sdL7+T7.5p Binary SDSS J1416+1348AB}",
      journal = {\apj},
         year = 2020,
        month = dec,
       volume = {905},
       number = {1},
          eid = {46},
        pages = {46},
          doi = {10.3847/1538-4357/abbee2},
archivePrefix = {arXiv},
       eprint = {2010.01224},
 primaryClass = {astro-ph.SR},
       adsurl = {https://ui.adsabs.harvard.edu/abs/2020ApJ...905...46G}
}

@ARTICLE{Gonzales2022ApJ...938...56G,
       author = {{Gonzales}, Eileen C. and {Burningham}, Ben and {Faherty}, Jacqueline K. and {Lewis}, Nikole K. and {Visscher}, Channon and {Marley}, Mark},
        title = "{A Comparative L-dwarf Sample Exploring the Interplay between Atmospheric Assumptions and Data Properties}",
      journal = {\apj},
         year = 2022,
        month = oct,
       volume = {938},
       number = {1},
          eid = {56},
        pages = {56},
          doi = {10.3847/1538-4357/ac8f2a},
archivePrefix = {arXiv},
       eprint = {2209.02754},
 primaryClass = {astro-ph.EP},
       adsurl = {https://ui.adsabs.harvard.edu/abs/2022ApJ...938...56G}
}

@ARTICLE{Lueber2022ApJ...930..136L,
       author = {{Lueber}, Anna and {Kitzmann}, Daniel and {Bowler}, Brendan P. and {Burgasser}, Adam J. and {Heng}, Kevin},
        title = "{Retrieval Study of Brown Dwarfs across the L-T Sequence}",
      journal = {\apj},
         year = 2022,
        month = may,
       volume = {930},
       number = {2},
          eid = {136},
        pages = {136},
          doi = {10.3847/1538-4357/ac63b9},
archivePrefix = {arXiv},
       eprint = {2204.01330},
 primaryClass = {astro-ph.EP},
       adsurl = {https://ui.adsabs.harvard.edu/abs/2022ApJ...930..136L}
}

@INPROCEEDINGS{Burgasser2014ASInC..11....7B,
       author = {{Burgasser}, Adam J.},
        title = "{The SpeX Prism Library: 1000+ low-resolution, near-infrared spectra of ultracool M, L, T and Y dwarfs}",
    booktitle = {Astronomical Society of India Conference Series},
         year = 2014,
       series = {Astronomical Society of India Conference Series},
       volume = {11},
        month = jan,
        pages = {7-16},
          doi = {10.48550/arXiv.1406.4887},
archivePrefix = {arXiv},
       eprint = {1406.4887},
 primaryClass = {astro-ph.SR},
       adsurl = {https://ui.adsabs.harvard.edu/abs/2014ASInC..11....7B}
}

@INPROCEEDINGS{Skilling2006AIPC..872..321S,
       author = {{Skilling}, John},
        title = "{Calibration and Interpolation}",
    booktitle = {Bayesian Inference and Maximum Entropy Methods In Science and Engineering},
         year = 2006,
       editor = {{Mohammad-Djafari}, Ali},
       series = {American Institute of Physics Conference Series},
       volume = {872},
        month = nov,
        pages = {321-330},
          doi = {10.1063/1.2423290},
       adsurl = {https://ui.adsabs.harvard.edu/abs/2006AIPC..872..321S}
}

@ARTICLE{Trotta2008ConPh..49...71T,
       author = {{Trotta}, Roberto},
        title = "{Bayes in the sky: Bayesian inference and model selection in cosmology}",
      journal = {Contemporary Physics},
         year = 2008,
        month = mar,
       volume = {49},
       number = {2},
        pages = {71-104},
          doi = {10.1080/00107510802066753},
archivePrefix = {arXiv},
       eprint = {0803.4089},
 primaryClass = {astro-ph},
       adsurl = {https://ui.adsabs.harvard.edu/abs/2008ConPh..49...71T}
}

@ARTICLE{Kirkpatrick2021ApJS,
       author = {{Kirkpatrick}, J. Davy and {Gelino}, Christopher R. and {Faherty}, Jacqueline K. and {Meisner}, Aaron M. and {Caselden}, Dan and {Schneider}, Adam C. and {Marocco}, Federico and {Cayago}, Alfred J. and {Smart}, R.~L. and {Eisenhardt}, Peter R. and {Kuchner}, Marc J. and {Wright}, Edward L. and {Cushing}, Michael C. and {Allers}, Katelyn N. and {Bardalez Gagliuffi}, Daniella C. and {Burgasser}, Adam J. and {Gagn{\'e}}, Jonathan and {Logsdon}, Sarah E. and {Martin}, Emily C. and {Ingalls}, James G. and {Lowrance}, Patrick J. and {Abrahams}, Ellianna S. and {Aganze}, Christian and {Gerasimov}, Roman and {Gonzales}, Eileen C. and {Hsu}, Chih-Chun and {Kamraj}, Nikita and {Kiman}, Rocio and {Rees}, Jon and {Theissen}, Christopher and {Ammar}, Kareem and {Andersen}, Nikolaj Stevnbak and {Beaulieu}, Paul and {Colin}, Guillaume and {Elachi}, Charles A. and {Goodman}, Samuel J. and {Gramaize}, L{\'e}opold and {Hamlet}, Leslie K. and {Hong}, Justin and {Jonkeren}, Alexander and {Khalil}, Mohammed and {Martin}, David W. and {Pendrill}, William and {Pumphrey}, Benjamin and {Rothermich}, Austin and {Sainio}, Arttu and {Stenner}, Andres and {Tanner}, Christopher and {Th{\'e}venot}, Melina and {Voloshin}, Nikita V. and {Walla}, Jim and {W{\k{e}}dracki}, Zbigniew and {Backyard Worlds: Planet 9 Collaboration}},
        title = "{The Field Substellar Mass Function Based on the Full-sky 20 pc Census of 525 L, T, and Y Dwarfs}",
      journal = {\apjs},
         year = 2021,
        month = mar,
       volume = {253},
       number = {1},
          eid = {7},
        pages = {7},
          doi = {10.3847/1538-4365/abd107},
archivePrefix = {arXiv},
       eprint = {2011.11616},
 primaryClass = {astro-ph.SR},
       adsurl = {https://ui.adsabs.harvard.edu/abs/2021ApJS..253....7K}
}

@INCOLLECTION{Lodders2006asup.book....1L,
       author = {{Lodders}, K. and {Fegley}, B., Jr.},
        title = "{Chemistry of Low Mass Substellar Objects}",
    booktitle = {Astrophysics Update 2},
         year = 2006,
       editor = {{Mason}, John W.},
        pages = {1},
          doi = {10.1007/3-540-30313-8_1},
       adsurl = {https://ui.adsabs.harvard.edu/abs/2006asup.book....1L}
}

@ARTICLE{Best2024ApJ...967..115B,
       author = {{Best}, William M.~J. and {Sanghi}, Aniket and {Liu}, Michael C. and {Magnier}, Eugene A. and {Dupuy}, Trent J.},
        title = "{A Volume-limited Sample of Ultracool Dwarfs. II. The Substellar Age and Mass Functions in the Solar Neighborhood}",
      journal = {\apj},
         year = 2024,
        month = jun,
       volume = {967},
       number = {2},
          eid = {115},
        pages = {115},
          doi = {10.3847/1538-4357/ad39ef},
archivePrefix = {arXiv},
       eprint = {2401.09535},
 primaryClass = {astro-ph.SR},
       adsurl = {https://ui.adsabs.harvard.edu/abs/2024ApJ...967..115B}
}

@ARTICLE{Morley2012ApJ...756..172M,
       author = {{Morley}, Caroline V. and {Fortney}, Jonathan J. and {Marley}, Mark S. and {Visscher}, Channon and {Saumon}, Didier and {Leggett}, S.~K.},
        title = "{Neglected Clouds in T and Y Dwarf Atmospheres}",
      journal = {\apj},
         year = 2012,
        month = sep,
       volume = {756},
       number = {2},
          eid = {172},
        pages = {172},
          doi = {10.1088/0004-637X/756/2/172},
archivePrefix = {arXiv},
       eprint = {1206.4313},
 primaryClass = {astro-ph.SR},
       adsurl = {https://ui.adsabs.harvard.edu/abs/2012ApJ...756..172M}
}

@ARTICLE{Malik2019AJ....157..170M,
       author = {{Malik}, Matej and {Kitzmann}, Daniel and {Mendon{\c{c}}a}, Jo{\~a}o M. and {Grimm}, Simon L. and {Marleau}, Gabriel-Dominique and {Linder}, Esther F. and {Tsai}, Shang-Min and {Heng}, Kevin},
        title = "{Self-luminous and Irradiated Exoplanetary Atmospheres Explored with HELIOS}",
      journal = {\aj},
         year = 2019,
        month = may,
       volume = {157},
       number = {5},
          eid = {170},
        pages = {170},
          doi = {10.3847/1538-3881/ab1084},
archivePrefix = {arXiv},
       eprint = {1903.06794},
 primaryClass = {astro-ph.EP},
       adsurl = {https://ui.adsabs.harvard.edu/abs/2019AJ....157..170M}
}

@ARTICLE{Marley2021ApJ...920...85M,
       author = {{Marley}, Mark S. and {Saumon}, Didier and {Visscher}, Channon and {Lupu}, Roxana and {Freedman}, Richard and {Morley}, Caroline and {Fortney}, Jonathan J. and {Seay}, Christopher and {Smith}, Adam J.~R.~W. and {Teal}, D.~J. and {Wang}, Ruoyan},
        title = "{The Sonora Brown Dwarf Atmosphere and Evolution Models. I. Model Description and Application to Cloudless Atmospheres in Rainout Chemical Equilibrium}",
      journal = {\apj},
         year = 2021,
        month = oct,
       volume = {920},
       number = {2},
          eid = {85},
        pages = {85},
          doi = {10.3847/1538-4357/ac141d},
archivePrefix = {arXiv},
       eprint = {2107.07434},
 primaryClass = {astro-ph.SR},
       adsurl = {https://ui.adsabs.harvard.edu/abs/2021ApJ...920...85M}
}

@ARTICLE{Mukherjee2024ApJ...963...73M,
       author = {{Mukherjee}, Sagnick and {Fortney}, Jonathan J. and {Morley}, Caroline V. and {Batalha}, Natasha E. and {Marley}, Mark S. and {Karalidi}, Theodora and {Visscher}, Channon and {Lupu}, Roxana and {Freedman}, Richard and {Gharib-Nezhad}, Ehsan},
        title = "{The Sonora Substellar Atmosphere Models. IV. Elf Owl: Atmospheric Mixing and Chemical Disequilibrium with Varying Metallicity and C/O Ratios}",
      journal = {\apj},
         year = 2024,
        month = mar,
       volume = {963},
       number = {1},
          eid = {73},
        pages = {73},
          doi = {10.3847/1538-4357/ad18c2},
archivePrefix = {arXiv},
       eprint = {2402.00756},
 primaryClass = {astro-ph.EP},
       adsurl = {https://ui.adsabs.harvard.edu/abs/2024ApJ...963...73M}
}

@book{breiman1984classification,
  title={Classification and Regression Trees},
  author={Breiman, L. and Friedman, J. and Stone, C.J. and Olshen, R.A.},
  isbn={9780412048418},
  lccn={83019708},
  url={https://books.google.ch/books?id=JwQx-WOmSyQC},
  year={1984},
  publisher={Taylor \& Francis}
}

@book{efron1994introduction,
  title={An Introduction to the Bootstrap},
  author={Efron, B. and Tibshirani, R.J.},
  isbn={9780412042317},
  lccn={93004489},
  series={Chapman \& Hall/CRC Monographs on Statistics \& Applied Probability},
  url={https://books.google.ch/books?id=gLlpIUxRntoC},
  year={1994},
  publisher={Taylor \& Francis}
}

@ARTICLE{Leggett2024arXiv240906158L,
       author = {{Leggett}, Sandy K.},
        title = "{Y Dwarfs: The Challenge of Discovering the Coldest Substellar Population in the Solar Neighborhood}",
      journal = {arXiv e-prints},
         year = 2024,
        month = sep,
          eid = {arXiv:2409.06158},
        pages = {arXiv:2409.06158},
          doi = {10.48550/arXiv.2409.06158},
archivePrefix = {arXiv},
       eprint = {2409.06158},
 primaryClass = {astro-ph.EP},
       adsurl = {https://ui.adsabs.harvard.edu/abs/2024arXiv240906158L}
}

@ARTICLE{Leggett2023ApJ...959...86L,
       author = {{Leggett}, S.~K. and {Tremblin}, Pascal},
        title = "{The First Y Dwarf Data from JWST Show that Dynamic and Diabatic Processes Regulate Cold Brown Dwarf Atmospheres}",
      journal = {\apj},
         year = 2023,
        month = dec,
       volume = {959},
       number = {2},
          eid = {86},
        pages = {86},
          doi = {10.3847/1538-4357/acfdad},
archivePrefix = {arXiv},
       eprint = {2309.14567},
 primaryClass = {astro-ph.SR},
       adsurl = {https://ui.adsabs.harvard.edu/abs/2023ApJ...959...86L}
}

@ARTICLE{Lacy2023ApJ...950....8L,
       author = {{Lacy}, Brianna and {Burrows}, Adam},
        title = "{Self-consistent Models of Y Dwarf Atmospheres with Water Clouds and Disequilibrium Chemistry}",
      journal = {\apj},
         year = 2023,
        month = jun,
       volume = {950},
       number = {1},
          eid = {8},
        pages = {8},
          doi = {10.3847/1538-4357/acc8cb},
archivePrefix = {arXiv},
       eprint = {2303.16295},
 primaryClass = {astro-ph.EP},
       adsurl = {https://ui.adsabs.harvard.edu/abs/2023ApJ...950....8L}
}

@ARTICLE{Morley2014ApJ...787...78M,
       author = {{Morley}, Caroline V. and {Marley}, Mark S. and {Fortney}, Jonathan J. and {Lupu}, Roxana and {Saumon}, Didier and {Greene}, Tom and {Lodders}, Katharina},
        title = "{Water Clouds in Y Dwarfs and Exoplanets}",
      journal = {\apj},
         year = 2014,
        month = may,
       volume = {787},
       number = {1},
          eid = {78},
        pages = {78},
          doi = {10.1088/0004-637X/787/1/78},
archivePrefix = {arXiv},
       eprint = {1404.0005},
 primaryClass = {astro-ph.SR},
       adsurl = {https://ui.adsabs.harvard.edu/abs/2014ApJ...787...78M}
}

@ARTICLE{Lee2025A&A...695A.111L,
       author = {{Lee}, Elspeth K.~H. and {Ohno}, Kazumasa},
        title = "{Three-dimensional dynamical evolution of cloud particle microphysics in sub-stellar atmospheres: I. Description and exploring Y-dwarf atmospheric variability}",
      journal = {\aap},
         year = 2025,
        month = mar,
       volume = {695},
          eid = {A111},
        pages = {A111},
          doi = {10.1051/0004-6361/202452922},
archivePrefix = {arXiv},
       eprint = {2411.10305},
 primaryClass = {astro-ph.EP},
       adsurl = {https://ui.adsabs.harvard.edu/abs/2025A&A...695A.111L}
}

@ARTICLE{Cushing2011ApJ...743...50C,
       author = {{Cushing}, Michael C. and {Kirkpatrick}, J. Davy and {Gelino}, Christopher R. and {Griffith}, Roger L. and {Skrutskie}, Michael F. and {Mainzer}, A. and {Marsh}, Kenneth A. and {Beichman}, Charles A. and {Burgasser}, Adam J. and {Prato}, Lisa A. and {Simcoe}, Robert A. and {Marley}, Mark S. and {Saumon}, D. and {Freedman}, Richard S. and {Eisenhardt}, Peter R. and {Wright}, Edward L.},
        title = "{The Discovery of Y Dwarfs using Data from the Wide-field Infrared Survey Explorer (WISE)}",
      journal = {\apj},
         year = 2011,
        month = dec,
       volume = {743},
       number = {1},
          eid = {50},
        pages = {50},
          doi = {10.1088/0004-637X/743/1/50},
archivePrefix = {arXiv},
       eprint = {1108.4678},
 primaryClass = {astro-ph.SR},
       adsurl = {https://ui.adsabs.harvard.edu/abs/2011ApJ...743...50C}
}

@ARTICLE{Kirkpatrick2012ApJ...753..156K,
       author = {{Kirkpatrick}, J. Davy and {Gelino}, Christopher R. and {Cushing}, Michael C. and {Mace}, Gregory N. and {Griffith}, Roger L. and {Skrutskie}, Michael F. and {Marsh}, Kenneth A. and {Wright}, Edward L. and {Eisenhardt}, Peter R. and {McLean}, Ian S. and {Mainzer}, Amanda K. and {Burgasser}, Adam J. and {Tinney}, C.~G. and {Parker}, Stephen and {Salter}, Graeme},
        title = "{Further Defining Spectral Type ``Y'' and Exploring the Low-mass End of the Field Brown Dwarf Mass Function}",
      journal = {\apj},
         year = 2012,
        month = jul,
       volume = {753},
       number = {2},
          eid = {156},
        pages = {156},
          doi = {10.1088/0004-637X/753/2/156},
archivePrefix = {arXiv},
       eprint = {1205.2122},
 primaryClass = {astro-ph.SR},
       adsurl = {https://ui.adsabs.harvard.edu/abs/2012ApJ...753..156K}
}

@ARTICLE{Tu2024ApJ...976...82T,
       author = {{Tu}, Zhijun and {Wang}, Shu and {Liu}, Jifeng},
        title = "{Physical Parameters and Properties of 20 Cold Brown Dwarfs in JWST}",
      journal = {\apj},
         year = 2024,
        month = nov,
       volume = {976},
       number = {1},
          eid = {82},
        pages = {82},
          doi = {10.3847/1538-4357/ad815a},
archivePrefix = {arXiv},
       eprint = {2409.19191},
 primaryClass = {astro-ph.SR},
       adsurl = {https://ui.adsabs.harvard.edu/abs/2024ApJ...976...82T}
}

@ARTICLE{Zalesky2022ApJ...936...44Z,
       author = {{Zalesky}, Joseph A. and {Saboi}, Kezman and {Line}, Michael R. and {Zhang}, Zhoujian and {Schneider}, Adam C. and {Liu}, Michael C. and {Best}, William M.~J. and {Marley}, Mark S.},
        title = "{A Uniform Retrieval Analysis of Ultra-cool Dwarfs. IV. A Statistical Census from 50 Late-T Dwarfs}",
      journal = {\apj},
         year = 2022,
        month = sep,
       volume = {936},
       number = {1},
          eid = {44},
        pages = {44},
          doi = {10.3847/1538-4357/ac786c},
archivePrefix = {arXiv},
       eprint = {2206.01199},
 primaryClass = {astro-ph.SR},
       adsurl = {https://ui.adsabs.harvard.edu/abs/2022ApJ...936...44Z}
}

@ARTICLE{Calamari2024ApJ...963...67C,
       author = {{Calamari}, Emily and {Faherty}, Jacqueline K. and {Visscher}, Channon and {Gemma}, Marina E. and {Burningham}, Ben and {Rothermich}, Austin},
        title = "{Predicting Cloud Conditions in Substellar Mass Objects Using Ultracool Dwarf Companions}",
      journal = {\apj},
         year = 2024,
        month = mar,
       volume = {963},
       number = {1},
          eid = {67},
        pages = {67},
          doi = {10.3847/1538-4357/ad1f6d},
       adsurl = {https://ui.adsabs.harvard.edu/abs/2024ApJ...963...67C}
}

@ARTICLE{NIRSpec2022A&A...661A..80J,
       author = {{Jakobsen}, P. and {Ferruit}, P. and {Alves de Oliveira}, C. and {Arribas}, S. and {Bagnasco}, G. and {Barho}, R. and {Beck}, T.~L. and {Birkmann}, S. and {B{\"o}ker}, T. and {Bunker}, A.~J. and {Charlot}, S. and {de Jong}, P. and {de Marchi}, G. and {Ehrenwinkler}, R. and {Falcolini}, M. and {Fels}, R. and {Franx}, M. and {Franz}, D. and {Funke}, M. and {Giardino}, G. and {Gnata}, X. and {Holota}, W. and {Honnen}, K. and {Jensen}, P.~L. and {Jentsch}, M. and {Johnson}, T. and {Jollet}, D. and {Karl}, H. and {Kling}, G. and {K{\"o}hler}, J. and {Kolm}, M. -G. and {Kumari}, N. and {Lander}, M.~E. and {Lemke}, R. and {L{\'o}pez-Caniego}, M. and {L{\"u}tzgendorf}, N. and {Maiolino}, R. and {Manjavacas}, E. and {Marston}, A. and {Maschmann}, M. and {Maurer}, R. and {Messerschmidt}, B. and {Moseley}, S.~H. and {Mosner}, P. and {Mott}, D.~B. and {Muzerolle}, J. and {Pirzkal}, N. and {Pittet}, J. -F. and {Plitzke}, A. and {Posselt}, W. and {Rapp}, B. and {Rauscher}, B.~J. and {Rawle}, T. and {Rix}, H. -W. and {R{\"o}del}, A. and {Rumler}, P. and {Sabbi}, E. and {Salvignol}, J. -C. and {Schmid}, T. and {Sirianni}, M. and {Smith}, C. and {Strada}, P. and {te Plate}, M. and {Valenti}, J. and {Wettemann}, T. and {Wiehe}, T. and {Wiesmayer}, M. and {Willott}, C.~J. and {Wright}, R. and {Zeidler}, P. and {Zincke}, C.},
        title = "{The Near-Infrared Spectrograph (NIRSpec) on the James Webb Space Telescope. I. Overview of the instrument and its capabilities}",
      journal = {\aap},
         year = 2022,
        month = may,
       volume = {661},
          eid = {A80},
        pages = {A80},
          doi = {10.1051/0004-6361/202142663},
archivePrefix = {arXiv},
       eprint = {2202.03305},
 primaryClass = {astro-ph.IM},
       adsurl = {https://ui.adsabs.harvard.edu/abs/2022A&A...661A..80J}
}

@ARTICLE{MIRI2015PASP..127..584R,
       author = {{Rieke}, G.~H. and {Wright}, G.~S. and {B{\"o}ker}, T. and {Bouwman}, J. and {Colina}, L. and {Glasse}, Alistair and {Gordon}, K.~D. and {Greene}, T.~P. and {G{\"u}del}, Manuel and {Henning}, Th. and {Justtanont}, K. and {Lagage}, P. -O. and {Meixner}, M.~E. and {N{\o}rgaard-Nielsen}, H. -U. and {Ray}, T.~P. and {Ressler}, M.~E. and {van Dishoeck}, E.~F. and {Waelkens}, C.},
        title = "{The Mid-Infrared Instrument for the James Webb Space Telescope, I: Introduction}",
      journal = {\pasp},
         year = 2015,
        month = jul,
       volume = {127},
       number = {953},
        pages = {584},
          doi = {10.1086/682252},
archivePrefix = {arXiv},
       eprint = {1508.02294},
 primaryClass = {astro-ph.IM},
       adsurl = {https://ui.adsabs.harvard.edu/abs/2015PASP..127..584R}
}

@ARTICLE{JWST2023PASP..135f8001G,
       author = {{Gardner}, Jonathan P. and {Mather}, John C. and {Abbott}, Randy and {Abell}, James S. and {Abernathy}, Mark and {Abney}, Faith E. and {Abraham}, John G. and {Abraham}, Roberto and {Abul-Huda}, Yasin M. and {Acton}, Scott and {Adams}, Cynthia K. and {Adams}, Evan and {Adler}, David S. and {Adriaensen}, Maarten and {Aguilar}, Jonathan Albert and {Ahmed}, Mansoor and {Ahmed}, Nasif S. and {Ahmed}, Tanjira and {Albat}, R{\"u}deger and {Albert}, Lo{\"\i}c and {Alberts}, Stacey and {Aldridge}, David and {Allen}, Mary Marsha and {Allen}, Shaune S. and {Altenburg}, Martin and {Altunc}, Serhat and {Alvarez}, Jose Lorenzo and {{\'A}lvarez-M{\'a}rquez}, Javier and {Alves de Oliveira}, Catarina and {Ambrose}, Leslie L. and {Anandakrishnan}, Satya M. and {Andersen}, Gregory C. and {Anderson}, Harry James and {Anderson}, Jay and {Anderson}, Kristen and {Anderson}, Sara M. and {Aprea}, Julio and {Archer}, Benita J. and {Arenberg}, Jonathan W. and {Argyriou}, Ioannis and {Arribas}, Santiago and {Artigau}, {\'E}tienne and {Arvai}, Amanda Rose and {Atcheson}, Paul and {Atkinson}, Charles B. and {Averbukh}, Jesse and {Aymergen}, Cagatay and {Bacinski}, John J. and {Baggett}, Wayne E. and {Bagnasco}, Giorgio and {Baker}, Lynn L. and {Balzano}, Vicki Ann and {Banks}, Kimberly A. and {Baran}, David A. and {Barker}, Elizabeth A. and {Barrett}, Larry K. and {Barringer}, Bruce O. and {Barto}, Allison and {Bast}, William and {Baudoz}, Pierre and {Baum}, Stefi and {Beatty}, Thomas G. and {Beaulieu}, Mathilde and {Bechtold}, Kathryn and {Beck}, Tracy and {Beddard}, Megan M. and {Beichman}, Charles and {Bellagama}, Larry and {Bely}, Pierre and {Berger}, Timothy W. and {Bergeron}, Louis E. and {Bernier}, Antoine-Darveau and {Bertch}, Maria D. and {Beskow}, Charlotte and {Betz}, Laura E. and {Biagetti}, Carl P. and {Birkmann}, Stephan and {Bjorklund}, Kurt F. and {Blackwood}, James D. and {Blazek}, Ronald Paul and {Blossfeld}, Stephen and {Bluth}, Marcel and {Boccaletti}, Anthony and {Boegner}, Martin E., Jr. and {Bohlin}, Ralph C. and {Boia}, John Joseph and {B{\"o}ker}, Torsten and {Bonaventura}, N. and {Bond}, Nicholas A. and {Bosley}, Kari Ann and {Boucarut}, Rene A. and {Bouchet}, Patrice and {Bouwman}, Jeroen and {Bower}, Gary and {Bowers}, Ariel S. and {Bowers}, Charles W. and {Boyce}, Leslye A. and {Boyer}, Christine T. and {Boyer}, Martha L. and {Boyer}, Michael and {Boyer}, Robert and {Bradley}, Larry D. and {Brady}, Gregory R. and {Brandl}, Bernhard R. and {Brannen}, Judith L. and {Breda}, David and {Bremmer}, Harold G. and {Brennan}, David and {Bresnahan}, Pamela A. and {Bright}, Stacey N. and {Broiles}, Brian J. and {Bromenschenkel}, Asa and {Brooks}, Brian H. and {Brooks}, Keira J. and {Brown}, Bob and {Brown}, Bruce and {Brown}, Thomas M. and {Bruce}, Barry W. and {Bryson}, Jonathan G. and {Bujanda}, Edwin D. and {Bullock}, Blake M. and {Bunker}, A.~J. and {Bureo}, Rafael and {Burt}, Irving J. and {Bush}, James Aaron and {Bushouse}, Howard A. and {Bussman}, Marie C. and {Cabaud}, Olivier and {Cale}, Steven and {Calhoon}, Charles D. and {Calvani}, Humberto and {Canipe}, Alicia M. and {Caputo}, Francis M. and {Cara}, Mihai and {Carey}, Larkin and {Case}, Michael Eli and {Cesari}, Thaddeus and {Cetorelli}, Lee D. and {Chance}, Don R. and {Chandler}, Lynn and {Chaney}, Dave and {Chapman}, George N. and {Charlot}, S. and {Chayer}, Pierre and {Cheezum}, Jeffrey I. and {Chen}, Bin and {Chen}, Christine H. and {Cherinka}, Brian and {Chichester}, Sarah C. and {Chilton}, Zachary S. and {Chittiraibalan}, Dharini and {Clampin}, Mark and {Clark}, Charles R. and {Clark}, Kerry W. and {Clark}, Stephanie M. and {Claybrooks}, Edward E. and {Cleveland}, Keith A. and {Cohen}, Andrew L. and {Cohen}, Lester M. and {Col{\'o}n}, Knicole D. and {Coleman}, Benee L. and {Colina}, Luis and {Comber}, Brian J. and {Comeau}, Thomas M. and {Comer}, Thomas and {Conde Reis}, Alain and {Connolly}, Dennis C. and {Conroy}, Kyle E. and {Contos}, Adam R. and {Contreras}, James and {Cook}, Neil J. and {Cooper}, James L. and {Cooper}, Rachel Aviva and {Correia}, Michael F. and {Correnti}, Matteo and {Cossou}, Christophe and {Costanza}, Brian F. and {Coulais}, Alain and {Cox}, Colin R. and {Coyle}, Ray T. and {Cracraft}, Misty M. and {Crew}, Keith A. and {Curtis}, Gary J. and {Cusveller}, Bianca and {Da Costa Maciel}, Cleyciane and {Dailey}, Christopher T. and {Daugeron}, Fr{\'e}d{\'e}ric and {Davidson}, Greg S. and {Davies}, James E. and {Davis}, Katherine Anne and {Davis}, Michael S. and {Day}, Ratna and {de Chambure}, Daniel and {de Jong}, Pauline and {De Marchi}, Guido and {Dean}, Bruce H. and {Decker}, John E. and {Delisa}, Amy S. and {Dell}, Lawrence C. and {Dellagatta}, Gail and {Dembinska}, Franciszka and {Demosthenes}, Sandor and {Dencheva}, Nadezhda M. and {Deneu}, Philippe and {DePriest}, William W. and {Deschenes}, Jeremy and {Dethienne}, Nathalie and {Detre}, {\"O}rs Hunor and {Diaz}, Rosa Izela and {Dicken}, Daniel and {DiFelice}, Audrey S. and {Dillman}, Matthew and {Disharoon}, Maureen O. and {Dixon}, William V. and {Doggett}, Jesse B. and {Dominguez}, Keisha L. and {Donaldson}, Thomas S. and {Doria-Warner}, Cristina M. and {Santos}, Tony Dos and {Doty}, Heather and {Douglas}, Robert E., Jr. and {Doyon}, Ren{\'e} and {Dressler}, Alan and {Driggers}, Jennifer and {Driggers}, Phillip A. and {Dunn}, Jamie L. and {DuPrie}, Kimberly C. and {Dupuis}, Jean and {Durning}, John and {Dutta}, Sanghamitra B. and {Earl}, Nicholas M. and {Eccleston}, Paul and {Ecobichon}, Pascal and {Egami}, Eiichi and {Ehrenwinkler}, Ralf and {Eisenhamer}, Jonathan D. and {Eisenhower}, Michael and {Eisenstein}, Daniel J. and {El Hamel}, Zaky and {Elie}, Michelle L. and {Elliott}, James and {Elliott}, Kyle Wesley and {Engesser}, Michael and {Espinoza}, N{\'e}stor and {Etienne}, Odessa and {Etxaluze}, Mireya and {Evans}, Leah and {Fabreguettes}, Luce and {Falcolini}, Massimo and {Falini}, Patrick R. and {Fatig}, Curtis and {Feeney}, Matthew and {Feinberg}, Lee D. and {Fels}, Raymond and {Ferdous}, Nazma and {Ferguson}, Henry C. and {Ferrarese}, Laura and {Ferreira}, Marie-H{\'e}l{\'e}ne and {Ferruit}, Pierre and {Ferry}, Malcolm and {Filippazzo}, Joseph Charles and {Firre}, Daniel and {Fix}, Mees and {Flagey}, Nicolas and {Flanagan}, Kathryn A. and {Fleming}, Scott W. and {Florian}, Michael and {Flynn}, James R. and {Foiadelli}, Luca and {Fontaine}, Mark R. and {Fontanella}, Erin Marie and {Forshay}, Peter Randolph and {Fortner}, Elizabeth A. and {Fox}, Ori D. and {Framarini}, Alexandro P. and {Francisco}, John I. and {Franck}, Randy and {Franx}, Marijn and {Franz}, David E. and {Friedman}, Scott D. and {Friend}, Katheryn E. and {Frost}, James R. and {Fu}, Henry and {Fullerton}, Alexander W. and {Gaillard}, Lionel and {Galkin}, Sergey and {Gallagher}, Ben and {Galyer}, Anthony D. and {Garc{\'\i}a Mar{\'\i}n}, Macarena and {Gardner}, Lisa E. and {Garland}, Dennis and {Garrett}, Bruce Albert and {Gasman}, Danny and {G{\'a}sp{\'a}r}, Andr{\'a}s and {Gastaud}, Ren{\'e} and {Gaudreau}, Daniel and {Gauthier}, Peter Timothy and {Geers}, Vincent and {Geithner}, Paul H. and {Gennaro}, Mario and {Gerber}, John and {Gereau}, John C. and {Giampaoli}, Robert and {Giardino}, Giovanna and {Gibbons}, Paul C. and {Gilbert}, Karoline and {Gilman}, Larry and {Girard}, Julien H. and {Giuliano}, Mark E. and {Gkountis}, Konstantinos and {Glasse}, Alistair and {Glassmire}, Kirk Zachary and {Glauser}, Adrian Michael and {Glazer}, Stuart D. and {Goldberg}, Joshua and {Golimowski}, David A. and {Gonzaga}, Shireen P. and {Gordon}, Karl D. and {Gordon}, Shawn J. and {Goudfrooij}, Paul and {Gough}, Michael J. and {Graham}, Adrian J. and {Grau}, Christopher M. and {Green}, Joel David and {Greene}, Gretchen R. and {Greene}, Thomas P. and {Greenfield}, Perry E. and {Greenhouse}, Matthew A. and {Greve}, Thomas R. and {Greville}, Edgar M. and {Grimaldi}, Stefano and {Groe}, Frank E. and {Groebner}, Andrew and {Grumm}, David M. and {Grundy}, Timothy and {G{\"u}del}, Manuel and {Guillard}, Pierre and {Guldalian}, John and {Gunn}, Christopher A. and {Gurule}, Anthony and {Gutman}, Irvin Meyer and {Guy}, Paul D. and {Guyot}, Benjamin and {Hack}, Warren J. and {Haderlein}, Peter and {Hagan}, James B. and {Hagedorn}, Andria and {Hainline}, Kevin and {Haley}, Craig and {Hami}, Maryam and {Hamilton}, Forrest Clifford and {Hammann}, Jeffrey and {Hammel}, Heidi B. and {Hanley}, Christopher J. and {Hansen}, Carl August and {Hardy}, Bruce and {Harnisch}, Bernd and {Harr}, Michael Hunter and {Harris}, Pamela and {Hart}, Jessica Ann and {Hartig}, George F. and {Hasan}, Hashima and {Hashim}, Kathleen Marie and {Hashimoto}, Ryan and {Haskins}, Sujee J. and {Hawkins}, Robert Edward and {Hayden}, Brian and {Hayden}, William L. and {Healy}, Mike and {Hecht}, Karen and {Heeg}, Vince J. and {Hejal}, Reem and {Helm}, Kristopher A. and {Hengemihle}, Nicholas J. and {Henning}, Thomas and {Henry}, Alaina and {Henry}, Ronald L. and {Henshaw}, Katherine and {Hernandez}, Scarlin and {Herrington}, Donald C. and {Heske}, Astrid and {Hesman}, Brigette Emily and {Hickey}, David L. and {Hilbert}, Bryan N. and {Hines}, Dean C. and {Hinz}, Michael R. and {Hirsch}, Michael and {Hitcho}, Robert S. and {Hodapp}, Klaus and {Hodge}, Philip E. and {Hoffman}, Melissa and {Holfeltz}, Sherie T. and {Holler}, Bryan Jason and {Hoppa}, Jennifer Rose and {Horner}, Scott and {Howard}, Joseph M. and {Howard}, Richard J. and {Huber}, Jean M. and {Hunkeler}, Joseph S. and {Hunter}, Alexander and {Hunter}, David Gavin and {Hurd}, Spencer W. and {Hurst}, Brendan J. and {Hutchings}, John B. and {Hylan}, Jason E. and {Ignat}, Luminita Ilinca and {Illingworth}, Garth and {Irish}, Sandra M. and {Isaacs}, John C., III and {Jackson}, Wallace C., Jr. and {Jaffe}, Daniel T. and {Jahic}, Jasmin and {Jahromi}, Amir and {Jakobsen}, Peter and {James}, Bryan and {James}, John C. and {James}, LeAndrea Rae and {Jamieson}, William Brian and {Jandra}, Raymond D. and {Jayawardhana}, Ray and {Jedrzejewski}, Robert and {Jeffers}, Basil S. and {Jensen}, Peter and {Joanne}, Egges and {Johns}, Alan T. and {Johnson}, Carl A. and {Johnson}, Eric L. and {Johnson}, Patricia and {Johnson}, Phillip Stephen and {Johnson}, Thomas K. and {Johnson}, Timothy W. and {Johnstone}, Doug and {Jollet}, Delphine and {Jones}, Danny P. and {Jones}, Gregory S. and {Jones}, Olivia C. and {Jones}, Ronald A. and {Jones}, Vicki and {Jordan}, Ian J. and {Jordan}, Margaret E. and {Jue}, Reginald and {Jurkowski}, Mark H. and {Justis}, Grant and {Justtanont}, Kay and {Kaleida}, Catherine C. and {Kalirai}, Jason S. and {Kalmanson}, Phillip Cabrales and {Kaltenegger}, Lisa and {Kammerer}, Jens and {Kan}, Samuel K. and {Kanarek}, Graham Childs and {Kao}, Shaw-Hong and {Karakla}, Diane M. and {Karl}, Hermann and {Kassin}, Susan A. and {Kauffman}, David D. and {Kavanagh}, Patrick and {Kelley}, Leigh L. and {Kelly}, Douglas M. and {Kendrew}, Sarah and {Kennedy}, Herbert V. and {Kenny}, Deborah A. and {Keski-Kuha}, Ritva A. and {Keyes}, Charles D. and {Khan}, Ali and {Kidwell}, Richard C. and {Kimble}, Randy A. and {King}, James S. and {King}, Richard C. and {Kinzel}, Wayne M. and {Kirk}, Jeffrey R. and {Kirkpatrick}, Marc E. and {Klaassen}, Pamela and {Klingemann}, Lana and {Klintworth}, Paul U. and {Knapp}, Bryan Adam and {Knight}, Scott and {Knollenberg}, Perry J. and {Knutsen}, Daniel Mark and {Koehler}, Robert and {Koekemoer}, Anton M. and {Kofler}, Earl T. and {Kontson}, Vicki L. and {Kovacs}, Aiden Rose and {Kozhurina-Platais}, Vera and {Krause}, Oliver and {Kriss}, Gerard A. and {Krist}, John and {Kristoffersen}, Monica R. and {Krogel}, Claudia and {Krueger}, Anthony P. and {Kulp}, Bernard A. and {Kumari}, Nimisha and {Kwan}, Sandy W. and {Kyprianou}, Mark and {Labador}, Aurora Gadiano and {Labiano}, {\'A}lvaro and {Lafreni{\`e}re}, David and {Lagage}, Pierre-Olivier and {Laidler}, Victoria G. and {Laine}, Benoit and {Laird}, Simon and {Lajoie}, Charles-Philippe and {Lallo}, Matthew D. and {Lam}, May Yen and {LaMassa}, Stephanie Marie and {Lambros}, Scott D. and {Lampenfield}, Richard Joseph and {Lander}, Matthew Ed and {Langston}, James Hutton and {Larson}, Kirsten and {Larson}, Melora and {LaVerghetta}, Robert Joseph and {Law}, David R. and {Lawrence}, Jon F. and {Lee}, David W. and {Lee}, Janice and {Lee}, Yat-Ning Paul and {Leisenring}, Jarron and {Leveille}, Michael Dunlap and {Levenson}, Nancy A. and {Levi}, Joshua S. and {Levine}, Marie B. and {Lewis}, Dan and {Lewis}, Jake and {Lewis}, Nikole and {Libralato}, Mattia and {Lidon}, Norbert and {Liebrecht}, Paula Louisa and {Lightsey}, Paul and {Lilly}, Simon and {Lim}, Frederick C. and {Lim}, Pey Lian and {Ling}, Sai-Kwong and {Link}, Lisa J. and {Link}, Miranda Nicole and {Lipinski}, Jamie L. and {Liu}, XiaoLi and {Lo}, Amy S. and {Lobmeyer}, Lynette and {Logue}, Ryan M. and {Long}, Chris A. and {Long}, Douglas R. and {Long}, Ilana D. and {Long}, Knox S. and {L{\'o}pez-Caniego}, Marcos and {Lotz}, Jennifer M. and {Love-Pruitt}, Jennifer M. and {Lubskiy}, Michael and {Luers}, Edward B. and {Luetgens}, Robert A. and {Luevano}, Annetta J. and {Lui}, Sarah Marie G. Flores and {Lund}, James M., III and {Lundquist}, Ray A. and {Lunine}, Jonathan and {L{\"u}tzgendorf}, Nora and {Lynch}, Richard J. and {MacDonald}, Alex J. and {MacDonald}, Kenneth and {Macias}, Matthew J. and {Macklis}, Keith I. and {Maghami}, Peiman and {Maharaja}, Rishabh Y. and {Maiolino}, Roberto and {Makrygiannis}, Konstantinos G. and {Malla}, Sunita Giri and {Malumuth}, Eliot M. and {Manjavacas}, Elena and {Marini}, Andrea and {Marrione}, Amanda and {Marston}, Anthony and {Martel}, Andr{\'e} R. and {Martin}, Didier and {Martin}, Peter G. and {Martinez}, Kristin L. and {Maschmann}, Marc and {Masci}, Gregory L. and {Masetti}, Margaret E. and {Maszkiewicz}, Michael and {Matthews}, Gary and {Matuskey}, Jacob E. and {McBrayer}, Glen A. and {McCarthy}, Donald W. and {McCaughrean}, Mark J. and {McClare}, Leslie A. and {McClare}, Michael D. and {McCloskey}, John C. and {McClurg}, Taylore D. and {McCoy}, Martin and {McElwain}, Michael W. and {McGregor}, Roy D. and {McGuffey}, Douglas B. and {McKay}, Andrew G. and {McKenzie}, William K. and {McLean}, Brian and {McMaster}, Matthew and {McNeil}, Warren and {De Meester}, Wim and {Mehalick}, Kimberly L. and {Meixner}, Margaret and {Mel{\'e}ndez}, Marcio and {Menzel}, Michael P. and {Menzel}, Michael T. and {Merz}, Matthew and {Mesterharm}, David D. and {Meyer}, Michael R. and {Meyett}, Michele L. and {Meza}, Luis E. and {Midwinter}, Calvin and {Milam}, Stefanie N. and {Miller}, Jay Todd and {Miller}, William C. and {Miskey}, Cherie L. and {Misselt}, Karl and {Mitchell}, Eileen P. and {Mohan}, Martin and {Montoya}, Emily E. and {Moran}, Michael J. and {Morishita}, Takahiro and {Moro-Mart{\'\i}n}, Amaya and {Morrison}, Debra L. and {Morrison}, Jane and {Morse}, Ernie C. and {Moschos}, Michael and {Moseley}, S.~H. and {Mosier}, Gary E. and {Mosner}, Peter and {Mountain}, Matt and {Muckenthaler}, Jason S. and {Mueller}, Donald G. and {Mueller}, Migo and {Muhiem}, Daniella and {M{\"u}hlmann}, Prisca and {Mullally}, Susan Elizabeth and {Mullen}, Stephanie M. and {Munger}, Alan J. and {Murphy}, Jess and {Murray}, Katherine T. and {Muzerolle}, James C. and {Mycroft}, Matthew and {Myers}, Andrew and {Myers}, Carey R. and {Myers}, Fred Richard R. and {Myers}, Richard and {Myrick}, Kaila and {Nagle}, Adrian F., IV and {Nayak}, Omnarayani and {Naylor}, Bret and {Neff}, Susan G. and {Nelan}, Edmund P. and {Nella}, John and {Nguyen}, Duy Tuong and {Nguyen}, Michael N. and {Nickson}, Bryony and {Nidhiry}, John Joseph and {Niedner}, Malcolm B. and {Nieto-Santisteban}, Maria and {Nikolov}, Nikolay K. and {Nishisaka}, Mary Ann and {Noriega-Crespo}, Alberto and {Nota}, Antonella and {O'Mara}, Robyn C. and {Oboryshko}, Michael and {O'Brien}, Marcus B. and {Ochs}, William R. and {Offenberg}, Joel D. and {Ogle}, Patrick Michael and {Ohl}, Raymond G. and {Olmsted}, Joseph Hamden and {Osborne}, Shannon Barbara and {O'Shaughnessy}, Brian Patrick and {{\"O}stlin}, G{\"o}ran and {O'Sullivan}, Brian and {Otor}, O. Justin and {Ottens}, Richard and {Ouellette}, Nathalie N. -Q. and {Outlaw}, Daria J. and {Owens}, Beverly A. and {Pacifici}, Camilla and {Page}, James Christophe and {Paranilam}, James G. and {Park}, Sang and {Parrish}, Keith A. and {Paschal}, Laura and {Patapis}, Polychronis and {Patel}, Jignasha and {Patrick}, Keith and {Pattishall}, Robert A., Jr. and {Paul}, Douglas William and {Paul}, Shirley J. and {Pauly}, Tyler Andrew and {Pavlovsky}, Cheryl M. and {Pe{\~n}a-Guerrero}, Maria and {Pedder}, Andrew H. and {Peek}, Matthew Weldon and {Pelham}, Patricia A. and {Penanen}, Konstantin and {Perriello}, Beth A. and {Perrin}, Marshall D. and {Perrine}, Richard F. and {Perrygo}, Chuck and {Peslier}, Muriel and {Petach}, Michael and {Peterson}, Karla A. and {Pfarr}, Tom and {Pierson}, James M. and {Pietraszkiewicz}, Martin and {Pilchen}, Guy and {Pipher}, Judy L. and {Pirzkal}, Norbert and {Pitman}, Joseph T. and {Player}, Danielle M. and {Plesha}, Rachel and {Plitzke}, Anja and {Pohner}, John A. and {Poletis}, Karyn Konstantin and {Pollizzi}, Joseph A. and {Polster}, Ethan and {Pontius}, James T. and {Pontoppidan}, Klaus and {Porges}, Susana C. and {Potter}, Gregg D. and {Prescott}, Stephen and {Proffitt}, Charles R. and {Pueyo}, Laurent and {Quispe Neira}, Irma Aracely and {Radich}, Armando and {Rager}, Reiko T. and {Rameau}, Julien and {Ramey}, Deborah D. and {Ramos Alarcon}, Rafael and {Rampini}, Riccardo and {Rapp}, Robert and {Rashford}, Robert A. and {Rauscher}, Bernard J. and {Ravindranath}, Swara and {Rawle}, Timothy and {Rawlings}, Tynika N. and {Ray}, Tom and {Regan}, Michael W. and {Rehm}, Brian and {Rehm}, Kenneth D. and {Reid}, Neill and {Reis}, Carl A. and {Renk}, Florian and {Reoch}, Tom B. and {Ressler}, Michael and {Rest}, Armin W. and {Reynolds}, Paul J. and {Richon}, Joel G. and {Richon}, Karen V. and {Ridgaway}, Michael and {Riedel}, Adric Richard and {Rieke}, George H. and {Rieke}, Marcia J. and {Rifelli}, Richard E. and {Rigby}, Jane R. and {Riggs}, Catherine S. and {Ringel}, Nancy J. and {Ritchie}, Christine E. and {Rix}, Hans-Walter and {Robberto}, Massimo and {Robinson}, Gregory L. and {Robinson}, Michael S. and {Robinson}, Orion and {Rock}, Frank W. and {Rodriguez}, David R. and {Rodr{\'\i}guez del Pino}, Bruno and {Roellig}, Thomas and {Rohrbach}, Scott O. and {Roman}, Anthony J. and {Romelfanger}, Frederick J. and {Romo}, Felipe P., Jr. and {Rosales}, Jose J. and {Rose}, Perry and {Roteliuk}, Anthony F. and {Roth}, Marc N. and {Rothwell}, Braden Quinn and {Rouzaud}, Sylvain and {Rowe}, Jason and {Rowlands}, Neil and {Roy}, Arpita and {Royer}, Pierre and {Rui}, Chunlei and {Rumler}, Peter and {Rumpl}, William and {Russ}, Melissa L. and {Ryan}, Michael B. and {Ryan}, Richard M. and {Saad}, Karl and {Sabata}, Modhumita and {Sabatino}, Rick and {Sabbi}, Elena and {Sabelhaus}, Phillip A. and {Sabia}, Stephen and {Sahu}, Kailash C. and {Saif}, Babak N. and {Salvignol}, Jean-Christophe and {Samara-Ratna}, Piyal and {Samuelson}, Bridget S. and {Sanders}, Felicia A. and {Sappington}, Bradley and {Sargent}, B.~A. and {Sauer}, Arne and {Savadkin}, Bruce J. and {Sawicki}, Marcin and {Schappell}, Tina M. and {Scheffer}, Caroline and {Scheithauer}, Silvia and {Scherer}, Ron and {Schiff}, Conrad and {Schlawin}, Everett and {Schmeitzky}, Olivier and {Schmitz}, Tyler S. and {Schmude}, Donald J. and {Schneider}, Analyn and {Schreiber}, J{\"u}rgen and {Schroeven-Deceuninck}, Hilde and {Schultz}, John J. and {Schwab}, Ryan and {Schwartz}, Curtis H. and {Scoccimarro}, Dario and {Scott}, John F. and {Scott}, Michelle B. and {Seaton}, Bonita L. and {Seely}, Bruce S. and {Seery}, Bernard and {Seidleck}, Mark and {Sembach}, Kenneth and {Shanahan}, Clare Elizabeth and {Shaughnessy}, Bryan and {Shaw}, Richard A. and {Shay}, Christopher Michael and {Sheehan}, Even and {Sheth}, Kartik and {Shih}, Hsin-Yi and {Shivaei}, Irene and {Siegel}, Noah and {Sienkiewicz}, Matthew G. and {Simmons}, Debra D. and {Simon}, Bernard P. and {Sirianni}, Marco and {Sivaramakrishnan}, Anand and {Slade}, Jeffrey E. and {Sloan}, G.~C. and {Slocum}, Christine E. and {Slowinski}, Steven E. and {Smith}, Corbett T. and {Smith}, Eric P. and {Smith}, Erin C. and {Smith}, Koby and {Smith}, Robert and {Smith}, Stephanie J. and {Smolik}, John L. and {Soderblom}, David R. and {Sohn}, Sangmo Tony and {Sokol}, Jeff and {Sonneborn}, George and {Sontag}, Christopher D. and {Sooy}, Peter R. and {Soummer}, Remi and {Southwood}, Dana M. and {Spain}, Kay and {Sparmo}, Joseph and {Speer}, David T. and {Spencer}, Richard and {Sprofera}, Joseph D. and {Stallcup}, Scott S. and {Stanley}, Marcia K. and {Stansberry}, John A. and {Stark}, Christopher C. and {Starr}, Carl W. and {Stassi}, Diane Y. and {Steck}, Jane A. and {Steeley}, Christine D. and {Stephens}, Matthew A. and {Stephenson}, Ralph J. and {Stewart}, Alphonso C. and {Stiavelli}, Massimo and {}, Stockman, Hervey Jr. and {Strada}, Paolo and {Straughn}, Amber N. and {Streetman}, Scott and {Strickland}, David Kendal and {Strobele}, Jingping F. and {Stuhlinger}, Martin and {Stys}, Jeffrey Edward and {Such}, Miguel and {Sukhatme}, Kalyani and {Sullivan}, Joseph F. and {Sullivan}, Pamela C. and {Sumner}, Sandra M. and {Sun}, Fengwu and {Sunnquist}, Benjamin Dale and {Swade}, Daryl Allen and {Swam}, Michael S. and {Swenton}, Diane F. and {Swoish}, Robby A. and {Tam Litten}, Oi In and {Tamas}, Laszlo and {Tao}, Andrew and {Taylor}, David K. and {Taylor}, Joanna M. and {te Plate}, Maurice and {Van Tea}, Mason and {Teague}, Kelly K. and {Telfer}, Randal C. and {Temim}, Tea and {Texter}, Scott C. and {Thatte}, Deepashri G. and {Thompson}, Christopher Lee and {Thompson}, Linda M. and {Thomson}, Shaun R. and {Thronson}, Harley and {Tierney}, C.~M. and {Tikkanen}, Tuomo and {Tinnin}, Lee and {Tippet}, William Thomas and {Todd}, Connor William and {Tran}, Hien D. and {Trauger}, John and {Trejo}, Edwin Gregorio and {Vinh Truong}, Justin Hoang and {Tsukamoto}, Christine L. and {Tufail}, Yasir and {Tumlinson}, Jason and {Tustain}, Samuel and {Tyra}, Harrison and {Ubeda}, Leonardo and {Underwood}, Kelli and {Uzzo}, Michael A. and {Vaclavik}, Steven and {Valenduc}, Frida and {Valenti}, Jeff A. and {Van Campen}, Julie and {van de Wetering}, Inge and {Van Der Marel}, Roeland P. and {van Haarlem}, Remy and {Vandenbussche}, Bart and {van Dishoeck}, Ewine F. and {Vanterpool}, Dona D. and {Vernoy}, Michael R. and {Vila Costas}, Maria Bego{\~n}a and {Volk}, Kevin and {Voorzaat}, Piet and {Voyton}, Mark F. and {Vydra}, Ekaterina and {Waddy}, Darryl J. and {Waelkens}, Christoffel and {Wahlgren}, Glenn Michael and {Walker}, Frederick E., Jr. and {Wander}, Michel and {Warfield}, Christine K. and {Warner}, Gerald and {Wasiak}, Francis C. and {Wasiak}, Matthew F. and {Wehner}, James and {Weiler}, Kevin R. and {Weilert}, Mark and {Weiss}, Stanley B. and {Wells}, Martyn and {Welty}, Alan D. and {Wheate}, Lauren and {Wheeler}, Thomas P. and {White}, Christy L. and {Whitehouse}, Paul and {Whiteleather}, Jennifer Margaret and {Whitman}, William Russell and {Williams}, Christina C. and {Willmer}, Christopher N.~A. and {Willott}, Chris J. and {Willoughby}, Scott P. and {Wilson}, Andrew and {Wilson}, Debra and {Wilson}, Donna V. and {Windhorst}, Rogier and {Wislowski}, Emily Christine and {Wolfe}, David J. and {Wolfe}, Michael A. and {Wolff}, Schuyler and {Wondel}, Amancio and {Woo}, Cindy and {Woods}, Robert T. and {Worden}, Elaine and {Workman}, William and {Wright}, Gillian S. and {Wu}, Carl and {Wu}, Chi-Rai and {Wun}, Dakin D. and {Wymer}, Kristen B. and {Yadetie}, Thomas and {Yan}, Isabelle C. and {Yang}, Keith C. and {Yates}, Kayla L. and {Yeager}, Christopher R. and {Yerger}, Ethan John and {Young}, Erick T. and {Young}, Gary and {Yu}, Gene and {Yu}, Susan and {Zak}, Dean S. and {Zeidler}, Peter and {Zepp}, Robert and {Zhou}, Julia and {Zincke}, Christian A. and {Zonak}, Stephanie and {Zondag}, Elisabeth},
        title = "{The James Webb Space Telescope Mission}",
      journal = {\pasp},
         year = 2023,
        month = jun,
       volume = {135},
       number = {1048},
          eid = {068001},
        pages = {068001},
          doi = {10.1088/1538-3873/acd1b5},
archivePrefix = {arXiv},
       eprint = {2304.04869},
 primaryClass = {astro-ph.IM},
       adsurl = {https://ui.adsabs.harvard.edu/abs/2023PASP..135f8001G}
}

@ARTICLE{Best2020AJ....159..257B,
       author = {{Best}, William M.~J. and {Liu}, Michael C. and {Magnier}, Eugene A. and {Dupuy}, Trent J.},
        title = "{The Hawaii Infrared Parallax Program. IV. A Comprehensive Parallax Survey of L0-T8 Dwarfs with UKIRT}",
      journal = {\aj},
         year = 2020,
        month = jun,
       volume = {159},
       number = {6},
          eid = {257},
        pages = {257},
          doi = {10.3847/1538-3881/ab84f4},
archivePrefix = {arXiv},
       eprint = {2010.15850},
 primaryClass = {astro-ph.SR},
       adsurl = {https://ui.adsabs.harvard.edu/abs/2020AJ....159..257B}
}

@ARTICLE{Tinney2003AJ....126..975T,
       author = {{Tinney}, C.~G. and {Burgasser}, Adam J. and {Kirkpatrick}, J. Davy},
        title = "{Infrared Parallaxes for Methane T Dwarfs}",
      journal = {\aj},
         year = 2003,
        month = aug,
       volume = {126},
       number = {2},
        pages = {975-992},
          doi = {10.1086/376481},
archivePrefix = {arXiv},
       eprint = {astro-ph/0304339},
 primaryClass = {astro-ph},
       adsurl = {https://ui.adsabs.harvard.edu/abs/2003AJ....126..975T}
}

@ARTICLE{Kirkpatrick2011ApJS..197...19K,
       author = {{Kirkpatrick}, J. Davy and {Cushing}, Michael C. and {Gelino}, Christopher R. and {Griffith}, Roger L. and {Skrutskie}, Michael F. and {Marsh}, Kenneth A. and {Wright}, Edward L. and {Mainzer}, A. and {Eisenhardt}, Peter R. and {McLean}, Ian S. and {Thompson}, Maggie A. and {Bauer}, James M. and {Benford}, Dominic J. and {Bridge}, Carrie R. and {Lake}, Sean E. and {Petty}, Sara M. and {Stanford}, S.~A. and {Tsai}, Chao-Wei and {Bailey}, Vanessa and {Beichman}, Charles A. and {Bloom}, Joshua S. and {Bochanski}, John J. and {Burgasser}, Adam J. and {Capak}, Peter L. and {Cruz}, Kelle L. and {Hinz}, Philip M. and {Kartaltepe}, Jeyhan S. and {Knox}, Russell P. and {Manohar}, Swarnima and {Masters}, Daniel and {Morales-Calder{\'o}n}, Maria and {Prato}, Lisa A. and {Rodigas}, Timothy J. and {Salvato}, Mara and {Schurr}, Steven D. and {Scoville}, Nicholas Z. and {Simcoe}, Robert A. and {Stapelfeldt}, Karl R. and {Stern}, Daniel and {Stock}, Nathan D. and {Vacca}, William D.},
        title = "{The First Hundred Brown Dwarfs Discovered by the Wide-field Infrared Survey Explorer (WISE)}",
      journal = {\apjs},
         year = 2011,
        month = dec,
       volume = {197},
       number = {2},
          eid = {19},
        pages = {19},
          doi = {10.1088/0067-0049/197/2/19},
archivePrefix = {arXiv},
       eprint = {1108.4677},
 primaryClass = {astro-ph.SR},
       adsurl = {https://ui.adsabs.harvard.edu/abs/2011ApJS..197...19K}
}

@ARTICLE{Tinney2014ApJ...796...39T,
       author = {{Tinney}, C.~G. and {Faherty}, Jacqueline K. and {Kirkpatrick}, J. Davy and {Cushing}, Mike and {Morley}, Caroline V. and {Wright}, Edward L.},
        title = "{The Luminosities of the Coldest Brown Dwarfs}",
      journal = {\apj},
         year = 2014,
        month = nov,
       volume = {796},
       number = {1},
          eid = {39},
        pages = {39},
          doi = {10.1088/0004-637X/796/1/39},
archivePrefix = {arXiv},
       eprint = {1410.0746},
 primaryClass = {astro-ph.SR},
       adsurl = {https://ui.adsabs.harvard.edu/abs/2014ApJ...796...39T}
}

@ARTICLE{Meisner2020ApJ...889...74M,
       author = {{Meisner}, Aaron M. and {Caselden}, Dan and {Kirkpatrick}, J. Davy and {Marocco}, Federico and {Gelino}, Christopher R. and {Cushing}, Michael C. and {Eisenhardt}, Peter R.~M. and {Wright}, Edward L. and {Faherty}, Jacqueline K. and {Koontz}, Renata and {Marchese}, Elijah J. and {Khalil}, Mohammed and {Fowler}, John W. and {Schlafly}, Edward F.},
        title = "{Expanding the Y Dwarf Census with Spitzer Follow-up of the Coldest CatWISE Solar Neighborhood Discoveries}",
      journal = {\apj},
         year = 2020,
        month = feb,
       volume = {889},
       number = {2},
          eid = {74},
        pages = {74},
          doi = {10.3847/1538-4357/ab6215},
archivePrefix = {arXiv},
       eprint = {1911.12372},
 primaryClass = {astro-ph.SR},
       adsurl = {https://ui.adsabs.harvard.edu/abs/2020ApJ...889...74M}
}

@ARTICLE{Schneider2015ApJ...804...92S,
       author = {{Schneider}, Adam C. and {Cushing}, Michael C. and {Kirkpatrick}, J. Davy and {Gelino}, Christopher R. and {Mace}, Gregory N. and {Wright}, Edward L. and {Eisenhardt}, Peter R. and {Skrutskie}, M.~F. and {Griffith}, Roger L. and {Marsh}, Kenneth A.},
        title = "{Hubble Space Telescope Spectroscopy of Brown Dwarfs Discovered with the Wide-field Infrared Survey Explorer}",
      journal = {\apj},
         year = 2015,
        month = may,
       volume = {804},
       number = {2},
          eid = {92},
        pages = {92},
          doi = {10.1088/0004-637X/804/2/92},
archivePrefix = {arXiv},
       eprint = {1502.05365},
 primaryClass = {astro-ph.SR},
       adsurl = {https://ui.adsabs.harvard.edu/abs/2015ApJ...804...92S}
}

@INPROCEEDINGS{Cushing2014ASSL..401..113C,
       author = {{Cushing}, Michael C.},
        title = "{Ultracool Objects: L, T, and Y Dwarfs}",
    booktitle = {50 Years of Brown Dwarfs},
         year = 2014,
       editor = {{Joergens}, Viki},
       series = {Astrophysics and Space Science Library},
       volume = {401},
        month = jan,
        pages = {113},
          doi = {10.1007/978-3-319-01162-2_7},
       adsurl = {https://ui.adsabs.harvard.edu/abs/2014ASSL..401..113C}
}

@ARTICLE{Mace2013ApJS..205....6M,
       author = {{Mace}, Gregory N. and {Kirkpatrick}, J. Davy and {Cushing}, Michael C. and {Gelino}, Christopher R. and {Griffith}, Roger L. and {Skrutskie}, Michael F. and {Marsh}, Kenneth A. and {Wright}, Edward L. and {Eisenhardt}, Peter R. and {McLean}, Ian S. and {Thompson}, Maggie A. and {Mix}, Katholeen and {Bailey}, Vanessa and {Beichman}, Charles A. and {Bloom}, Joshua S. and {Burgasser}, Adam J. and {Fortney}, Jonathan J. and {Hinz}, Philip M. and {Knox}, Russell P. and {Lowrance}, Patrick J. and {Marley}, Mark S. and {Morley}, Caroline V. and {Rodigas}, Timothy J. and {Saumon}, Didier and {Sheppard}, Scott S. and {Stock}, Nathan D.},
        title = "{A Study of the Diverse T Dwarf Population Revealed by WISE}",
      journal = {\apjs},
         year = 2013,
        month = mar,
       volume = {205},
       number = {1},
          eid = {6},
        pages = {6},
          doi = {10.1088/0067-0049/205/1/6},
archivePrefix = {arXiv},
       eprint = {1301.3913},
 primaryClass = {astro-ph.SR},
       adsurl = {https://ui.adsabs.harvard.edu/abs/2013ApJS..205....6M}
}

@ARTICLE{Tinney2012ApJ...759...60T,
       author = {{Tinney}, C.~G. and {Faherty}, Jacqueline K. and {Kirkpatrick}, J. Davy and {Wright}, Edward L. and {Gelino}, Christopher R. and {Cushing}, Michael C. and {Griffith}, Roger L. and {Salter}, Graeme},
        title = "{WISE J163940.83-684738.6: A Y Dwarf Identified by Methane Imaging}",
      journal = {\apj},
         year = 2012,
        month = nov,
       volume = {759},
       number = {1},
          eid = {60},
        pages = {60},
          doi = {10.1088/0004-637X/759/1/60},
archivePrefix = {arXiv},
       eprint = {1209.6123},
 primaryClass = {astro-ph.SR},
       adsurl = {https://ui.adsabs.harvard.edu/abs/2012ApJ...759...60T}
}

@ARTICLE{Tinney2018ApJS..236...28T,
       author = {{Tinney}, C.~G. and {Kirkpatrick}, J. Davy and {Faherty}, Jacqueline K. and {Mace}, Gregory N. and {Cushing}, Mike and {Gelino}, Christopher R. and {Burgasser}, Adam J. and {Sheppard}, Scott S. and {Wright}, Edward L.},
        title = "{New Y and T Dwarfs from WISE Identified by Methane Imaging}",
      journal = {\apjs},
         year = 2018,
        month = jun,
       volume = {236},
       number = {2},
          eid = {28},
        pages = {28},
          doi = {10.3847/1538-4365/aabad3},
archivePrefix = {arXiv},
       eprint = {1804.00362},
 primaryClass = {astro-ph.SR},
       adsurl = {https://ui.adsabs.harvard.edu/abs/2018ApJS..236...28T}
}

@ARTICLE{Burningham2013MNRAS.433..457B,
       author = {{Burningham}, Ben and {Cardoso}, C.~V. and {Smith}, L. and {Leggett}, S.~K. and {Smart}, R.~L. and {Mann}, A.~W. and {Dhital}, S. and {Lucas}, P.~W. and {Tinney}, C.~G. and {Pinfield}, D.~J. and {Zhang}, Z. and {Morley}, C. and {Saumon}, D. and {Aller}, K. and {Littlefair}, S.~P. and {Homeier}, D. and {Lodieu}, N. and {Deacon}, N. and {Marley}, M.~S. and {van Spaandonk}, L. and {Baker}, D. and {Allard}, F. and {Andrei}, A.~H. and {Canty}, J. and {Clarke}, J. and {Day-Jones}, A.~C. and {Dupuy}, T. and {Fortney}, J.~J. and {Gomes}, J. and {Ishii}, M. and {Jones}, H.~R.~A. and {Liu}, M. and {Magazz{\'u}}, A. and {Marocco}, F. and {Murray}, D.~N. and {Rojas-Ayala}, B. and {Tamura}, M.},
        title = "{76 T dwarfs from the UKIDSS LAS: benchmarks, kinematics and an updated space density}",
      journal = {\mnras},
         year = 2013,
        month = jul,
       volume = {433},
       number = {1},
        pages = {457-497},
          doi = {10.1093/mnras/stt740},
archivePrefix = {arXiv},
       eprint = {1304.7246},
 primaryClass = {astro-ph.SR},
       adsurl = {https://ui.adsabs.harvard.edu/abs/2013MNRAS.433..457B}
}

@ARTICLE{Strauss1999ApJ...522L..61S,
       author = {{Strauss}, Michael A. and {Fan}, Xiaohui and {Gunn}, James E. and {Leggett}, S.~K. and {Geballe}, T.~R. and {Pier}, Jeffrey R. and {Lupton}, Robert H. and {Knapp}, G.~R. and {Annis}, James and {Brinkmann}, J. and {Crocker}, J.~H. and {Csabai}, Istv{\'a}n and {Fukugita}, Masataka and {Golimowski}, David A. and {Harris}, Frederick H. and {Hennessy}, G.~S. and {Hindsley}, Robert B. and {Ivezi{\'c} }, {\v{Z}}eljko and {Kent}, Stephen and {Lamb}, D.~Q. and {Munn}, Jeffrey A. and {Newberg}, Heidi Jo and {Rechenmacher}, Ron and {Schneider}, Donald P. and {Smith}, J. Allyn and {Stoughton}, Chris and {Tucker}, Douglas L. and {Waddell}, Patrick and {York}, Donald G.},
        title = "{The Discovery of a Field Methane Dwarf from Sloan Digital Sky Survey Commissioning Data}",
      journal = {\apjl},
         year = 1999,
        month = sep,
       volume = {522},
       number = {1},
        pages = {L61-L64},
          doi = {10.1086/312218},
archivePrefix = {arXiv},
       eprint = {astro-ph/9905391},
 primaryClass = {astro-ph},
       adsurl = {https://ui.adsabs.harvard.edu/abs/1999ApJ...522L..61S}
}

@ARTICLE{Thompson2013PASP..125..809T,
       author = {{Thompson}, Maggie A. and {Kirkpatrick}, J. Davy and {Mace}, Gregory N. and {Cushing}, Michael C. and {Gelino}, Christopher R. and {Griffith}, Roger L. and {Skrutskie}, Michael F. and {Eisenhardt}, Peter R.~M. and {Wright}, Edward L. and {Marsh}, Kenneth A. and {Mix}, Katholeen J. and {Beichman}, Charles A. and {Faherty}, Jacqueline K. and {Toloza}, Odette and {Ferrara}, Jocelyn and {Apodaca}, Brian and {McLean}, Ian S. and {Bloom}, Joshua S.},
        title = "{Nearby M, L, and T Dwarfs Discovered by the Wide-field Infrared Survey Explorer (WISE)}",
      journal = {\pasp},
         year = 2013,
        month = jul,
       volume = {125},
       number = {929},
        pages = {809},
          doi = {10.1086/671426},
archivePrefix = {arXiv},
       eprint = {1305.4590},
 primaryClass = {astro-ph.SR},
       adsurl = {https://ui.adsabs.harvard.edu/abs/2013PASP..125..809T}
}

@ARTICLE{Tsvetanov2000ApJ...531L..61T,
       author = {{Tsvetanov}, Zlatan I. and {Golimowski}, David A. and {Zheng}, Wei and {Geballe}, T.~R. and {Leggett}, S.~K. and {Ford}, Holland C. and {Davidsen}, Arthur F. and {Uomoto}, Alan and {Fan}, Xiaohui and {Knapp}, G.~R. and {Strauss}, Michael A. and {Brinkmann}, J. and {Lamb}, D.~Q. and {Newberg}, Heidi Jo and {Rechenmacher}, Ron and {Schneider}, Donald P. and {York}, Donald G. and {Lupton}, Robert H. and {Pier}, Jeffrey R. and {Annis}, James and {Csabai}, Istv{\'a}n and {Hindsley}, Robert B. and {Ivesi{\'c}}, {\v{Z}}eljko and {Munn}, Jeffrey A. and {Thakar}, Aniruddha R. and {Waddell}, Patrick},
        title = "{The Discovery of a Second Field Methane Brown Dwarf from Sloan Digital Sky Survey Commissioning Data}",
      journal = {\apjl},
         year = 2000,
        month = mar,
       volume = {531},
       number = {1},
        pages = {L61-L65},
          doi = {10.1086/312515},
archivePrefix = {arXiv},
       eprint = {astro-ph/0001062},
 primaryClass = {astro-ph},
       adsurl = {https://ui.adsabs.harvard.edu/abs/2000ApJ...531L..61T}
}

\onecolumn

\begin{appendix}

\section{Figures}
\label{sec:appendix_figures}

Fig.~\ref{fig:grid_RvP} represents the real vs. predicted (RvP) analysis from the creation of the model grid, which is then applied for the feature importance analysis of JWST NIRSpec and MIRI modes.
Same as in Fig.~\ref{fig:Case_BF_spectrum} for the case study Y1-dwarf WISEPAJ1541-22, the posterior retrieval fits and residuals for the preferred models of the remaining objects are represented in Fig.~\ref{fig:bestfits}, while corresponding thermal profiles are shown in Fig.~\ref{fig:tp}.  

\begin{figure*}[h]
    \centering
    \includegraphics[width=0.82\linewidth]{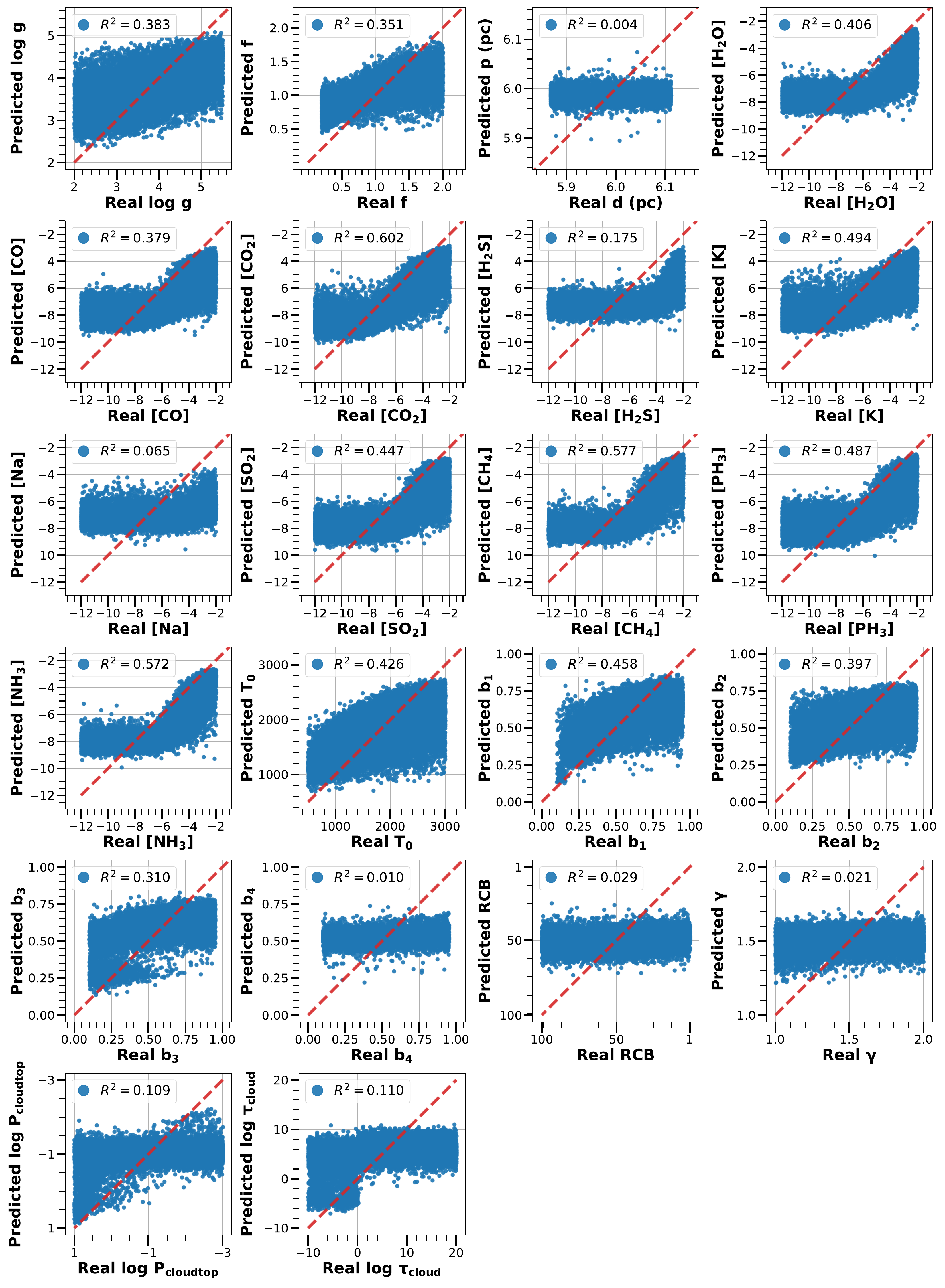}
    \caption{Real vs. predicted (RvP) comparison for all included model parameters. The RvP originates from training and testing the random forest on the same grid. No noise is assumed. In each panel, the red dashed line indicates perfect agreement.}
    \label{fig:grid_RvP}
\end{figure*}

\begin{figure*}[h]
    \centering
    \includegraphics[width=0.9\linewidth]{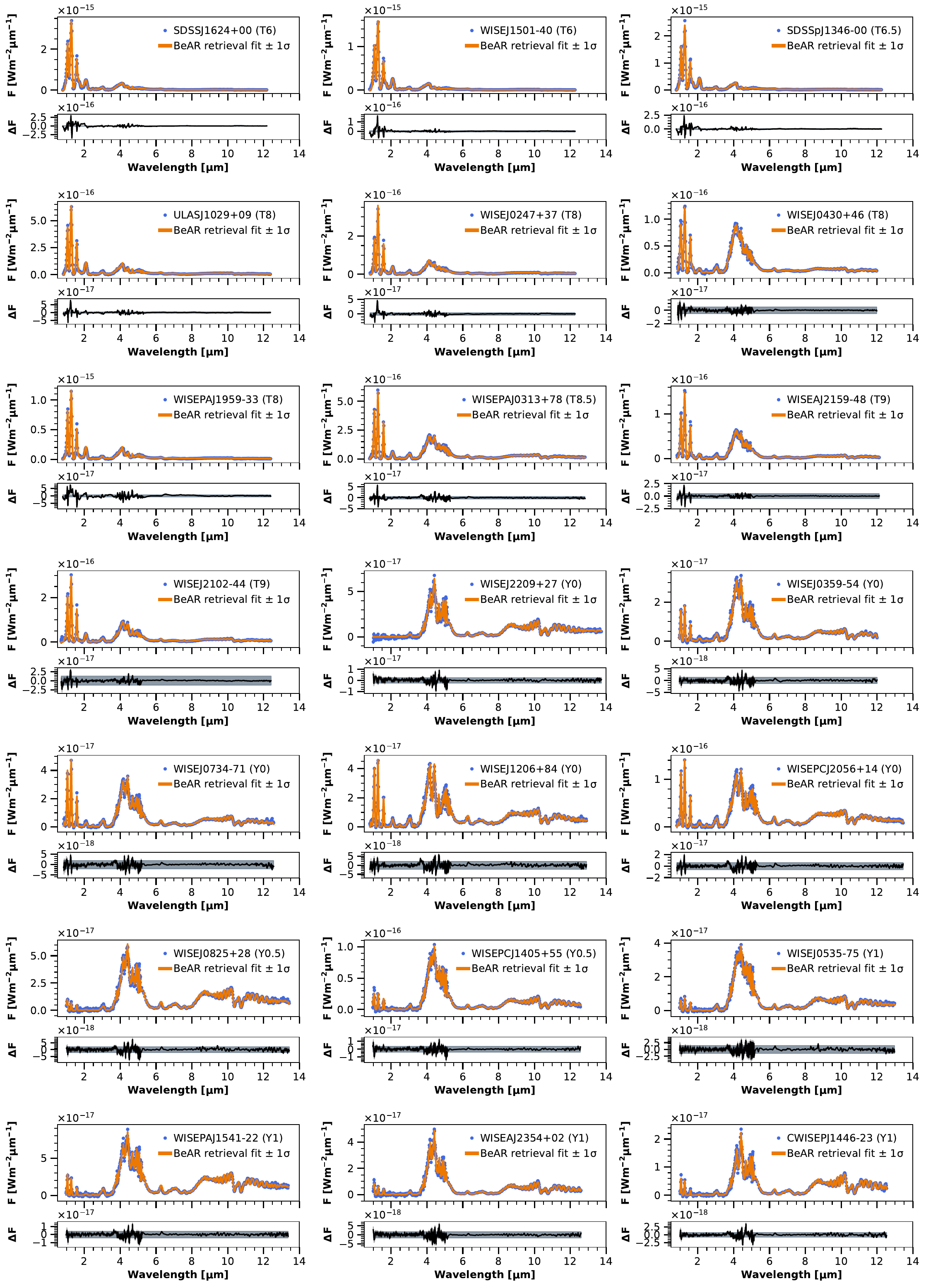}
    \caption{Posterior retrieval median fits $F$ (orange line) and residuals $\Delta$F (black line) associated with the free-chemistry retrieval analyses for the curated brown dwarfs, analogous to Fig.~\ref{fig:Case_BF_spectrum}, using the preferred cloud model by Bayesian model selection. The retrieved error inflations are represented by a gray horizontal bars. The JWST NIRSpec and MIRI data are shown as blue dots with associated uncertainties.}
    \label{fig:bestfits}
\end{figure*}

\begin{figure*}[h]
    \centering
    \includegraphics[width=0.73\linewidth]{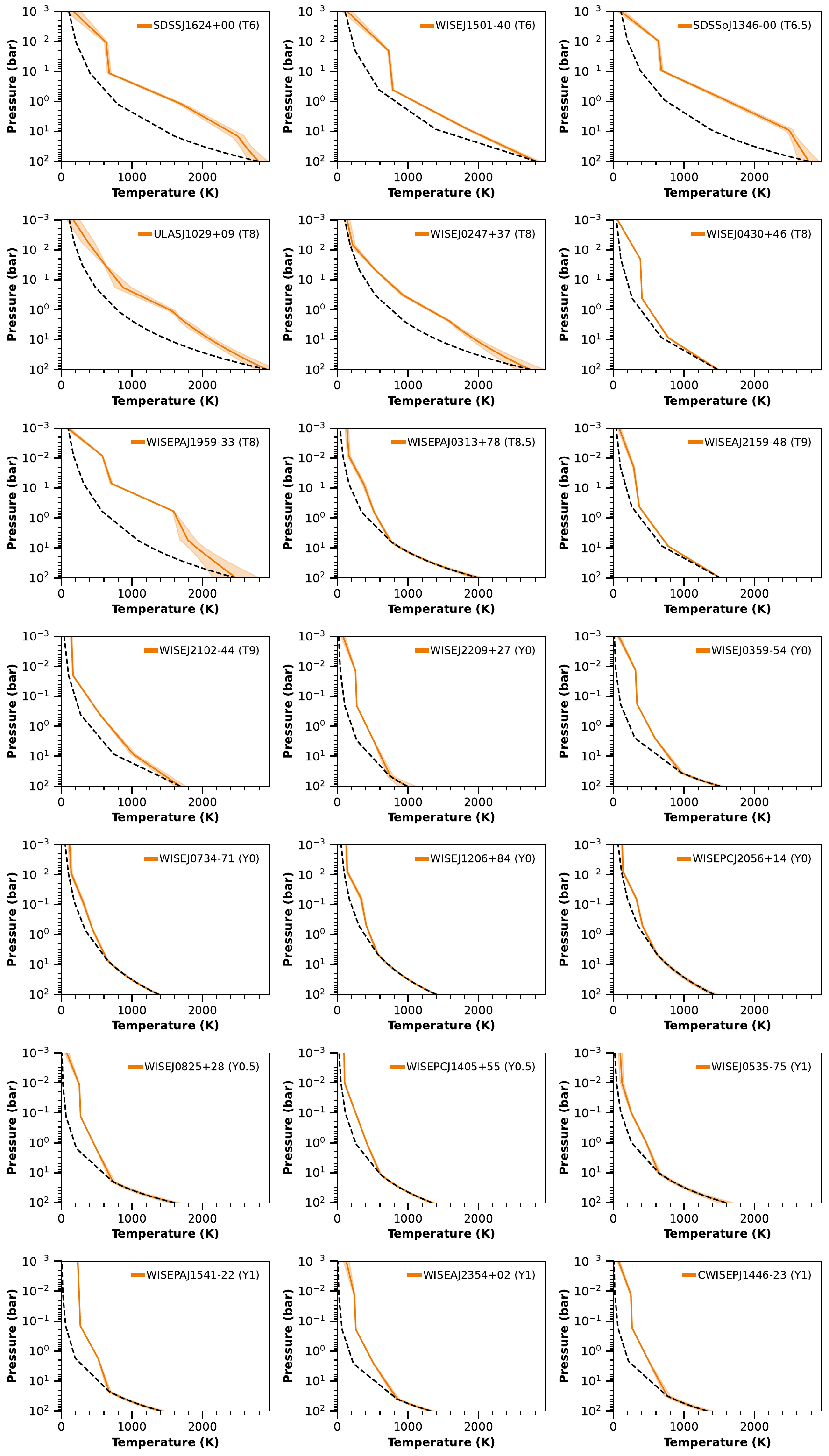}
    \caption{Retrieved median temperature–pressure profiles and associated 1-$\sigma$ uncertainties for the free-chemistry retrieval analyses of the curated brown dwarfs, analogous to Fig.~\ref{fig:Case_corner}, using the preferred cloud model by Bayesian model selection. Adiabatic profiles (black dashed line) are indicated for comparison.}
    \label{fig:tp}
\end{figure*}

\section{Posterior tables}
\label{sec:appendix_tables}
For completeness, Tabs.~\ref{tab:BeAR_Post1_CF} to \ref{tab:BeAR_Post2_G} list the posterior parameters constrained from the \bear atmospheric retrievals, both using a cloud-free and a gray cloud model. Tab.~\ref{tab:HELA_Post} then records all posterior parameters originating from the random forest retrieval study using the \texttt{HELA} framework, which is individually trained on each model grid.

\begin{table*}[h]
\centering
\caption{Summary of the cloud-free \bear retrieval outcomes for the curated sample of late-T and Y dwarfs. Listed are the following parameters: Spectral type (SpT), effective temperature (\teff), surface gravity (\logg), distance to the object (d), its radius (R), and molecular volume mixing ratios for water ($\log_{10} \text{H}_2\text{O}$), methane ($\log_{10} \text{CH}_4$), carbon monoxide ($\log_{10} \text{CO}$), and carbon dioxide ($\log_{10} \text{CO}_2$).}
\resizebox{\textwidth}{!}{
\begin{tabular}{lcccccccccc}
\hline
Source & SpT & \teff \, (K) & \logg \, (cm/s$^2$) & d (pc) & R (R$_J$) & $\log_{10} \text{H}_2\text{O}$ & $\log_{10} \text{CH}_4$ & $\log_{10} \text{CO}$ & $\log_{10} \text{CO}_2$\\
\hline
SDSSJ1624+00 & T6 & $992_{-7}^{+7}$ & $4.89_{-0.10}^{+0.11}$ & $10.89_{-0.13}^{+0.13}$ & $0.80_{-0.02}^{+0.02}$ & $-3.77_{-0.03}^{+0.03}$ & $-4.41_{-0.05}^{+0.05}$ & $-4.25_{-0.08}^{+0.08}$ & $-7.72_{-0.09}^{+0.08}$ \\
WISEJ1501-40 & T6 & $984_{-7}^{+7}$ & $4.24_{-0.08}^{+0.07}$ & $13.76_{-0.37}^{+0.37}$ & $0.68_{-0.02}^{+0.02}$ & $-3.29_{-0.03}^{+0.03}$ & $-3.72_{-0.04}^{+0.04}$ & $-3.52_{-0.09}^{+0.09}$ & $-6.65_{-0.06}^{+0.06}$ \\
SDSSpJ1346-00 & T6.5 & $932_{-8}^{+8}$ & $4.85_{-0.12}^{+0.12}$ & $14.46_{-0.41}^{+0.42}$ & $1.02_{-0.03}^{+0.04}$ & $-3.62_{-0.03}^{+0.04}$ & $-4.35_{-0.05}^{+0.05}$ & $-3.44_{-0.10}^{+0.10}$ & $-6.77_{-0.08}^{+0.08}$ \\
ULASJ1029+09 & T8 & $733_{-10}^{+11}$ & $4.55_{-0.48}^{+0.06}$ & $14.60_{-0.32}^{+0.31}$ & $0.90_{-0.03}^{+0.03}$ & $-3.31_{-0.09}^{+0.04}$ & $-4.11_{-0.25}^{+0.05}$ & $-3.61_{-0.18}^{+0.10}$ & $-7.24_{-0.15}^{+0.10}$ \\
WISEJ0247+37 & T8 & $683_{-3}^{+3}$ & $4.66_{-0.06}^{+0.06}$ & $15.45_{-0.36}^{+0.36}$ & $0.82_{-0.02}^{+0.02}$ & $-3.32_{-0.03}^{+0.03}$ & $-4.13_{-0.03}^{+0.03}$ & $-4.87_{-0.04}^{+0.04}$ & $-8.44_{-0.08}^{+0.07}$ \\
WISEJ0430+46 & T8 & $571_{-1}^{+1}$ & $4.92_{-0.05}^{+0.05}$ & $10.42_{-0.26}^{+0.27}$ & $0.75_{-0.02}^{+0.02}$ & $-3.34_{-0.02}^{+0.02}$ & $-3.70_{-0.03}^{+0.03}$ & $-5.59_{-0.02}^{+0.02}$ & $-8.50_{-0.06}^{+0.06}$ \\
WISEPAJ1959-33 & T8 & $736_{-6}^{+6}$ & $4.06_{-0.04}^{+0.07}$ & $11.93_{-0.25}^{+0.25}$ & $0.99_{-0.03}^{+0.03}$ & $-3.44_{-0.03}^{+0.03}$ & $-4.31_{-0.03}^{+0.03}$ & $-3.93_{-0.06}^{+0.06}$ & $-7.58_{-0.06}^{+0.06}$ \\
WISEPAJ0313+78 & T8.5 & $589_{-3}^{+2}$ & $5.49_{-0.01}^{+0.01}$ & $7.37_{-0.14}^{+0.14}$ & $0.88_{-0.02}^{+0.02}$ & $-2.78_{-0.02}^{+0.01}$ & $-2.80_{-0.01}^{+0.01}$ & $-4.51_{-0.03}^{+0.03}$ & $-7.28_{-0.04}^{+0.04}$ \\
WISEAJ2159-48 & T9 & $574_{-1}^{+1}$ & $5.46_{-0.06}^{+0.03}$ & $13.57_{-0.42}^{+0.42}$ & $0.85_{-0.03}^{+0.03}$ & $-2.98_{-0.03}^{+0.02}$ & $-3.15_{-0.03}^{+0.02}$ & $-5.31_{-0.03}^{+0.03}$ & $-8.26_{-0.09}^{+0.08}$ \\
WISEJ2102-44 & T9 & $597_{-3}^{+4}$ & $5.49_{-0.01}^{+0.01}$ & $10.76_{-0.19}^{+0.19}$ & $0.83_{-0.02}^{+0.02}$ & $-2.79_{-0.02}^{+0.02}$ & $-2.76_{-0.01}^{+0.01}$ & $-4.28_{-0.03}^{+0.03}$  & $-7.11_{-0.04}^{+0.04}$ \\
WISEJ2209+27 & Y0 & $353_{-2}^{+2}$ & $4.45_{-0.08}^{+0.08}$ & $6.18_{-0.07}^{+0.07}$ & $1.03_{-0.02}^{+0.02}$ & $-3.16_{-0.04}^{+0.04}$ & $-3.15_{-0.04}^{+0.04}$ & $-4.91_{-0.04}^{+0.04}$ & $-7.98_{-0.05}^{+0.05}$ \\
WISEJ0359-54 & Y0 & $463_{-2}^{+2}$ & $4.80_{-0.07}^{+0.07}$ & $13.60_{-0.33}^{+0.33}$ & $0.93_{-0.02}^{+0.03}$ & $-3.14_{-0.04}^{+0.04}$ & $-3.31_{-0.04}^{+0.04}$ & $-4.86_{-0.04}^{+0.04}$ & $-8.15_{-0.06}^{+0.06}$ \\
WISEJ0734-71 & Y0 & $480_{-2}^{+2}$ & $5.10_{-0.07}^{+0.05}$ & $13.44_{-0.29}^{+0.29}$ & $0.94_{-0.02}^{+0.02}$ & $-2.89_{-0.03}^{+0.03}$ & $-2.94_{-0.04}^{+0.03}$ & $-4.59_{-0.03}^{+0.03}$ & $-7.68_{-0.05}^{+0.05}$ \\
WISEJ1206+84 & Y0 & $451_{-2}^{+2}$ & $5.04_{-0.07}^{+0.06}$ & $11.82_{-0.27}^{+0.27}$ & $1.10_{-0.03}^{+0.03}$ & $-2.70_{-0.04}^{+0.05}$ & $-2.81_{-0.04}^{+0.04}$ & $-3.51_{-0.05}^{+0.05}$ & $-6.82_{-0.05}^{+0.05}$ \\
WISEPCJ2056+14 & Y0 & $459_{-2}^{+2}$ & $5.04_{-0.12}^{+0.08}$ & $7.10_{-0.10}^{+0.10}$ & $1.09_{-0.02}^{+0.02}$ & $-2.79_{-0.06}^{+0.05}$ & $-2.85_{-0.07}^{+0.05}$ & $-4.00_{-0.06}^{+0.05}$ & $-7.29_{-0.08}^{+0.06}$ \\
WISEJ0825+28 & Y0.5 & $379_{-6}^{+35}$ & $4.32_{-0.09}^{+0.09}$ & $6.55_{-0.08}^{+0.08}$ & $1.07_{-0.02}^{+0.02}$ & $-3.15_{-0.03}^{+0.04}$ & $-3.26_{-0.04}^{+0.04}$ & $-5.18_{-0.04}^{+0.04}$ & $-8.33_{-0.06}^{+0.06}$ \\
WISEPCJ1405+55 & Y0.5 & $399_{-3}^{+5}$ & $4.67_{-0.06}^{+0.06}$ & $6.33_{-0.09}^{+0.09}$ & $0.98_{-0.02}^{+0.02}$ & $-3.21_{-0.03}^{+0.03}$ & $-3.27_{-0.04}^{+0.04}$ & $-5.59_{-0.04}^{+0.04}$ & $-8.61_{-0.07}^{+0.06}$ \\
WISEJ0535-75 & Y1 & $400_{-2}^{+3}$ & $4.27_{-0.09}^{+0.09}$ & $13.78_{-0.22}^{+0.21}$ & $1.40_{-0.02}^{+0.01}$ & $-3.33_{-0.03}^{+0.03}$ & $-3.54_{-0.04}^{+0.04}$ & $-5.96_{-0.04}^{+0.04}$ & $-9.08_{-0.11}^{+0.10}$ \\
WISEPAJ1541-22 & Y1 & $399_{-2}^{+2}$ & $4.71_{-0.12}^{+0.12}$ & $5.99_{-0.06}^{+0.06}$ & $1.05_{-0.02}^{+0.02}$ & $-2.98_{-0.05}^{+0.06}$ & $-2.99_{-0.08}^{+0.08}$ & $-4.85_{-0.06}^{+0.06}$  & $-7.80_{-0.07}^{+0.08}$ \\
CWISEPJ1047+54 & Y1 & $430_{-7}^{+6}$ & $5.06_{-0.07}^{+0.06}$ & $14.82_{-0.92}^{+0.86}$ & $0.94_{-0.06}^{+0.05}$ & $-2.56_{-0.05}^{+0.05}$ & $-2.69_{-0.05}^{+0.05}$ & $-3.79_{-0.05}^{+0.05}$ & $-6.67_{-0.05}^{+0.05}$ \\
WISEAJ2354+02 & Y1 & $382_{-4}^{+5}$ & $4.76_{-0.07}^{+0.07}$ & $7.66_{-0.17}^{+0.18}$ & $0.86_{-0.02}^{+0.02}$ & $-3.00_{-0.03}^{+0.03}$ & $-3.10_{-0.04}^{+0.04}$ & $-5.27_{-0.04}^{+0.03}$ & $-8.06_{-0.05}^{+0.05}$ \\
CWISEPJ1446-23 & Y1 & $372_{-4}^{+5}$ & $4.87_{-0.11}^{+0.10}$ & $9.67_{-0.41}^{+0.41}$ & $0.91_{-0.04}^{+0.04}$ & $-2.78_{-0.06}^{+0.06}$ & $-2.71_{-0.06}^{+0.06}$ & $-4.48_{-0.06}^{+0.06}$ & $-7.27_{-0.07}^{+0.07}$ \\
\hline
\end{tabular}}
\label{tab:BeAR_Post1_CF}
\end{table*}

\begin{table*}[h]
\centering
\caption{Summary of the cloud-free \bear retrieval outcomes for the curated sample of late-T and Y dwarfs, analogous to Tab.~\ref{tab:BeAR_Post1_CF}. Listed are the following parameters: Spectral type (SpT) and molecular volume mixing ratios for hydrogen sulfide ($\log_{10} \text{H}_2\text{S}$), potassium ($\log_{10} \text{K}$), sulfur dioxide ($\log_{10} \text{SO}_2$), phosphine ($\log_{10} \text{PH}_3$), and ammonia ($\log_{10} \text{NH}_3$). Additionally, the radiative-convective boundary ($\log_{10}$ RCB) and adiabatic index ($\gamma$) are recorded.}
\begin{tabular}{lccccccccc}
\hline
Source & SpT  & $\log_{10} \text{H}_2\text{S}$ & $\log_{10} \text{K}$ & $\log_{10} \text{SO}_2$ &  $\log_{10} \text{PH}_3$ & $\log_{10} \text{NH}_3$ & $\log_{10}$ RCB & $\gamma$\\
\hline
SDSSJ1624+00 & T6 & $-9.18_{-1.78}^{+1.79}$ & $-6.36_{-0.03}^{+0.03}$ & $-8.37_{-2.37}^{+1.29}$ & $-7.45_{-0.16}^{+0.13}$ & $-9.44_{-1.60}^{+1.59}$ & $1.12_{-0.08}^{+0.09}$ & $1.06_{-0.03}^{+0.03}$ \\
WISEJ1501-40 & T6 &  $-8.92_{-2.11}^{+2.09}$ & $-6.33_{-0.03}^{+0.03}$ & $-6.24_{-1.22}^{+0.21}$ & $-9.25_{-1.85}^{+1.75}$ & $-9.05_{-1.98}^{+1.88}$  & $1.84_{-0.16}^{+0.16}$ & $1.49_{-0.33}^{+0.33}$ \\
SDSSpJ1346-00 & T6.5 &  $-9.48_{-1.63}^{+1.75}$ & $-6.25_{-0.03}^{+0.03}$ & $-6.70_{-0.23}^{+0.16}$ & $-7.44_{-0.24}^{+0.20}$ & $-8.59_{-2.22}^{+1.56}$ & $0.94_{-0.09}^{+0.09}$ & $1.05_{-0.03}^{+0.02}$ \\
ULASJ1029+09 & T8 &  $-9.22_{-1.81}^{+1.89}$ & $-6.31_{-0.05}^{+0.07}$ & $-6.64_{-0.18}^{+0.13}$ & $-7.63_{-0.26}^{+0.21}$ & $-6.16_{-0.75}^{+0.25}$ & $0.02_{-0.01}^{+0.69}$ & $1.16_{-0.02}^{+0.01}$ \\
WISEJ0247+37 & T8 & $-9.13_{-1.69}^{+1.78}$ & $-6.20_{-0.03}^{+0.03}$ & $-7.60_{-1.43}^{+0.28}$ & $-10.77_{-0.76}^{+0.80}$ & $-6.52_{-0.16}^{+0.13}$ & $0.36_{-0.05}^{+0.05}$ & $1.17_{-0.03}^{+0.02}$ \\
WISEJ0430+46 & T8 &  $-4.31_{-0.05}^{+0.05}$ & $-8.03_{-0.04}^{+0.04}$ & $-8.96_{-2.10}^{+1.61}$ & $-10.45_{-1.04}^{+1.09}$ & $-4.78_{-0.03}^{+0.03}$ & $1.84_{-0.16}^{+0.16}$ & $1.50_{-0.33}^{+0.34}$ \\
WISEPAJ1959-33 & T8 &  $-9.59_{-1.55}^{+1.58}$ & $-6.33_{-0.04}^{+0.04}$ & $-10.16_{-1.21}^{+1.28}$ & $-7.28_{-0.08}^{+0.07}$ & $-6.40_{-0.17}^{+0.14}$ & $0.70_{-0.06}^{+0.06}$ & $1.12_{-0.07}^{+0.06}$ \\
WISEPAJ0313+78 & T8.5 &  $-8.82_{-2.17}^{+2.11}$ & $-7.53_{-0.03}^{+0.03}$ & $-9.41_{-1.72}^{+1.78}$ & $-8.77_{-2.15}^{+1.36}$ & $-4.53_{-0.03}^{+0.03}$ & $0.72_{-0.03}^{+0.05}$ & $1.32_{-0.00}^{+0.01}$ \\
WISEAJ2159-48 & T9 &  $-8.94_{-2.05}^{+2.12}$ & $-7.56_{-0.04}^{+0.04}$ & $-9.41_{-1.75}^{+1.74}$ & $-7.43_{-0.10}^{+0.09}$ & $-4.39_{-0.05}^{+0.03}$ & $0.80_{-0.05}^{+0.05}$ & $1.34_{-0.00}^{+0.00}$ \\
WISEJ2102-44 & T9 &  $-8.87_{-2.09}^{+2.11}$ & $-7.46_{-0.04}^{+0.04}$ & $-8.42_{-2.40}^{+2.09}$ & $-7.02_{-0.77}^{+0.20}$ & $-4.41_{-0.03}^{+0.03}$ & $0.61_{-0.07}^{+0.08}$ & $1.31_{-0.00}^{+0.00}$ \\
WISEJ2209+27 & Y0 &  $-9.09_{-1.95}^{+2.04}$ & $-8.92_{-2.02}^{+1.99}$ & $-9.50_{-1.64}^{+1.55}$ & $-10.16_{-1.23}^{+1.21}$ & $-4.56_{-0.05}^{+0.05}$ & $1.64_{-0.07}^{+0.08}$ & $1.58_{-0.35}^{+0.27}$ \\
WISEJ0359-54 & Y0 & $-8.24_{-2.47}^{+2.56}$ & $-7.69_{-0.07}^{+0.06}$ & $-9.58_{-1.59}^{+1.53}$ & $-7.78_{-0.11}^{+0.10}$ & $-4.61_{-0.05}^{+0.05}$ & $1.52_{-0.06}^{+0.05}$ & $1.77_{-0.09}^{+0.11}$ \\
WISEJ0734-71 & Y0  & $-8.78_{-2.16}^{+2.24}$ & $-8.21_{-0.08}^{+0.09}$ & $-9.56_{-1.65}^{+1.68}$ & $-7.44_{-0.17}^{+0.12}$ & $-4.39_{-0.04}^{+0.04}$ & $0.81_{-0.05}^{+0.05}$ & $1.40_{-0.01}^{+0.01}$ \\
WISEJ1206+84 & Y0  & $-4.34_{-1.40}^{+0.21}$ & $-8.28_{-0.06}^{+0.07}$ & $-9.38_{-1.76}^{+1.80}$ & $-6.96_{-0.10}^{+0.09}$ & $-4.15_{-0.06}^{+0.06}$ & $0.42_{-0.04}^{+0.05}$ & $1.33_{-0.01}^{+0.01}$ \\
WISEPCJ2056+14 & Y0  & $-8.65_{-2.30}^{+2.35}$ & $-8.18_{-0.11}^{+0.18}$ & $-9.70_{-1.56}^{+1.62}$ & $-7.05_{-0.12}^{+0.10}$ & $-4.32_{-0.08}^{+0.06}$ & $0.73_{-0.06}^{+0.05}$ & $1.41_{-0.01}^{+0.02}$ \\
WISEJ0825+28 & Y0.5 &  $-9.04_{-1.92}^{+2.00}$ & $-9.85_{-1.40}^{+1.24}$ & $-9.23_{-1.83}^{+1.53}$ & $-10.38_{-1.07}^{+1.13}$ & $-4.71_{-0.05}^{+0.05}$ & $1.25_{-0.05}^{+0.06}$ & $1.95_{-0.05}^{+0.03}$ \\
WISEPCJ1405+55 & Y0.5  & $-6.16_{-3.93}^{+1.28}$ & $-8.41_{-1.48}^{+0.27}$ & $-10.04_{-1.28}^{+1.34}$ & $-9.40_{-1.65}^{+0.95}$ & $-4.60_{-0.04}^{+0.04}$ & $1.04_{-0.04}^{+0.04}$ & $1.57_{-0.03}^{+0.04}$ \\
WISEJ0535-75 & Y1 &  $-4.72_{-0.15}^{+0.13}$ & $-7.75_{-0.18}^{+0.13}$ & $-9.06_{-1.97}^{+1.47}$ & $-10.58_{-0.94}^{+1.00}$ & $-4.83_{-0.05}^{+0.05}$ & $0.98_{-0.08}^{+0.09}$ & $1.66_{-0.09}^{+0.13}$ \\
WISEPAJ1541-22 & Y1  & $-4.94_{-4.29}^{+0.46}$ & $-8.02_{-0.39}^{+0.16}$ & $-9.67_{-1.52}^{+1.62}$ & $-10.36_{-1.10}^{+1.14}$ & $-4.56_{-0.07}^{+0.07}$ & $1.39_{-0.07}^{+0.08}$ & $1.97_{-0.04}^{+0.02}$ \\
CWISEPJ1047+54 & Y1  & $-8.80_{-2.16}^{+2.19}$ & $-10.64_{-0.92}^{+0.97}$ & $-9.46_{-1.72}^{+1.64}$ & $-10.00_{-1.36}^{+1.41}$ & $-4.02_{-0.06}^{+0.06}$ & $0.52_{-0.03}^{+0.04}$ & $1.39_{-0.01}^{+0.01}$ \\
WISEAJ2354+02 & Y1 & $-8.77_{-2.15}^{+2.22}$ & $-10.09_{-1.26}^{+1.22}$ & $-9.81_{-1.46}^{+1.50}$ & $-8.04_{-0.23}^{+0.15}$ & $-4.39_{-0.04}^{+0.04}$ & $1.56_{-0.06}^{+0.05}$ & $1.95_{-0.07}^{+0.04}$ \\
CWISEPJ1446-23 & Y1 & $-8.65_{-2.25}^{+2.35}$ & $-9.94_{-1.37}^{+1.32}$ & $-9.31_{-1.79}^{+1.75}$ & $-10.44_{-1.04}^{+1.11}$ & $-4.20_{-0.07}^{+0.07}$ & $1.46_{-0.06}^{+0.06}$ & $1.93_{-0.07}^{+0.05}$ \\
\hline
\end{tabular}
\label{tab:BeAR_Post2_CF}
\end{table*}

\begin{table*}[h]
\centering
\caption{Summary of the gray-cloud \bear retrieval outcomes for the curated sample of late-T and Y dwarfs. Listed are the following parameters: Spectral type (SpT), effective temperature (\teff), surface gravity (\logg), distance to the object (d), its radius (R), and molecular volume mixing ratios for water ($\log_{10} \text{H}_2\text{O}$), methane ($\log_{10} \text{CH}_4$), carbon monoxide ($\log_{10} \text{CO}$), carbon dioxide ($\log_{10} \text{CO}_2$), and hydrogen sulfide ($\log_{10} \text{H}_2\text{S}$).}
\resizebox{\textwidth}{!}{
\begin{tabular}{lccccccccccc}
\hline
Source & SpT & \teff \, (K) & \logg \, (cm/s$^2$) & d (pc) & R (R$_J$) & $\log_{10} \text{H}_2\text{O}$ & $\log_{10} \text{CH}_4$ & $\log_{10} \text{CO}$ & $\log_{10} \text{CO}_2$ & $\log_{10} \text{H}_2\text{S}$\\
\hline
SDSSJ1624+00 & T6 & $991_{-7}^{+7}$ & $4.90_{-0.09}^{+0.10}$ & $10.90_{-0.12}^{+0.12}$ & $0.80_{-0.02}^{+0.02}$ & $-3.77_{-0.03}^{+0.03}$ & $-4.41_{-0.04}^{+0.05}$ & $-4.25_{-0.08}^{+0.08}$ & $-7.71_{-0.09}^{+0.08}$ & $-9.17_{-1.75}^{+1.76}$ \\
WISEJ1501-40 & T6 & $982_{-7}^{+7}$ & $4.24_{-0.07}^{+0.07}$ & $13.75_{-0.38}^{+0.38}$ & $0.68_{-0.02}^{+0.02}$ & $-3.29_{-0.03}^{+0.03}$ & $-3.72_{-0.04}^{+0.04}$ & $-3.52_{-0.08}^{+0.09}$ & $-6.65_{-0.06}^{+0.06}$ & $-8.92_{-2.09}^{+2.08}$ \\
SDSSpJ1346-00 & T6.5 & $932_{-8}^{+8}$ & $4.84_{-0.11}^{+0.12}$ & $14.48_{-0.42}^{+0.42}$ & $1.02_{-0.03}^{+0.04}$ & $-3.62_{-0.03}^{+0.03}$ & $-4.34_{-0.05}^{+0.05}$ & $-3.43_{-0.10}^{+0.10}$ & $-6.77_{-0.08}^{+0.08}$ & $-9.48_{-1.63}^{+1.71}$ \\
ULASJ1029+09 & T8 & $742_{-7}^{+7}$ & $4.03_{-0.02}^{+0.03}$ & $14.60_{-0.31}^{+0.30}$ & $0.88_{-0.03}^{+0.03}$ & $-3.44_{-0.03}^{+0.03}$ & $-4.39_{-0.02}^{+0.02}$ & $-3.85_{-0.07}^{+0.07}$ & $-7.45_{-0.08}^{+0.07}$ & $-9.31_{-1.67}^{+1.73}$ \\
WISEJ0247+37 & T8 & $689_{-4}^{+4}$ & $4.69_{-0.06}^{+0.07}$ & $15.46_{-0.41}^{+0.41}$ & $0.81_{-0.02}^{+0.02}$ & $-3.39_{-0.02}^{+0.03}$ & $-4.16_{-0.03}^{+0.03}$ & $-4.94_{-0.04}^{+0.04}$ & $-8.40_{-0.07}^{+0.07}$ & $-9.16_{-1.81}^{+1.87}$ \\
WISEJ0430+46 & T8 & $567_{-2}^{+2}$ & $4.87_{-0.05}^{+0.06}$ & $10.43_{-0.24}^{+0.23}$ & $0.76_{-0.02}^{+0.02}$ & $-3.37_{-0.02}^{+0.03}$ & $-3.73_{-0.03}^{+0.03}$ & $-5.63_{-0.03}^{+0.03}$ & $-8.58_{-0.07}^{+0.07}$ & $-4.59_{-0.14}^{+0.15}$ \\
WISEPAJ1959-33 & T8 & $736_{-6}^{+6}$ & $4.06_{-0.04}^{+0.06}$ & $11.93_{-0.25}^{+0.26}$ & $0.99_{-0.03}^{+0.03}$ & $-3.44_{-0.03}^{+0.03}$ & $-4.31_{-0.03}^{+0.03}$ & $-3.93_{-0.06}^{+0.06}$ & $-7.59_{-0.06}^{+0.06}$ & $-9.61_{-1.57}^{+1.67}$ \\
WISEPAJ0313+78 & T8.5 & $580_{-2}^{+2}$ & $5.20_{-0.06}^{+0.07}$ & $7.38_{-0.14}^{+0.13}$ & $0.91_{-0.02}^{+0.02}$ & $-2.74_{-0.03}^{+0.03}$ & $-2.85_{-0.04}^{+0.04}$ & $-4.37_{-0.04}^{+0.04}$ & $-7.39_{-0.05}^{+0.05}$ & $-8.26_{-2.44}^{+2.45}$ \\
WISEAJ2159-48 & T9 & $551_{-2}^{+2}$ & $4.96_{-0.05}^{+0.05}$ & $13.53_{-0.31}^{+0.27}$ & $0.92_{-0.02}^{+0.02}$ & $-3.16_{-0.02}^{+0.02}$ & $-3.39_{-0.02}^{+0.02}$ & $-5.40_{-0.02}^{+0.02}$ & $-8.59_{-0.09}^{+0.08}$ & $-8.89_{-1.83}^{+1.85}$ \\
WISEJ2102-44 & T9 & $592_{-3}^{+3}$ & $4.70_{-0.07}^{+0.08}$ & $10.76_{-0.18}^{+0.19}$ & $0.85_{-0.02}^{+0.02}$ & $-2.87_{-0.03}^{+0.03}$ & $-3.04_{-0.04}^{+0.04}$ & $-4.28_{-0.03}^{+0.03}$ & $-7.42_{-0.05}^{+0.05}$ & $-8.71_{-2.13}^{+2.05}$ \\
WISEJ2209+27 & Y0 & $353_{-2}^{+2}$ & $4.45_{-0.08}^{+0.08}$ & $6.18_{-0.07}^{+0.07}$ & $1.03_{-0.02}^{+0.02}$ & $-3.16_{-0.03}^{+0.04}$ & $-3.15_{-0.04}^{+0.04}$ & $-4.91_{-0.04}^{+0.04}$ & $-7.98_{-0.05}^{+0.05}$ & $-9.03_{-1.93}^{+1.98}$ \\
WISEJ0359-54 & Y0 & $462_{-2}^{+2}$ & $4.80_{-0.07}^{+0.07}$ & $13.62_{-0.34}^{+0.33}$ & $0.93_{-0.03}^{+0.03}$ & $-3.14_{-0.03}^{+0.04}$ & $-3.31_{-0.04}^{+0.04}$ & $-4.86_{-0.04}^{+0.04}$ & $-8.15_{-0.06}^{+0.06}$ & $-8.17_{-2.50}^{+2.49}$ \\
WISEJ0734-71 & Y0 & $468_{-2}^{+2}$ & $4.34_{-0.04}^{+0.04}$ & $13.37_{-0.17}^{+0.17}$ & $0.99_{-0.02}^{+0.02}$ & $-3.15_{-0.02}^{+0.02}$ & $-3.31_{-0.02}^{+0.02}$ & $-4.79_{-0.02}^{+0.02}$ & $-8.08_{-0.03}^{+0.03}$ & $-8.91_{-1.58}^{+1.61}$ \\
WISEJ1206+84 & Y0 & $456_{-2}^{+2}$ & $5.33_{-0.07}^{+0.07}$ & $11.83_{-0.28}^{+0.27}$ & $1.07_{-0.03}^{+0.03}$ & $-2.62_{-0.04}^{+0.04}$ & $-2.65_{-0.04}^{+0.04}$ & $-3.44_{-0.04}^{+0.04}$ & $-6.68_{-0.05}^{+0.04}$ & $-8.10_{-2.63}^{+2.58}$ \\
WISEPCJ2056+14 & Y0 & $444_{-4}^{+4}$ & $4.96_{-0.12}^{+0.10}$ & $7.10_{-0.09}^{+0.09}$ & $1.17_{-0.03}^{+0.03}$ & $-2.87_{-0.06}^{+0.06}$ & $-2.92_{-0.07}^{+0.06}$ & $-4.08_{-0.06}^{+0.05}$ & $-7.35_{-0.07}^{+0.07}$ & $-8.81_{-2.07}^{+2.09}$ \\
WISEJ0825+28 & Y0.5 & $378_{-6}^{+35}$ & $4.32_{-0.09}^{+0.09}$ & $6.55_{-0.08}^{+0.08}$ & $1.07_{-0.02}^{+0.02}$ & $-3.15_{-0.03}^{+0.04}$ & $-3.26_{-0.04}^{+0.04}$ & $-5.18_{-0.04}^{+0.04}$ & $-8.32_{-0.06}^{+0.06}$ & $-9.01_{-1.99}^{+2.01}$ \\
WISEPCJ1405+55 & Y0.5 & $394_{-3}^{+3}$ & $4.62_{-0.00}^{+0.00}$ & $6.43_{-0.00}^{+0.00}$ & $0.93_{-0.00}^{+0.00}$ & $-3.29_{-0.00}^{+0.00}$ & $-3.29_{-0.00}^{+0.00}$ & $-5.58_{-0.00}^{+0.00}$ & $-8.60_{-0.00}^{+0.00}$ & $-6.85_{-0.00}^{+0.00}$ \\
WISEJ0535-75 & Y1 & $399_{-3}^{+3}$ & $4.30_{-0.09}^{+0.09}$ & $13.69_{-0.24}^{+0.24}$ & $1.40_{-0.01}^{+0.01}$ & $-3.34_{-0.03}^{+0.03}$ & $-3.54_{-0.04}^{+0.04}$ & $-5.98_{-0.06}^{+0.05}$ & $-9.07_{-0.11}^{+0.10}$ & $-4.71_{-0.14}^{+0.12}$ \\
WISEPAJ1541-22 & Y1 & $350_{-4}^{+6}$ & $4.74_{-0.08}^{+0.09}$ & $5.98_{-0.06}^{+0.05}$ & $1.07_{-0.05}^{+0.03}$ & $-2.97_{-0.04}^{+0.04}$ & $-2.92_{-0.04}^{+0.05}$ & $-4.80_{-0.04}^{+0.04}$ & $-7.73_{-0.05}^{+0.05}$ & $-8.63_{-2.01}^{+2.16}$ \\
CWISEPJ1047+54 & Y1 & $430_{-6}^{+5}$ & $5.34_{-0.08}^{+0.07}$ & $14.79_{-0.84}^{+0.84}$ & $0.92_{-0.05}^{+0.05}$ & $-2.41_{-0.05}^{+0.05}$ & $-2.50_{-0.05}^{+0.04}$ & $-3.62_{-0.05}^{+0.04}$ & $-6.48_{-0.05}^{+0.05}$ & $-8.64_{-2.24}^{+2.27}$ \\
WISEAJ2354+02 & Y1 & $381_{-4}^{+5}$ & $4.76_{-0.07}^{+0.06}$ & $7.67_{-0.17}^{+0.18}$ & $0.86_{-0.02}^{+0.02}$ & $-3.00_{-0.03}^{+0.03}$ & $-3.10_{-0.04}^{+0.04}$ & $-5.27_{-0.04}^{+0.04}$ & $-8.06_{-0.05}^{+0.05}$ & $-8.69_{-2.15}^{+2.12}$ \\
CWISEPJ1446-23 & Y1 & $372_{-3}^{+5}$ & $4.88_{-0.11}^{+0.10}$ & $9.67_{-0.42}^{+0.40}$ & $0.91_{-0.04}^{+0.04}$ & $-2.77_{-0.05}^{+0.05}$ & $-2.71_{-0.06}^{+0.06}$ & $-4.47_{-0.05}^{+0.06}$ & $-7.27_{-0.07}^{+0.06}$ & $-8.57_{-2.28}^{+2.25}$ \\
\hline
\end{tabular}}
\label{tab:BeAR_Post1_G}
\end{table*}

\begin{table*}[h]
\centering
\caption{Summary of the gray-cloud \bear retrieval outcomes for the curated sample of late-T and Y dwarfs, analogous to Tab.~\ref{tab:BeAR_Post1_G}. Listed are the following parameters: Spectral type (SpT) and molecular volume mixing ratios for potassium ($\log_{10} \text{K}$), sulfur dioxide ($\log_{10} \text{SO}_2$), phosphine ($\log_{10} \text{PH}_3$), and ammonia ($\log_{10} \text{NH}_3$). Additionally, the cloud optical depth ($\tau_\mathrm{c}$), cloud top pressure ($\log p_{\mathrm{t}}$), the radiative-convective boundary ($\log_{10}$ RCB), and adiabatic index ($\gamma$) are recorded.}
\resizebox{\textwidth}{!}{
\begin{tabular}{lccccccccccc}
\hline
Source & SpT  & $\log_{10} \text{K}$ & $\log_{10} \text{SO}_2$ &  $\log_{10} \text{PH}_3$ & $\log_{10} \text{NH}_3$ & $\tau_\mathrm{c}$ & $\log p_{\mathrm{t}}$ & $\log_{10}$ RCB & $\gamma$\\
\hline
SDSSJ1624+00 & T6 &  $-6.36_{-0.03}^{+0.03}$ & $-8.18_{-2.47}^{+1.12}$ & $-7.45_{-0.16}^{+0.13}$ & $-9.44_{-1.62}^{+1.61}$ & $9.45_{-8.16}^{+6.74}$ & $-1.46_{-1.00}^{+1.51}$ & $1.12_{-0.08}^{+0.09}$ & $1.06_{-0.03}^{+0.03}$ \\
WISEJ1501-40 & T6 & $-6.32_{-0.03}^{+0.03}$ & $-6.23_{-0.79}^{+0.20}$ & $-9.17_{-1.93}^{+1.62}$ & $-9.05_{-1.96}^{+1.90}$ & $8.33_{-12.25}^{+8.78}$ & $-1.97_{-0.70}^{+0.86}$ & $1.84_{-0.16}^{+0.16}$ & $1.50_{-0.34}^{+0.33}$ \\
SDSSpJ1346-00 & T6.5  & $-6.25_{-0.03}^{+0.03}$ & $-6.71_{-0.23}^{+0.17}$ & $-7.43_{-0.23}^{+0.19}$ & $-8.60_{-2.15}^{+1.54}$ & $9.24_{-8.43}^{+6.94}$ & $-1.49_{-0.98}^{+1.52}$ & $0.93_{-0.08}^{+0.09}$ & $1.04_{-0.03}^{+0.02}$ \\
ULASJ1029+09 & T8  & $-6.21_{-0.04}^{+0.04}$ & $-6.85_{-0.09}^{+0.08}$ & $-7.63_{-0.16}^{+0.14}$ & $-7.23_{-1.39}^{+0.43}$ & $8.87_{-7.56}^{+6.88}$ & $-1.36_{-1.08}^{+1.47}$ & $0.79_{-0.05}^{+0.04}$ & $1.14_{-0.06}^{+0.05}$ \\
WISEJ0247+37 & T8 &  $-6.28_{-0.03}^{+0.03}$ & $-7.22_{-0.17}^{+0.14}$ & $-10.78_{-0.79}^{+0.88}$ & $-6.78_{-0.24}^{+0.18}$ & $8.87_{-7.65}^{+7.22}$ & $-1.39_{-1.07}^{+1.51}$ & $1.74_{-0.07}^{+0.07}$ & $1.53_{-0.28}^{+0.28}$ \\
WISEJ0430+46 & T8  & $-7.97_{-0.05}^{+0.04}$ & $-9.44_{-1.70}^{+1.67}$ & $-10.52_{-0.97}^{+1.02}$ & $-4.82_{-0.04}^{+0.04}$ & $3.15_{-0.65}^{+0.50}$ & $-0.85_{-0.28}^{+0.14}$ & $1.84_{-0.16}^{+0.15}$ & $1.51_{-0.33}^{+0.32}$ \\
WISEPAJ1959-33 & T8  & $-6.33_{-0.04}^{+0.03}$ & $-10.13_{-1.24}^{+1.25}$ & $-7.28_{-0.08}^{+0.07}$ & $-6.39_{-0.16}^{+0.13}$ & $8.64_{-7.79}^{+7.44}$ & $-1.26_{-1.16}^{+1.48}$ & $0.70_{-0.06}^{+0.07}$ & $1.11_{-0.07}^{+0.07}$ \\
WISEPAJ0313+78 & T8.5  & $-7.38_{-0.05}^{+0.05}$ & $-9.68_{-1.48}^{+1.54}$ & $-10.09_{-1.23}^{+1.24}$ & $-4.38_{-0.06}^{+0.05}$ & $18.89_{-1.12}^{+0.91}$ & $0.26_{-0.03}^{+0.03}$ & $0.74_{-0.05}^{+0.04}$ & $1.54_{-0.03}^{+0.03}$ \\
WISEAJ2159-48 & T9  & $-7.51_{-0.03}^{+0.02}$ & $-9.22_{-1.64}^{+1.51}$ & $-7.95_{-0.15}^{+0.12}$ & $-4.53_{-0.03}^{+0.03}$ & $0.93_{-0.10}^{+0.11}$ & $-0.61_{-0.03}^{+0.03}$ & $1.85_{-0.15}^{+0.14}$ & $1.48_{-0.30}^{+0.30}$ \\
WISEJ2102-44 & T9  & $-7.45_{-0.07}^{+0.06}$ & $-8.60_{-2.26}^{+1.83}$ & $-9.54_{-1.60}^{+1.52}$ & $-4.53_{-0.05}^{+0.05}$ & $10.20_{-1.09}^{+1.38}$ & $0.12_{-0.10}^{+0.08}$ & $1.84_{-0.16}^{+0.16}$ & $1.50_{-0.32}^{+0.32}$ \\
WISEJ2209+27 & Y0  & $-8.85_{-2.06}^{+1.99}$ & $-9.58_{-1.59}^{+1.58}$ & $-10.11_{-1.19}^{+1.13}$ & $-4.56_{-0.05}^{+0.05}$ & $9.75_{-11.93}^{+7.11}$ & $-1.80_{-0.79}^{+1.20}$ & $1.64_{-0.07}^{+0.07}$ & $1.59_{-0.31}^{+0.27}$ \\
WISEJ0359-54 & Y0 &  $-7.69_{-0.07}^{+0.06}$ & $-9.57_{-1.54}^{+1.57}$ & $-7.77_{-0.11}^{+0.10}$ & $-4.61_{-0.05}^{+0.05}$ & $7.97_{-12.31}^{+8.35}$ & $-2.04_{-0.62}^{+0.65}$ & $1.52_{-0.06}^{+0.05}$ & $1.78_{-0.09}^{+0.10}$ \\
WISEJ0734-71 & Y0  & $-8.17_{-0.04}^{+0.04}$ & $-6.83_{-0.30}^{+0.17}$ & $-7.65_{-0.07}^{+0.07}$ & $-4.68_{-0.03}^{+0.03}$ & $0.77_{-0.07}^{+0.05}$ & $-0.24_{-0.04}^{+0.04}$ & $1.85_{-0.13}^{+0.12}$ & $1.52_{-0.26}^{+0.26}$ \\
WISEJ1206+84 & Y0  & $-8.05_{-0.09}^{+0.11}$ & $-9.08_{-2.02}^{+1.94}$ & $-6.72_{-0.09}^{+0.08}$ & $-4.11_{-0.05}^{+0.05}$ & $8.13_{-1.25}^{+1.27}$ & $-0.13_{-0.08}^{+0.07}$ & $0.60_{-0.06}^{+0.06}$ & $1.41_{-0.02}^{+0.02}$ \\
WISEPCJ2056+14 & Y0  & $-8.13_{-0.10}^{+0.11}$ & $-9.71_{-1.45}^{+1.53}$ & $-7.24_{-0.12}^{+0.11}$ & $-4.35_{-0.07}^{+0.07}$ & $0.54_{-0.09}^{+0.11}$ & $-0.48_{-0.07}^{+0.06}$ & $0.62_{-0.05}^{+0.05}$ & $1.37_{-0.01}^{+0.01}$ \\
WISEJ0825+28 & Y0.5  & $-9.86_{-1.40}^{+1.24}$ & $-9.25_{-1.81}^{+1.56}$ & $-10.41_{-1.04}^{+1.17}$ & $-4.71_{-0.05}^{+0.05}$ & $6.15_{-10.07}^{+9.41}$ & $-2.07_{-0.61}^{+0.64}$ & $1.25_{-0.06}^{+0.05}$ & $1.96_{-0.05}^{+0.03}$ \\
WISEPCJ1405+55 & Y0.5  & $-8.26_{-0.00}^{+0.00}$ & $-8.05_{-0.00}^{+0.00}$ & $-8.95_{-0.00}^{+0.00}$ & $-4.66_{-0.00}^{+0.00}$ & $5.35_{-0.00}^{+0.00}$ & $-0.78_{-0.00}^{+0.00}$ & $1.11_{-0.00}^{+0.00}$ & $1.65_{-0.00}^{+0.00}$ \\
WISEJ0535-75 & Y1 & $-7.77_{-0.17}^{+0.13}$ & $-9.27_{-1.73}^{+1.53}$ & $-10.60_{-0.90}^{+0.97}$ & $-4.84_{-0.04}^{+0.05}$ & $5.78_{-2.65}^{+6.78}$ & $-0.83_{-1.55}^{+0.82}$ & $0.92_{-0.15}^{+0.11}$ & $1.64_{-0.08}^{+0.12}$ \\
WISEPAJ1541-22 & Y1  & $-8.29_{-1.04}^{+0.27}$ & $-9.75_{-1.27}^{+1.29}$ & $-10.38_{-0.92}^{+0.97}$ & $-4.49_{-0.05}^{+0.05}$ & $5.31_{-3.01}^{+2.85}$ & $-0.33_{-0.06}^{+0.04}$ & $1.32_{-0.04}^{+0.05}$ & $1.46_{-0.04}^{+0.03}$ \\
CWISEPJ1047+54 & Y1 & $-10.25_{-1.13}^{+1.23}$ & $-9.54_{-1.63}^{+1.70}$ & $-9.91_{-1.39}^{+1.44}$ & $-3.83_{-0.06}^{+0.06}$ & $13.44_{-0.92}^{+1.39}$ & $-0.12_{-0.07}^{+0.06}$ & $0.62_{-0.04}^{+0.05}$ & $1.47_{-0.02}^{+0.02}$ \\
WISEAJ2354+02 & Y1  & $-10.16_{-1.20}^{+1.28}$ & $-9.79_{-1.47}^{+1.47}$ & $-8.03_{-0.22}^{+0.14}$ & $-4.39_{-0.04}^{+0.04}$ & $7.42_{-11.16}^{+7.87}$ & $-1.96_{-0.68}^{+0.75}$ & $1.55_{-0.05}^{+0.05}$ & $1.95_{-0.06}^{+0.04}$ \\
CWISEPJ1446-23 & Y1  & $-9.91_{-1.39}^{+1.31}$ & $-9.31_{-1.81}^{+1.76}$ & $-10.48_{-1.00}^{+1.09}$ & $-4.19_{-0.06}^{+0.07}$ & $5.23_{-10.27}^{+9.98}$ & $-2.03_{-0.63}^{+0.66}$ & $1.46_{-0.06}^{+0.06}$ & $1.93_{-0.07}^{+0.05}$ \\
\hline
\end{tabular}}
\label{tab:BeAR_Post2_G}
\end{table*}

\clearpage

\begin{longtable}{llcccccccc}
\caption{Summary of the random forest retrieval outcomes using the \texttt{HELA framework}. Posterior parameters are listed separately for each model on which the algorithm has been trained. Therefore, posteriors are provided only for parameters that have been included in the specific model grid used for training.}\label{tab:HELA_Post} \\
\hline \hline
Source & SpT & \teff & \logg & [M/H] & (C/O)$_{\odot}$ & $\log\kappa_{zz}$ &
$f_{sed}$ & adiab. index & Radius \\ 
&  & (K) & (cm/s$^2$) & & & (cm$^2$/s) & & & (R$_J$) \\
\hline
\endfirsthead
\hline
\multicolumn{10}{c}{\textbf{Helios}} \\
\hline
SDSSJ1624+00 & T6 & 750$^{+49}_{-50}$ & 4.73$^{+0.65}_{-0.73}$ & 0.0$^{+0.0}_{-0.0}$ & 0.5$^{+0.0}_{-0.0}$ & \nodata & \nodata & \nodata & 1.06$^{+0.23}_{-0.21}$ \\ 
WISEJ1501-40 & T6 & 684$^{+116}_{-84}$ & 4.76$^{+0.64}_{-0.56}$ & 0.0$^{+0.0}_{-0.0}$ & 0.5$^{+0.0}_{-0.0}$ & \nodata & \nodata & \nodata & 0.88$^{+0.14}_{-0.12}$ \\ 
SDSSpJ1346-00 & T6.5 & 742$^{+57}_{-42}$ & 4.61$^{+0.59}_{-0.63}$ & 0.0$^{+0.0}_{-0.0}$ & 0.5$^{+0.0}_{-0.0}$ & \nodata & \nodata & \nodata & 0.97$^{+0.27}_{-0.12}$ \\ 
ULASJ1029+09 & T8 & 621$^{+78}_{-21}$ & 4.55$^{+0.85}_{-0.75}$ & 0.0$^{+0.0}_{-0.0}$ & 0.5$^{+0.0}_{-0.0}$ & \nodata & \nodata & \nodata & 0.91$^{+0.17}_{-0.18}$ \\ 
WISEJ0247+37 & T8 & 568$^{+31}_{-68}$ & 4.65$^{+0.55}_{-0.65}$ & 0.0$^{+0.0}_{-0.0}$ & 0.5$^{+0.0}_{-0.0}$ & \nodata & \nodata & \nodata & 0.85$^{+0.19}_{-0.14}$ \\ 
WISEJ0430+46 & T8 & 548$^{+51}_{-48}$ & 4.90$^{+0.88}_{-1.01}$ & 0.0$^{+0.0}_{-0.0}$ & 0.5$^{+0.0}_{-0.0}$ & \nodata & \nodata & \nodata & 0.93$^{+0.28}_{-0.21}$ \\ 
WISEPAJ1959-33 & T8 & 722$^{+77}_{-98}$ & 4.32$^{+0.48}_{-0.52}$ & 0.0$^{+0.0}_{-0.0}$ & 0.5$^{+0.0}_{-0.0}$ & \nodata & \nodata & \nodata & 0.92$^{+0.13}_{-0.12}$ \\ 
WISEPAJ0313+78 & T8.5 & 672$^{+27}_{-72}$ & 4.35$^{+0.85}_{-0.75}$ & 0.0$^{+0.0}_{-0.0}$ & 0.5$^{+0.0}_{-0.0}$ & \nodata & \nodata & \nodata & 1.04$^{+0.14}_{-0.17}$ \\ 
WISEAJ2159-48 & T9 & 518$^{+81}_{-118}$ & 4.86$^{+0.79}_{-0.86}$ & 0.0$^{+0.0}_{-0.0}$ & 0.5$^{+0.0}_{-0.0}$ & \nodata & \nodata & \nodata & 0.94$^{+0.27}_{-0.23}$ \\ 
WISEJ2102-44 & T9 & 581$^{+18}_{-81}$ & 4.43$^{+0.77}_{-0.63}$ & 0.0$^{+0.0}_{-0.0}$ & 0.5$^{+0.0}_{-0.0}$ & \nodata & \nodata & \nodata & 0.98$^{+0.24}_{-0.25}$ \\ 
WISEJ2209+27 & Y0 & 470$^{+29}_{-70}$ & 4.35$^{+0.45}_{-0.55}$ & 0.0$^{+0.0}_{-0.0}$ & 0.5$^{+0.0}_{-0.0}$ & \nodata & \nodata & \nodata & 1.17$^{+0.19}_{-0.26}$ \\ 
WISEJ0359-54 & Y0 & 440$^{+59}_{-40}$ & 4.62$^{+0.98}_{-0.82}$ & 0.0$^{+0.0}_{-0.0}$ & 0.5$^{+0.0}_{-0.0}$ & \nodata & \nodata & \nodata & 1.09$^{+0.23}_{-0.20}$ \\ 
WISEJ0734-71 & Y0 & 450$^{+49}_{-50}$ & 4.58$^{+0.62}_{-0.78}$ & 0.0$^{+0.0}_{-0.0}$ & 0.5$^{+0.0}_{-0.0}$ & \nodata & \nodata & \nodata & 1.09$^{+0.24}_{-0.19}$ \\ 
WISEJ1206+84 & Y0 & 480$^{+42}_{-80}$ & 4.34$^{+0.46}_{-0.54}$ & 0.0$^{+0.0}_{-0.0}$ & 0.5$^{+0.0}_{-0.0}$ & \nodata & \nodata & \nodata & 1.11$^{+0.23}_{-0.30}$ \\ 
WISEPCJ2056+14 & Y0 & 566$^{+74}_{-66}$ & 4.26$^{+0.74}_{-0.66}$ & 0.0$^{+0.0}_{-0.0}$ & 0.5$^{+0.0}_{-0.0}$ & \nodata & \nodata & \nodata & 1.14$^{+0.23}_{-0.26}$ \\ 
WISEJ0825+28 & Y0.5 & 471$^{+28}_{-71}$ & 4.28$^{+0.42}_{-0.50}$ & 0.0$^{+0.0}_{-0.0}$ & 0.5$^{+0.0}_{-0.0}$ & \nodata & \nodata & \nodata & 1.21$^{+0.15}_{-0.14}$ \\ 
WISEPCJ1405+55 & Y0.5 & 503$^{+96}_{-103}$ & 4.36$^{+0.84}_{-0.76}$ & 0.0$^{+0.0}_{-0.0}$ & 0.5$^{+0.0}_{-0.0}$ & \nodata & \nodata & \nodata & 1.16$^{+0.19}_{-0.25}$ \\ 
WISEJ0535-75 & Y1 & 436$^{+63}_{-36}$ & 4.46$^{+0.56}_{-0.66}$ & 0.0$^{+0.0}_{-0.0}$ & 0.5$^{+0.0}_{-0.0}$ & \nodata & \nodata & \nodata & 1.15$^{+0.20}_{-0.24}$ \\ 
WISEPAJ1541-22 & Y1 & 520$^{+79}_{-110}$ & 4.23$^{+0.57}_{-0.63}$ & 0.0$^{+0.0}_{-0.0}$ & 0.5$^{+0.0}_{-0.0}$ & \nodata & \nodata & \nodata & 1.17$^{+0.19}_{-0.26}$ \\ 
CWISEPJ1047+54 & Y1 & 383$^{+16}_{-83}$ & 4.48$^{+0.99}_{-0.88}$ & 0.0$^{+0.0}_{-0.0}$ & 0.5$^{+0.0}_{-0.0}$ & \nodata & \nodata & \nodata & 1.07$^{+0.22}_{-0.21}$ \\ 
WISEAJ2354+02 & Y1 & 424$^{+75}_{-24}$ & 4.45$^{+0.75}_{-0.65}$ & 0.0$^{+0.0}_{-0.0}$ & 0.5$^{+0.0}_{-0.0}$ & \nodata & \nodata & \nodata & 1.16$^{+0.18}_{-0.25}$ \\ 
CWISEPJ1446-23 & Y1 & 396$^{+3}_{-83}$ & 4.34$^{+0.53}_{-0.74}$ & 0.0$^{+0.0}_{-0.0}$ & 0.5$^{+0.0}_{-0.0}$ & \nodata & \nodata & \nodata & 1.12$^{+0.22}_{-0.24}$ \\ 
\hline
\multicolumn{10}{c}{\textbf{Sonora Bobcat}} \\
\hline
SDSSJ1624+00 & T6 & 786$^{+13}_{-36}$ & 4.76$^{+0.24}_{-0.76}$ & 0.05$^{+0.45}_{-0.55}$ & 0.95$^{+0.05}_{-0.14}$ & \nodata & \nodata & \nodata & 0.99$^{+0.09}_{-0.09}$ \\ 
WISEJ1501-40 & T6 & 731$^{+68}_{-31}$ & 4.63$^{+0.87}_{-0.88}$ & 0.03$^{+0.03}_{-0.03}$ & 1.05$^{+0.05}_{-0.05}$ & \nodata & \nodata & \nodata & 0.84$^{+0.09}_{-0.09}$ \\ 
SDSSpJ1346-00 & T6.5 & 781$^{+18}_{-31}$ & 4.66$^{+0.51}_{-0.66}$ & 0.01$^{+0.49}_{-0.51}$ & 0.99$^{+0.01}_{-0.01}$ & \nodata & \nodata & \nodata & 0.94$^{+0.08}_{-0.08}$ \\ 
ULASJ1029+09 & T8 & 652$^{+47}_{-77}$ & 4.22$^{+0.78}_{-0.47}$ & 0.07$^{+0.43}_{-0.07}$ & 1.00$^{+0.00}_{-0.00}$ & \nodata & \nodata & \nodata & 0.84$^{+0.14}_{-0.11}$ \\ 
WISEJ0247+37 & T8 & 544$^{+55}_{-69}$ & 4.59$^{+0.91}_{-0.59}$ & 0.18$^{+0.32}_{-0.18}$ & 1.00$^{+0.00}_{-0.00}$ & \nodata & \nodata & \nodata & 0.91$^{+0.10}_{-0.12}$ \\ 
WISEJ0430+46 & T8 & 534$^{+40}_{-34}$ & 4.82$^{+0.68}_{-0.82}$ & -0.13$^{+0.13}_{-0.37}$ & 0.93$^{+0.07}_{-0.43}$ & \nodata & \nodata & \nodata & 0.96$^{+0.12}_{-0.17}$ \\ 
WISEPAJ1959-33 & T8 & 742$^{+58}_{-42}$ & 4.38$^{+0.62}_{-0.63}$ & 0.11$^{+0.39}_{-0.11}$ & 1.02$^{+0.02}_{-0.02}$ & \nodata & \nodata & \nodata & 0.89$^{+0.10}_{-0.14}$ \\ 
WISEPAJ0313+78 & T8.5 & 651$^{+83}_{-76}$ & 4.43$^{+0.57}_{-0.68}$ & 0.21$^{+0.29}_{-0.21}$ & 1.02$^{+0.02}_{-0.02}$ & \nodata & \nodata & \nodata & 1.09$^{+0.24}_{-0.24}$ \\ 
WISEAJ2159-48 & T9 & 508$^{+41}_{-58}$ & 4.70$^{+0.80}_{-0.70}$ & 0.09$^{+0.41}_{-0.09}$ & 1.06$^{+0.20}_{-0.06}$ & \nodata & \nodata & \nodata & 0.92$^{+0.11}_{-0.14}$ \\ 
WISEJ2102-44 & T9 & 572$^{+27}_{-47}$ & 4.30$^{+0.70}_{-0.55}$ & 0.19$^{+0.31}_{-0.19}$ & 0.99$^{+0.01}_{-0.01}$ & \nodata & \nodata & \nodata & 1.00$^{+0.18}_{-0.20}$ \\ 
WISEJ2209+27 & Y0 & 441$^{+33}_{-41}$ & 4.21$^{+0.54}_{-0.46}$ & 0.11$^{+0.39}_{-0.61}$ & 1.01$^{+0.01}_{-0.01}$ & \nodata & \nodata & \nodata & 1.13$^{+0.21}_{-0.22}$ \\ 
WISEJ0359-54 & Y0 & 442$^{+37}_{-42}$ & 4.25$^{+0.75}_{-0.50}$ & 0.09$^{+0.41}_{-0.59}$ & 1.02$^{+0.02}_{-0.02}$ & \nodata & \nodata & \nodata & 0.96$^{+0.24}_{-0.22}$ \\ 
WISEJ0734-71 & Y0 & 451$^{+48}_{-29}$ & 4.26$^{+0.74}_{-0.51}$ & 0.19$^{+0.31}_{-0.19}$ & 1.03$^{+0.03}_{-0.03}$ & \nodata & \nodata & \nodata & 0.96$^{+0.21}_{-0.23}$ \\ 
WISEJ1206+84 & Y0 & 462$^{+37}_{-37}$ & 4.23$^{+0.52}_{-0.48}$ & 0.18$^{+0.32}_{-0.18}$ & 1.02$^{+0.02}_{-0.02}$ & \nodata & \nodata & \nodata & 1.05$^{+0.21}_{-0.21}$ \\ 
WISEPCJ2056+14 & Y0 & 539$^{+49}_{-39}$ & 4.38$^{+0.62}_{-0.63}$ & -0.01$^{+0.51}_{-0.49}$ & 0.98$^{+0.02}_{-0.02}$ & \nodata & \nodata & \nodata & 1.16$^{+0.20}_{-0.25}$ \\ 
WISEJ0825+28 & Y0.5 & 446$^{+53}_{-46}$ & 4.21$^{+0.54}_{-0.46}$ & 0.07$^{+0.43}_{-0.57}$ & 1.01$^{+0.01}_{-0.01}$ & \nodata & \nodata & \nodata & 1.13$^{+0.23}_{-0.23}$ \\ 
WISEPCJ1405+55 & Y0.5 & 476$^{+48}_{-55}$ & 4.36$^{+0.64}_{-0.54}$ & 0.06$^{+0.44}_{-0.56}$ & 1.02$^{+0.02}_{-0.02}$ & \nodata & \nodata & \nodata & 1.17$^{+0.21}_{-0.22}$ \\ 
WISEJ0535-75 & Y1 & 426$^{+48}_{-51}$ & 4.20$^{+0.55}_{-0.45}$ & 0.05$^{+0.45}_{-0.55}$ & 1.00$^{+0.00}_{-0.00}$ & \nodata & \nodata & \nodata & 1.06$^{+0.23}_{-0.23}$ \\ 
WISEPAJ1541-22 & Y1 & 482$^{+67}_{-57}$ & 4.31$^{+0.69}_{-0.56}$ & 0.04$^{+0.46}_{-0.54}$ & 1.02$^{+0.02}_{-0.02}$ & \nodata & \nodata & \nodata & 1.14$^{+0.23}_{-0.24}$ \\ 
CWISEPJ1047+54 & Y1 & 406$^{+18}_{-31}$ & 4.10$^{+0.40}_{-0.35}$ & 0.21$^{+0.29}_{-0.21}$ & 1.01$^{+0.01}_{-0.01}$ & \nodata & \nodata & \nodata & 0.89$^{+0.15}_{-0.14}$ \\ 
WISEAJ2354+02 & Y1 & 417$^{+32}_{-42}$ & 4.21$^{+0.54}_{-0.46}$ & 0.13$^{+0.37}_{-0.23}$ & 1.00$^{+0.00}_{-0.00}$ & \nodata & \nodata & \nodata & 1.08$^{+0.24}_{-0.23}$ \\ 
CWISEPJ1446-23 & Y1 & 396$^{+28}_{-46}$ & 4.04$^{+0.46}_{-0.29}$ & 0.20$^{+0.30}_{-0.20}$ & 1.00$^{+0.00}_{-0.00}$ & \nodata & \nodata & \nodata & 1.05$^{+0.23}_{-0.22}$ \\
\hline
\multicolumn{10}{c}{\textbf{Sonora Elf Owl}} \\
\hline
SDSSJ1624+00 & T6 & 759$^{+40}_{-59}$ & 4.25$^{+1.00}_{-0.76}$ & -0.55$^{+0.55}_{-0.45}$ & 0.89$^{+0.61}_{-0.39}$ & 6.67$^{+2.33}_{-2.67}$ & \nodata & \nodata & 1.01$^{+0.14}_{-0.11}$ \\ 
WISEJ1501-40 & T6 & 741$^{+58}_{-91}$ & 4.61$^{+0.89}_{-0.86}$ & -0.34$^{+0.84}_{-0.66}$ & 1.01$^{+0.49}_{-0.51}$ & 6.68$^{+2.32}_{-2.68}$ & \nodata & \nodata & 0.86$^{+0.12}_{-0.12}$ \\ 
SDSSpJ1346-00 & T6.5 & 753$^{+46}_{-53}$ & 4.32$^{+0.93}_{-0.83}$ & -0.47$^{+0.47}_{-0.53}$ & 0.91$^{+0.59}_{-0.41}$ & 6.80$^{+2.20}_{-2.80}$ & \nodata & \nodata & 0.96$^{+0.14}_{-0.16}$ \\ 
ULASJ1029+09 & T8 & 664$^{+85}_{-89}$ & 4.64$^{+0.86}_{-0.89}$ & -0.11$^{+0.81}_{-0.89}$ & 1.07$^{+0.43}_{-0.57}$ & 6.40$^{+2.60}_{-2.40}$ & \nodata & \nodata & 0.88$^{+0.14}_{-0.14}$ \\ 
WISEJ0247+37 & T8 & 618$^{+81}_{-93}$ & 4.86$^{+0.64}_{-0.86}$ & 0.16$^{+0.54}_{-0.66}$ & 1.04$^{+0.46}_{-0.54}$ & 6.53$^{+2.47}_{-2.53}$ & \nodata & \nodata & 0.87$^{+0.13}_{-0.13}$ \\ 
WISEJ0430+46 & T8 & 613$^{+86}_{-88}$ & 4.86$^{+0.64}_{-0.86}$ & 0.03$^{+0.67}_{-1.03}$ & 0.90$^{+0.60}_{-0.40}$ & 6.49$^{+2.51}_{-2.49}$ & \nodata & \nodata & 0.89$^{+0.14}_{-0.14}$ \\ 
WISEPAJ1959-33 & T8 & 718$^{+81}_{-68}$ & 4.41$^{+0.84}_{-0.92}$ & -0.41$^{+0.41}_{-0.59}$ & 0.86$^{+0.64}_{-0.36}$ & 6.50$^{+2.50}_{-2.50}$ & \nodata & \nodata & 0.98$^{+0.17}_{-0.18}$ \\ 
WISEPAJ0313+78 & T8.5 & 664$^{+85}_{-89}$ & 4.56$^{+0.94}_{-0.81}$ & -0.35$^{+0.35}_{-0.65}$ & 0.76$^{+0.24}_{-0.26}$ & 5.75$^{+2.25}_{-3.75}$ & \nodata & \nodata & 1.10$^{+0.21}_{-0.20}$ \\ 
WISEAJ2159-48 & T9 & 599$^{+100}_{-74}$ & 4.93$^{+0.57}_{-0.68}$ & 0.28$^{+0.72}_{-0.78}$ & 1.00$^{+0.50}_{-0.50}$ & 6.52$^{+2.48}_{-2.52}$ & \nodata & \nodata & 0.87$^{+0.14}_{-0.13}$ \\ 
WISEJ2102-44 & T9 & 639$^{+110}_{-89}$ & 4.67$^{+0.83}_{-0.92}$ & -0.25$^{+0.75}_{-0.75}$ & 0.95$^{+0.55}_{-0.45}$ & 6.08$^{+2.92}_{-4.08}$ & \nodata & \nodata & 0.92$^{+0.17}_{-0.16}$ \\ 
WISEJ2209+27 & Y0 & 544$^{+30}_{-44}$ & 4.64$^{+0.61}_{-0.89}$ & 0.27$^{+0.73}_{-0.77}$ & 0.87$^{+0.63}_{-0.37}$ & 5.65$^{+2.35}_{-3.65}$ & \nodata & \nodata & 0.97$^{+0.18}_{-0.18}$ \\ 
WISEJ0359-54 & Y0 & 551$^{+48}_{-51}$ & 4.79$^{+0.71}_{-0.79}$ & 0.53$^{+0.47}_{-0.53}$ & 1.00$^{+0.50}_{-0.50}$ & 6.56$^{+2.44}_{-2.56}$ & \nodata & \nodata & 0.84$^{+0.10}_{-0.10}$ \\ 
WISEJ0734-71 & Y0 & 560$^{+39}_{-60}$ & 4.88$^{+0.62}_{-0.63}$ & 0.50$^{+0.50}_{-0.50}$ & 1.07$^{+0.43}_{-0.57}$ & 6.41$^{+2.59}_{-2.41}$ & \nodata & \nodata & 0.85$^{+0.11}_{-0.11}$ \\ 
WISEJ1206+84 & Y0 & 574$^{+75}_{-74}$ & 4.75$^{+0.75}_{-0.75}$ & 0.50$^{+0.50}_{-0.50}$ & 1.07$^{+0.43}_{-0.57}$ & 6.45$^{+2.55}_{-2.45}$ & \nodata & \nodata & 0.88$^{+0.14}_{-0.14}$ \\ 
WISEPCJ2056+14 & Y0 & 619$^{+80}_{-94}$ & 4.49$^{+0.76}_{-1.00}$ & -0.16$^{+0.66}_{-0.34}$ & 0.83$^{+0.67}_{-0.33}$ & 5.46$^{+2.54}_{-3.46}$ & \nodata & \nodata & 1.12$^{+0.22}_{-0.21}$ \\ 
WISEJ0825+28 & Y0.5 & 552$^{+47}_{-52}$ & 4.56$^{+0.69}_{-0.81}$ & 0.25$^{+0.45}_{-0.75}$ & 0.89$^{+0.61}_{-0.39}$ & 5.42$^{+2.58}_{-3.42}$ & \nodata & \nodata & 1.02$^{+0.21}_{-0.20}$ \\ 
WISEPCJ1405+55 & Y0.5 & 576$^{+73}_{-76}$ & 4.68$^{+0.82}_{-0.93}$ & -0.16$^{+0.66}_{-0.84}$ & 0.83$^{+0.67}_{-0.33}$ & 5.70$^{+2.30}_{-3.70}$ & \nodata & \nodata & 1.02$^{+0.23}_{-0.21}$ \\ 
WISEJ0535-75 & Y1 & 544$^{+30}_{-44}$ & 4.78$^{+0.72}_{-0.78}$ & 0.41$^{+0.59}_{-0.41}$ & 0.92$^{+0.58}_{-0.42}$ & 5.97$^{+3.03}_{-3.97}$ & \nodata & \nodata & 0.86$^{+0.12}_{-0.12}$ \\ 
WISEPAJ1541-22 & Y1 & 575$^{+74}_{-75}$ & 4.48$^{+0.77}_{-0.99}$ & -0.02$^{+0.52}_{-0.48}$ & 0.86$^{+0.64}_{-0.36}$ & 5.08$^{+2.92}_{-3.08}$ & \nodata & \nodata & 1.12$^{+0.22}_{-0.22}$ \\ 
CWISEPJ1047+54 & Y1 & 524$^{+25}_{-24}$ & 4.90$^{+0.60}_{-0.65}$ & 0.61$^{+0.39}_{-0.61}$ & 1.23$^{+1.27}_{-0.73}$ & 6.58$^{+2.42}_{-2.58}$ & \nodata & \nodata & 0.82$^{+0.08}_{-0.08}$ \\ 
WISEAJ2354+02 & Y1 & 535$^{+39}_{-35}$ & 4.79$^{+0.71}_{-0.79}$ & 0.44$^{+0.56}_{-0.44}$ & 0.92$^{+0.58}_{-0.42}$ & 5.89$^{+3.11}_{-3.89}$ & \nodata & \nodata & 0.86$^{+0.11}_{-0.11}$ \\ 
CWISEPJ1446-23 & Y1 & 521$^{+29}_{-21}$ & 4.77$^{+0.73}_{-0.77}$ & 0.60$^{+0.40}_{-0.60}$ & 1.20$^{+0.30}_{-0.70}$ & 6.06$^{+2.94}_{-2.06}$ & \nodata & \nodata & 0.82$^{+0.08}_{-0.08}$ \\ 
\hline
\multicolumn{10}{c}{\textbf{Lacy \& Burrows}} \\
\hline
SDSSJ1624+00 & T6 & 586$^{+13}_{-11}$ & 4.15$^{+0.6}_{-0.65}$ & -0.37$^{+0.37}_{-0.13}$ & \nodata & 5.19$^{+0.81}_{-0.81}$ & \nodata & \nodata & 1.31$^{+0.06}_{-0.06}$ \\ 
WISEJ1501-40 & T6 & 554$^{+45}_{-29}$ & 4.15$^{+0.6}_{-0.65}$ & -0.16$^{+0.16}_{-0.34}$ & \nodata & 5.65$^{+0.35}_{-0.35}$ & \nodata & \nodata & 1.10$^{+0.22}_{-0.22}$ \\ 
SDSSpJ1346-00 & T6.5 & 578$^{+21}_{-28}$ & 4.08$^{+0.67}_{-0.58}$ & -0.26$^{+0.26}_{-0.24}$ & \nodata & 5.79$^{+0.21}_{-0.21}$ & \nodata & \nodata & 1.27$^{+0.11}_{-0.11}$ \\ 
ULASJ1029+09 & T8 & 531$^{+68}_{-85}$ & 4.05$^{+0.45}_{-0.55}$ & -0.07$^{+0.57}_{-0.43}$ & \nodata & 5.52$^{+0.48}_{-0.48}$ & \nodata & \nodata & 1.04$^{+0.26}_{-0.17}$ \\ 
WISEJ0247+37 & T8 & 500$^{+49}_{-75}$ & 4.10$^{+0.40}_{-0.35}$ & -0.17$^{+0.17}_{-0.33}$ & \nodata & 5.24$^{+0.76}_{-0.76}$ & \nodata & \nodata & 0.97$^{+0.22}_{-0.14}$ \\ 
WISEJ0430+46 & T8 & 502$^{+47}_{-52}$ & 4.15$^{+0.60}_{-0.40}$ & -0.33$^{+0.33}_{-0.17}$ & \nodata & 5.07$^{+0.93}_{-0.93}$ & \nodata & \nodata & 0.99$^{+0.22}_{-0.16}$ \\ 
WISEPAJ1959-33 & T8 & 575$^{+24}_{-25}$ & 4.07$^{+0.68}_{-0.57}$ & -0.24$^{+0.24}_{-0.26}$ & \nodata & 5.75$^{+0.25}_{-0.25}$ & \nodata & \nodata & 1.25$^{+0.11}_{-0.10}$ \\ 
WISEPAJ0313+78 & T8.5 & 576$^{+23}_{-26}$ & 4.11$^{+0.64}_{-0.61}$ & -0.27$^{+0.27}_{-0.23}$ & \nodata & 5.01$^{+0.99}_{-5.01}$ & \nodata & \nodata & 1.29$^{+0.08}_{-0.09}$ \\ 
WISEAJ2159-48 & T9 & 478$^{+46}_{-53}$ & 4.19$^{+0.56}_{-0.44}$ & -0.16$^{+0.66}_{-0.34}$ & \nodata & 5.12$^{+0.88}_{-0.88}$ & \nodata & \nodata & 0.95$^{+0.15}_{-0.16}$ \\ 
WISEJ2102-44 & T9 & 532$^{+67}_{-57}$ & 4.10$^{+0.65}_{-0.60}$ & -0.06$^{+0.56}_{-0.44}$ & \nodata & 5.28$^{+0.72}_{-0.72}$ & \nodata & \nodata & 1.05$^{+0.25}_{-0.18}$ \\ 
WISEJ2209+27 & Y0 & 410$^{+39}_{-35}$ & 3.96$^{+0.54}_{-0.46}$ & -0.21$^{+0.21}_{-0.29}$ & \nodata & 4.86$^{+1.14}_{-4.86}$ & \nodata & \nodata & 1.18$^{+0.19}_{-0.25}$ \\ 
WISEJ0359-54 & Y0 & 411$^{+38}_{-36}$ & 3.89$^{+0.11}_{-0.39}$ & -0.31$^{+0.31}_{-0.19}$ & \nodata & 5.50$^{+0.50}_{-0.50}$ & \nodata & \nodata & 0.96$^{+0.21}_{-0.17}$ \\ 
WISEJ0734-71 & Y0 & 422$^{+27}_{-47}$ & 3.89$^{+0.36}_{-0.39}$ & -0.26$^{+0.26}_{-0.24}$ & \nodata & 5.56$^{+0.44}_{-0.44}$ & \nodata & \nodata & 0.97$^{+0.21}_{-0.15}$ \\ 
WISEJ1206+84 & Y0 & 416$^{+33}_{-41}$ & 3.86$^{+0.14}_{-0.36}$ & -0.13$^{+0.13}_{-0.37}$ & \nodata & 5.52$^{+0.48}_{-0.48}$ & \nodata & \nodata & 1.10$^{+0.23}_{-0.23}$ \\ 
WISEPCJ2056+14 & Y0 & 516$^{+58}_{-66}$ & 4.12$^{+0.63}_{-0.62}$ & -0.11$^{+0.61}_{-0.39}$ & \nodata & 5.43$^{+0.57}_{-0.57}$ & \nodata & \nodata & 1.26$^{+0.13}_{-0.11}$ \\ 
WISEJ0825+28 & Y0.5 & 415$^{+34}_{-40}$ & 3.98$^{+0.52}_{-0.23}$ & -0.26$^{+0.26}_{-0.24}$ & \nodata & 4.46$^{+1.54}_{-4.46}$ & \nodata & \nodata & 1.22$^{+0.16}_{-0.23}$ \\ 
WISEPCJ1405+55 & Y0.5 & 448$^{+26}_{-23}$ & 4.22$^{+0.68}_{-0.47}$ & -0.20$^{+0.20}_{-0.30}$ & \nodata & 3.52$^{+2.48}_{-3.52}$ & \nodata & \nodata & 1.26$^{+0.12}_{-0.16}$ \\ 
WISEJ0535-75 & Y1 & 405$^{+44}_{-30}$ & 4.00$^{+0.50}_{-0.50}$ & -0.30$^{+0.30}_{-0.20}$ & \nodata & 4.18$^{+1.82}_{-4.18}$ & \nodata & \nodata & 1.07$^{+0.24}_{-0.19}$ \\ 
WISEPAJ1541-22 & Y1 & 456$^{+43}_{-56}$ & 4.10$^{+0.40}_{-0.35}$ & -0.20$^{+0.20}_{-0.30}$ & \nodata & 4.47$^{+1.53}_{-4.47}$ & \nodata & \nodata & 1.26$^{+0.12}_{-0.16}$ \\ 
CWISEPJ1047+54 & Y1 & 349$^{+31}_{-49}$ & 4.10$^{+0.40}_{-0.35}$ & 0.15$^{+0.35}_{-0.15}$ & \nodata & 5.39$^{+0.61}_{-0.61}$ & \nodata & \nodata & 1.00$^{+0.19}_{-0.19}$ \\ 
WISEAJ2354+02 & Y1 & 390$^{+34}_{-40}$ & 4.07$^{+0.43}_{-0.57}$ & -0.19$^{+0.19}_{-0.31}$ & \nodata & 4.60$^{+1.40}_{-4.60}$ & \nodata & \nodata & 1.10$^{+0.21}_{-0.22}$ \\ 
CWISEPJ1446-23 & Y1 & 355$^{+42}_{-30}$ & 4.02$^{+0.48}_{-0.52}$ & 0.00$^{+0.50}_{-0.50}$ & \nodata & 5.05$^{+0.95}_{-0.95}$ & \nodata & \nodata & 1.09$^{+0.24}_{-0.24}$ \\ 
\hline
\multicolumn{10}{c}{\textbf{ATMO2020++ without PH3}} \\
\hline
SDSSJ1624+00 & T6 & 761$^{+38}_{-61}$ & 4.45$^{+0.55}_{-0.45}$ & -0.35$^{+0.65}_{-0.65}$ & \nodata & 6.11$^{+0.89}_{-1.71}$ & \nodata & 2.91$^{+1.04}_{-2.51}$ & 0.97$^{+0.10}_{-0.14}$ \\ 
WISEJ1501-40 & T6 & 608$^{+91}_{-108}$ & 4.37$^{+0.63}_{-0.87}$ & -0.19$^{+0.49}_{-0.31}$ & \nodata & 6.25$^{+1.75}_{-1.25}$ & \nodata & 2.79$^{+1.16}_{-2.67}$ & 0.94$^{+0.08}_{-0.22}$ \\ 
SDSSpJ1346-00 & T6.5 & 711$^{+88}_{-111}$ & 4.21$^{+0.79}_{-0.71}$ & -0.14$^{+0.44}_{-0.36}$ & \nodata & 6.58$^{+1.42}_{-1.58}$ & \nodata & 1.89$^{+2.06}_{-1.77}$ & 0.93$^{+0.09}_{-0.12}$ \\ 
ULASJ1029+09 & T8 & 537$^{+62}_{-87}$ & 4.33$^{+0.67}_{-0.83}$ & -0.12$^{+0.42}_{-0.38}$ & \nodata & 6.34$^{+1.66}_{-1.34}$ & \nodata & 2.28$^{+1.67}_{-2.16}$ & 1.00$^{+0.14}_{-0.18}$ \\ 
WISEJ0247+37 & T8 & 474$^{+25}_{-74}$ & 4.76$^{+0.74}_{-0.76}$ & -0.17$^{+0.47}_{-0.33}$ & \nodata & 5.49$^{+1.51}_{-1.49}$ & \nodata & 4.67$^{+7.73}_{-4.27}$ & 1.02$^{+0.18}_{-0.13}$ \\ 
WISEJ0430+46 & T8 & 480$^{+19}_{-48}$ & 4.98$^{+0.52}_{-0.48}$ & -0.24$^{+0.54}_{-0.26}$ & \nodata & 5.04$^{+0.96}_{-1.04}$ & \nodata & 6.43$^{+6.07}_{-5.18}$ & 1.03$^{+0.18}_{-0.13}$ \\ 
WISEPAJ1959-33 & T8 & 653$^{+46}_{-53}$ & 4.24$^{+0.76}_{-0.74}$ & -0.14$^{+0.44}_{-0.36}$ & \nodata & 6.52$^{+1.48}_{-1.52}$ & \nodata & 2.01$^{+1.94}_{-1.89}$ & 0.96$^{+0.12}_{-0.17}$ \\ 
WISEPAJ0313+78 & T8.5 & 587$^{+101}_{-87}$ & 4.34$^{+0.66}_{-0.84}$ & -0.36$^{+0.66}_{-0.64}$ & \nodata & 6.32$^{+1.39}_{-1.32}$ & \nodata & 2.29$^{+1.66}_{-2.17}$ & 1.24$^{+0.15}_{-0.24}$ \\ 
WISEAJ2159-48 & T9 & 454$^{+45}_{-54}$ & 4.89$^{+0.61}_{-0.69}$ & -0.21$^{+0.51}_{-0.79}$ & \nodata & 5.22$^{+0.78}_{-1.22}$ & \nodata & 5.73$^{+6.77}_{-4.98}$ & 1.00$^{+0.20}_{-0.11}$ \\ 
WISEJ2102-44 & T9 & 507$^{+92}_{-57}$ & 4.65$^{+0.35}_{-0.65}$ & -0.09$^{+0.39}_{-0.41}$ & \nodata & 5.70$^{+1.30}_{-0.70}$ & \nodata & 3.24$^{+0.71}_{-2.84}$ & 1.05$^{+0.20}_{-0.13}$ \\ 
WISEJ2209+27 & Y0 & 428$^{+71}_{-28}$ & 4.29$^{+0.34}_{-0.79}$ & -0.08$^{+0.38}_{-0.42}$ & \nodata & 6.42$^{+1.58}_{-1.42}$ & \nodata & 1.58$^{+0.41}_{-1.46}$ & 1.20$^{+0.16}_{-0.19}$ \\ 
WISEJ0359-54 & Y0 & 428$^{+21}_{-78}$ & 4.54$^{+0.46}_{-0.54}$ & -0.21$^{+0.51}_{-0.79}$ & \nodata & 5.91$^{+1.09}_{-1.16}$ & \nodata & 2.94$^{+1.01}_{-2.54}$ & 0.92$^{+0.27}_{-0.21}$ \\ 
WISEJ0734-71 & Y0 & 434$^{+15}_{-34}$ & 4.61$^{+0.68}_{-0.61}$ & -0.12$^{+0.42}_{-0.88}$ & \nodata & 5.77$^{+1.23}_{-1.77}$ & \nodata & 3.26$^{+5.78}_{-2.86}$ & 0.94$^{+0.24}_{-0.23}$ \\ 
WISEJ1206+84 & Y0 & 452$^{+47}_{-52}$ & 4.18$^{+0.32}_{-0.18}$ & 0.02$^{+0.28}_{-0.02}$ & \nodata & 6.64$^{+0.36}_{-0.64}$ & \nodata & 1.18$^{+0.07}_{-0.78}$ & 0.98$^{+0.21}_{-0.20}$ \\ 
WISEPCJ2056+14 & Y0 & 523$^{+76}_{-73}$ & 4.16$^{+0.46}_{-0.66}$ & -0.10$^{+0.40}_{-0.40}$ & \nodata & 6.69$^{+1.31}_{-1.69}$ & \nodata & 1.25$^{+0.64}_{-1.13}$ & 1.25$^{+0.15}_{-0.24}$ \\ 
WISEJ0825+28 & Y0.5 & 437$^{+47}_{-37}$ & 4.13$^{+0.37}_{-0.63}$ & -0.13$^{+0.43}_{-0.37}$ & \nodata & 6.74$^{+1.26}_{-0.74}$ & \nodata & 1.11$^{+0.14}_{-0.99}$ & 1.24$^{+0.15}_{-0.23}$ \\ 
WISEPCJ1405+55 & Y0.5 & 458$^{+41}_{-58}$ & 4.61$^{+0.39}_{-0.61}$ & -0.19$^{+0.49}_{-0.31}$ & \nodata & 5.77$^{+1.23}_{-0.77}$ & \nodata & 2.94$^{+1.01}_{-2.54}$ & 1.24$^{+0.13}_{-0.23}$ \\ 
WISEJ0535-75 & Y1 & 414$^{+35}_{-64}$ & 4.57$^{+0.43}_{-0.57}$ & -0.25$^{+0.55}_{-0.75}$ & \nodata & 5.85$^{+1.15}_{-1.19}$ & \nodata & 3.10$^{+0.85}_{-2.70}$ & 1.09$^{+0.24}_{-0.20}$ \\ 
WISEPAJ1541-22 & Y1 & 483$^{+16}_{-33}$ & 4.21$^{+0.79}_{-0.71}$ & -0.15$^{+0.45}_{-0.35}$ & \nodata & 6.57$^{+1.43}_{-1.57}$ & \nodata & 1.44$^{+2.51}_{-1.32}$ & 1.23$^{+0.15}_{-0.22}$ \\ 
CWISEPJ1047+54 & Y1 & 361$^{+38}_{-11}$ & 4.22$^{+0.78}_{-0.72}$ & -0.14$^{+0.44}_{-0.36}$ & \nodata & 6.55$^{+1.45}_{-1.55}$ & \nodata & 2.06$^{+1.89}_{-1.94}$ & 0.95$^{+0.25}_{-0.23}$ \\ 
WISEAJ2354+02 & Y1 & 408$^{+41}_{-58}$ & 4.61$^{+0.89}_{-0.61}$ & -0.20$^{+0.50}_{-0.80}$ & \nodata & 5.78$^{+1.22}_{-1.78}$ & \nodata & 3.50$^{+9.00}_{-3.10}$ & 1.08$^{+0.24}_{-0.28}$ \\ 
CWISEPJ1446-23 & Y1 & 381$^{+68}_{-31}$ & 4.22$^{+0.78}_{-0.72}$ & -0.20$^{+0.27}_{-0.30}$ & \nodata & 6.57$^{+1.43}_{-1.57}$ & \nodata & 1.83$^{+2.12}_{-1.71}$ & 1.00$^{+0.27}_{-0.29}$ \\ 
\hline
%\begin{tablenotes}
%\small
%\item[*] TBD
%\end{tablenotes}
\end{longtable}

\end{appendix}

\end{document}